\documentclass[iop]{emulateapj}

\newcommand{\mytilde}{\raise.19ex\hbox{$\scriptstyle\sim$}}
\newcommand{\symvec}[1]{\mbox{\boldmath${#1}$}}

\shorttitle{Cosmic Shear Measurement in DLS}
\shortauthors{Jee et al.}

\begin{document}

\title{COSMIC SHEAR RESULTS FROM DEEP LENS SURVEY - I:  JOINT CONSTRAINTS ON $\Omega_M$ and $\sigma_8$ \\ WITH A TWO-DIMENSIONAL ANALYSIS} 

\author{M. JAMES JEE\altaffilmark{1}, J. ANTHONY TYSON\altaffilmark{1}, MICHAEL D.  SCHNEIDER\altaffilmark{1, 2}, DAVID WITTMAN\altaffilmark{1}, SAMUEL SCHMIDT\altaffilmark{1}, 
STEFAN HILBERT\altaffilmark{3}}

\begin{abstract}
We present a cosmic shear study from the Deep Lens Survey (DLS), 
a deep {\it BVRz} multi-band imaging survey of five  4 sq. degree fields with
two National Optical Astronomy Observatory (NOAO) 4-meter telescopes at Kitt Peak and Cerro Tololo.
For both telescopes, the change of the point-spread-function (PSF) shape across the focal plane is complicated, 
and  the exposure-to-exposure variation of this position-dependent PSF change
is significant. We overcome this challenge by modeling the PSF
separately for individual exposures and CCDs with principal component analysis (PCA).
We find that stacking these PSFs reproduces the final PSF pattern on the mosaic image with high fidelity, and
the method successfully separates PSF-induced systematics from gravitational lensing effects.
We calibrate our shears and estimate the errors, utilizing an image simulator, which
generates sheared ground-based galaxy images from deep {\it Hubble Space Telescope} archival data with
a realistic atmospheric turbulence model.
For cosmological parameter constraints, we marginalize over shear calibration error, photometric redshift uncertainty, and the Hubble constant.
We use cosmology-dependent covariances for the Markov Chain Monte Carlo analysis and find that the role of
this varying covariance is critical in our parameter estimation.
Our current non-tomographic analysis alone constrains the $\Omega_M-\sigma_8$ likelihood
contour tightly, providing a joint constraint of $\Omega_M=0.262\pm0.051$ and $\sigma_8=0.868\pm0.071$.
We expect that a future DLS weak-lensing tomographic study will further tighten these constraints
because explicit treatment of the redshift dependence of cosmic shear more efficiently breaks the $\Omega_M-\sigma_8$ degeneracy.
Combining the current results with the Wilkinson Microwave Anisotropy Probe 7-year (WMAP7) likelihood data, we obtain $\Omega_M=0.278\pm0.018$ and $\sigma_8=0.815\pm0.020$.
\end{abstract}

\altaffiltext{1}{Department of Physics, University of California, Davis, One Shields Avenue, Davis, CA 95616}
\altaffiltext{2}{Lawrence Livermore National Laboratory, P.O. Box 808 L-210, Livermore, CA 94551}
\altaffiltext{3}{Kavli Institute of Particle Astrophysics and Cosmology (KIPAC), Stanford University, 452 Lomita Mall, Stanford, CA 94305, and
SLAC National Accelerator Laboratory, 2575 Sand Hill Road, M/S 29, Menlo Park, CA 94025
}


\section{INTRODUCTION \label{section_introduction}}

Weak gravitational lensing from large-scale structures in the universe, often called cosmic shear, allows one to address a number of 
critical issues in modern cosmology. Its application encompasses
the study of the universe's matter density and its fluctuation, probes of the footprints of non-Gaussianity in the
primordial density fluctuation, constraints on dark energy and its evolution, tests for modified gravity, etc.
The consensus on the critical role of cosmic shear studies triggered quite a few optical surveys  such as the
Canada-France-Hawaii-Telescope Legacy Survey (CFHT-LS; Hoekstra et al. 2006, Semboloni et al. 2006; Fu et al. 2008), 
the Red-sequence Cluster Survey (RCS; Hoekstra et al. 2002), the Cerro
Tololo Inter-American Observatory (CTIO) Lensing Survey (Jarvis et al. 2006), the Garching-Bonn Deep Survey (GaBoDS, Hetterscheidt et al. 2007), 
the VIRMOS-DESCART survey (VIRMOS, Van Waerbeke et al. 2005), the Deep Lens Survey (DLS, Tyson et al. 2001,  Wittman et al. 2006), etc.
The current surveys include the Dark Energy Survey (DES, The Dark Energy Survey Collaboration 2005), the KIlo-Degree Survey (KIDS; Verdoes Kleijn et al. 2011), 
 the Panoramic Survey Telesope and Rapid Response System (Pan-STARRS, Kaiser et al. 2010), etc.
Next-generation weak-lensing projects are the Euclid mission (Laureijs et al. 2010), the Wide Field Infrared Survey Telescope (WFIRST; Green et al. 2011), and the Large Synoptic Survey Telescope (LSST, LSST Science Collaborations et al. 2009).

Needless to say, great effort should be given to control of systematics
in both shear and photometric redshift measurements for these future surveys. 
The unprecedentedly small statistical errors will bring revolutionary advances to cosmology {\it only if }  
progress in shear calibration and control of catastrophic errors in photometric redshift estimation 
parallels the increase in statistical power.
The recent shear estimation challenges such as the Shear TEsting Programme (STEP, Massey et al. 2007; Heymans et al. 2006),
the GRavitational lEnsing Accuracy Testing (GREAT, Bridle et al. 2009; Kitching et al. 2012), etc. are 
concerted efforts to quantify bias in the current popular shear estimation methods and also to identify the
limitation of the current weak-lensing simulation methods. Similar efforts toward improvement of photometric 
redshift estimation, albeit less mature,
are also underway (e.g., Hildebrandt et al. 2010).

\begin{figure}
\includegraphics[width=8.8cm]{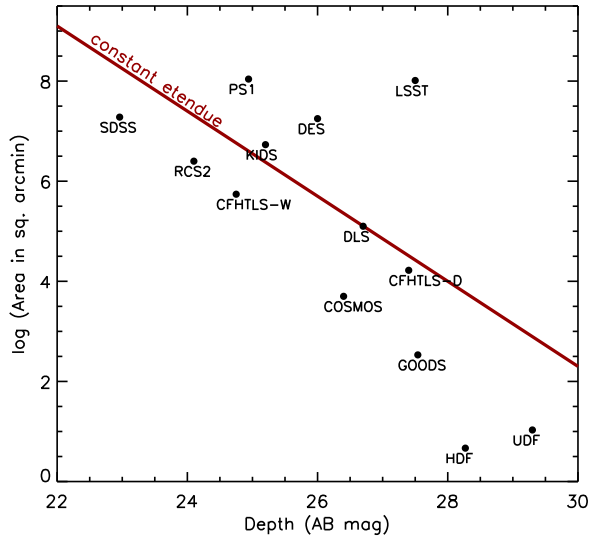}
\caption{Survey area and depth of various optical surveys. The red line represents the $A \Omega t=constant$ locus, where $A$, $\Omega$, and $t$
are the primary mirror area, field of view, and exposure time, respectively. DLS is the deepest optical survey to date among the current $\gtrsim10$ sq. degree surveys.
Depth is compared either in the {\it R} or {\it i} band. }
\label{fig_area_vs_depth}
\end{figure}

Both sky coverage and depth must be carefully balanced to maximize the scientific return from future cosmic shear surveys.
Large sky coverage is needed to minimize the contribution to the error from the sample variance. Deep
imaging is required to detect and measure the shapes of high redshift sources, which allows us to probe the
evolution of the cosmic structure over a significant fraction of the age of the universe.
The Deep Lens Survey (DLS; Tyson et al. 2001, Wittman et al. 2006) is designed as a precursor to these next generation cosmic shear surveys with emphasis on the
latter, reaching a mean source redshift of $z\sim1$ over 20 sq. degrees using the two
National Optical Astronomical Observatory (NOAO) Mayall and Blanco 4-meter telescopes.
Figure~\ref{fig_area_vs_depth} shows
the comparison of DLS sky coverage and depth with those of other optical surveys.
DLS is the deepest optical survey to date among the current $\gtrsim10$ sq. degree surveys.
Galaxy populations are dominated by faint blue galaxies when a survey
reaches or exceeds the depth of the DLS.  As no ground-based cosmic shear study with a comparable depth 
has been presented, the current cosmic shear analysis with the DLS is an important experiment, testing whether the shapes of the faint blue galaxy population smeared by atmospheric seeing can be reliably used for cosmic shear. We augment this experiment using image simulations with real galaxy images.
In addition, our seeing-matched photometry from the deep {\it BVRz} imaging 
in conjunction with a large spectroscopic sample allows us to stabilize our photometric redshift estimation and  to identify where
potential systematic errors lie in our results.
Reliable photometric redshifts are pivotal not only in the interpretation of the cosmic shear signal, but also in future application of the measurements to weak lensing tomography.

This paper is the first in a series of our DLS cosmic shear publications.
Here we mainly focus on the DLS systematics induced by the point-spread-function (PSF), 
the removal of the systematics with our principal component analysis (PCA) and ``StackFit'' methods, and
the two-dimensional (non-tomographic) analysis of the DLS cosmic shear signal.

The structure of this paper is as follows. In \textsection\ref{section_obs}, we describe our DLS data and analysis method including our detailed PSF modeling and 
shear calibration efforts.  The theoretical background of cosmic shear and our systematics control is presented in \textsection\ref{section_measurements}.
We discuss the study of cosmological parameter constraints in \textsection\ref{section_cosmology} and conclude in \textsection\ref{section_conclusion}.

\section{OBSERVATIONS \label{section_obs}}

\begin{deluxetable*}{cccccc}
\tabletypesize{\scriptsize}
\tablecaption{DLS Fields and Data}
\tablenum{1}
\tablehead{\colhead{Field Name} &  \colhead{RA} & \colhead{DEC}   &   \colhead{Median Seeing ($R$) } &  \colhead{ $\bar{z}_{source}$}  & \colhead{$n_{source}$}  \\ 
                     \colhead{                    } &  \colhead{} & \colhead{}  & \colhead{ (") }                     & \colhead{}          &   \colhead{(per sq. arcmin)} \\} 
\tablewidth{0pt}
\startdata
F1   & 00:53:25 &+12:33:55  & $0.96$ & 0.93 & 13.3 \\   
F2   & 09:18:00 &+30:00:00 &  $0.85$ & 1.07 & 20.5 \\
F3   & 05:20:00 &--49:00:00  &  $0.87$ & 1.15 & 16.0 \\
F4   & 10:52:00 &--05:00:00 &   $0.87$ & 1.08 & 14.3 \\
F5   & 13:55:00 &--10:00:00 &   $0.86$ & 1.07 & 15.8 \\
\enddata
\end{deluxetable*}

\subsection{Data} 
The detailed description of the DLS\footnote{http://dls.physics.ucdavis.edu.}  can be found in Wittman et al. (2006; 2012). Below we provide a brief summary of the survey and its data.

The DLS covers five $2\degr\times2\degr$ fields (hereafter F1-F5). F1 and F2 are in the northern sky, and observed with the Kitt Peak Mayall 4-m telescope/Mosaic Prime-Focus Imager (Muller et al. 1998). F3, F4, and F5, which are in the southern sky, were observed with the Cerro Tololo Blanco 4-m telescope/Mosaic Prime-Focus Imager. 
Table 1 lists the coordinates of the five fields.
 
Each Mosaic Imager provides a $\sim35\arcmin\times35\arcmin$ field of view with a $4\times2$ array of 2 k$\times$ 4 k CCDs ($\mytilde0\farcs26$ per pixel).
We divide each $2\degr\times2\degr$ DLS field into a grid of $3\times3$ array. Each $40\arcmin\times40\arcmin$ subfield, slightly larger than the camera field of view, was covered with dithers of $\mytilde200\arcsec$. 
The DLS data consists of 120 nights of $B$, $V$, $R$, and $z$ imaging. A priority was given to the $R$ filter, where we measure our lensing signal, whenever the seeing was better than $\sim0\farcs9$.
The mean cumulative exposure time in $R$ is about 18,000s per field whereas it is about 12,000s per field for each of the rest of the filters. The typical
exposure time per visit is about 900s.

\subsection{Reduction}

We applied initial bias, flat, and geometric distortion correction to the DLS data with the IRAF package {\tt MSCRED}. 
External astrometric calibration was performed by matching astronomical objects in each exposure to the USNO-B1 star catalog using the {\tt msccmatch} task. The residual uncertainty in the global coordinate system relative to the USNO-B1 catalog is less than $0.01\arcsec$. The mean rms error per object is $\mytilde0\farcs3$. The limiting factor
for this scatter per object is believed to be the internal accuracy of the USNO-B1 catalog.
Internal astrometric calibration between different epoch data was carried out using the common high S/N stars present in the overlapping region. Precise registration is
essential in precision weak-lensing analysis because a small $\mytilde0.5$ pixel error can create a noticeable correlation of object ellipticity over a large scale.
We verify that the mean rms error per object is less than $\mytilde0.1$ pixel and the scatter is isotropic, which indicates that the scatter is dominated by photon noise.

We found an initial non-negligible (10$-$20\%) residual flatfielding error in the final stack image after the application of the sky-flat correction. This is further 
refined to the 2$-$5\% level using the  ``\"ubercal" method (Padmanabhan et al. 2008).
Interested readers are referred to Wittman et al. (2012) for details of our ``ubercal" implementation and performance. 

Our team has developed two pipelines (Pipeline I and II) for the creation of the final mosaic.  Pipeline I is optimized for photometry and consists of independently implemented standalone programs 
 (Wittman et al. in preparation). It performs PSF-matched photometry to minimize the systematics in photometric redshift estimation (Schmidt \& Thorman 2012).
Pipeline II is optimized for weak lensing and controls the flow of the {\tt SCAMP} and {\tt SWARP} programs\footnote{available at http://www.astromatic.net.}. We process only R-band data with this second pipeline.
These two pipelines share the above procedures, but differ in that the weak-lensing pipeline uses the subset (with better seeing and less astrometric issues)  of the DLS data and creates a large $2\degr\times2\degr$ mosaic image per field whereas Pipeline I produces nine ($3\times3$) subfield images to cover each $2\degr\times2\degr$ field.
In \textsection\ref{section_psf_reconstruction} we describe this weak-lensing pipeline in detail in the context of the PSF reconstruction.

\subsection{PSF Reconstruction} \label{section_psf_reconstruction}

\begin{figure}
\includegraphics[width=8.5cm]{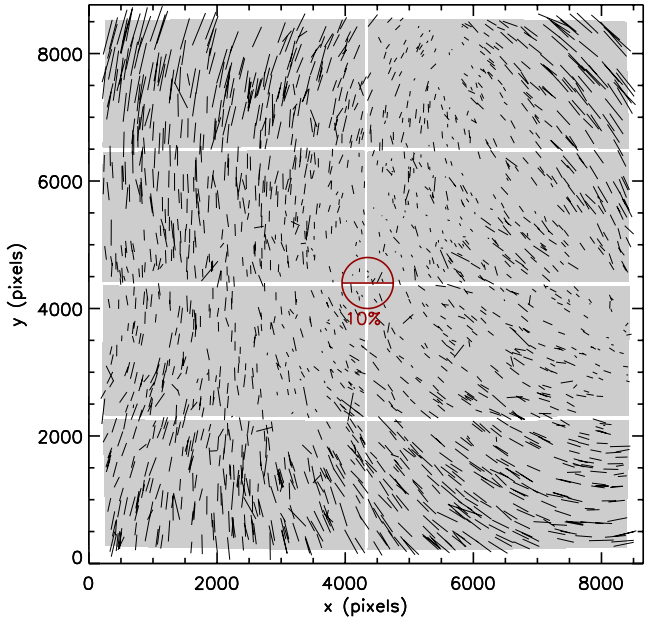}
\caption{Example of spatial variation of DLS PSF.  Although this particular pattern is observed on 24 February 2001 from the Blanco telescope, 
a similar degree of PSF variation complexity is commonly present in all of our DLS data. Each``whisker'' shows the direction and magnitude
of the stellar ellipticity at the location by its orientation and length, respectively. The red stick in the middle shows the size of 10\% ellipticity [i.e., $(a-b)/(a+b)=0.1$].
The eight shaded rectangles depict the eight CCDs of the camera. 
Here we did not clean up outliers (e.g., cosmic-ray hit stars, binary stars, etc.), and they do not represent real PSFs. A similar degree of PSF variation complexity is
commonly present in DLS data, albeit typically with a smaller amplitude.
\label{fig_psf_pattern_focal_plane}}
\end{figure}

The spatial variation of the PSF is substantial and complicated for both the Mayall and Blanco telescopes. 
An example of this PSF pattern is displayed in Figure~\ref{fig_psf_pattern_focal_plane}. Although this particular pattern is observed on 24 February 2001 from the Blanco telescope, 
a similar degree of PSF variation complexity is commonly present in all of our DLS data. It is difficult to interpolate the variation over the entire focal plane with a single set of polynomials. Thus,
polynomial interpolation should be limited to a smaller area, where the variation is slow and tractable. 
Hence, we choose to model the PSF variation on a CCD-by-CCD basis. This chipwise approach was investigated by Jee et al. (2011) for the LSST, where the small
$f$-ratio of the optics makes the potential aberration highly sensitive to CCD flatness, giving rise to a sudden, noticeable jump in PSF patterns across CCD boundaries.
For the Mayall and Blanco telescopes, we often found a somewhat smaller, but clear discontinuity across the CCD gaps, although in principle the relatively large $f$-ratio of the two telescopes should make the CCD-to-CCD flatness much less important.

\begin{figure*}
\includegraphics[width=8.8cm]{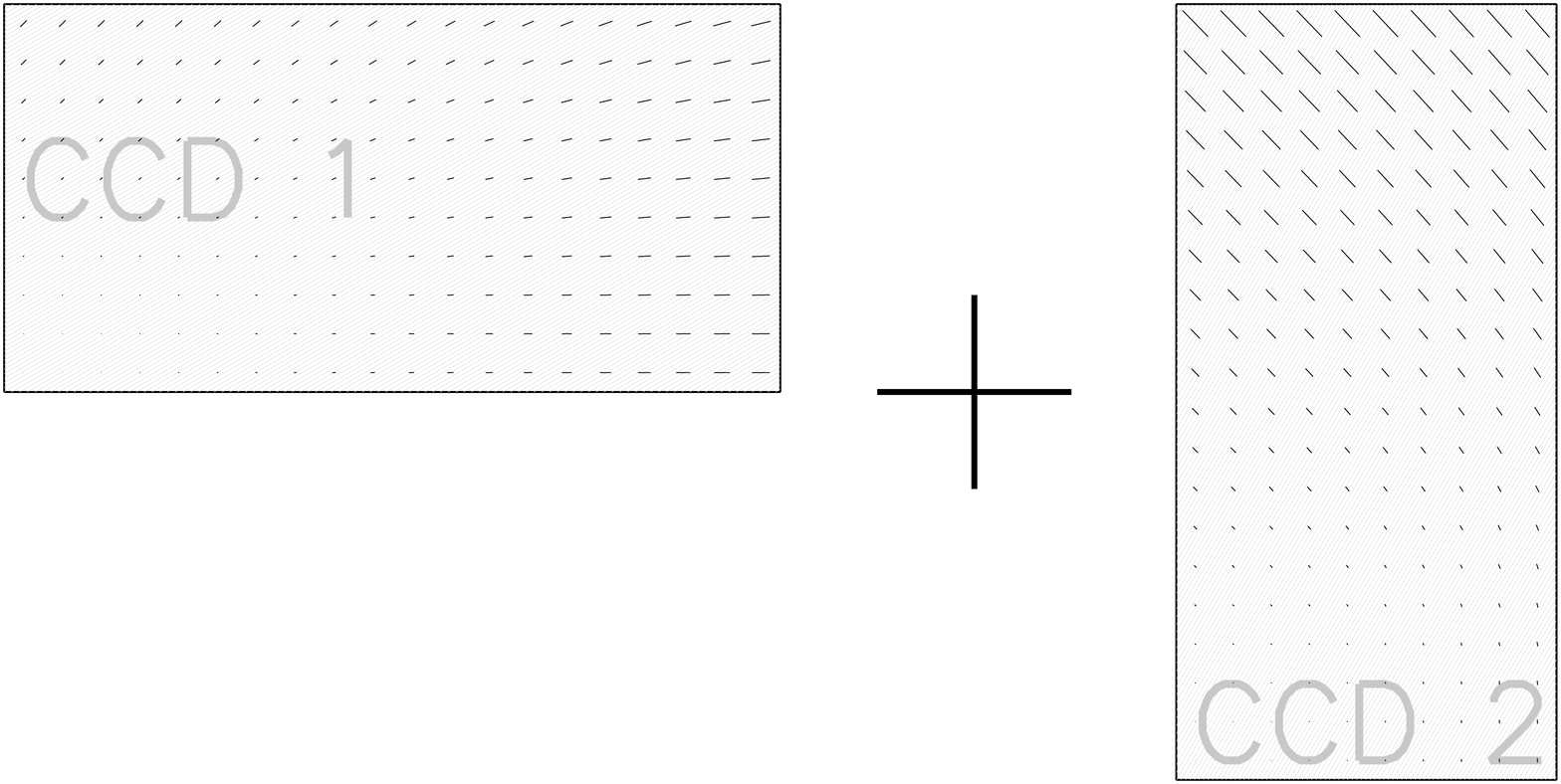}
\includegraphics[width=8.8cm]{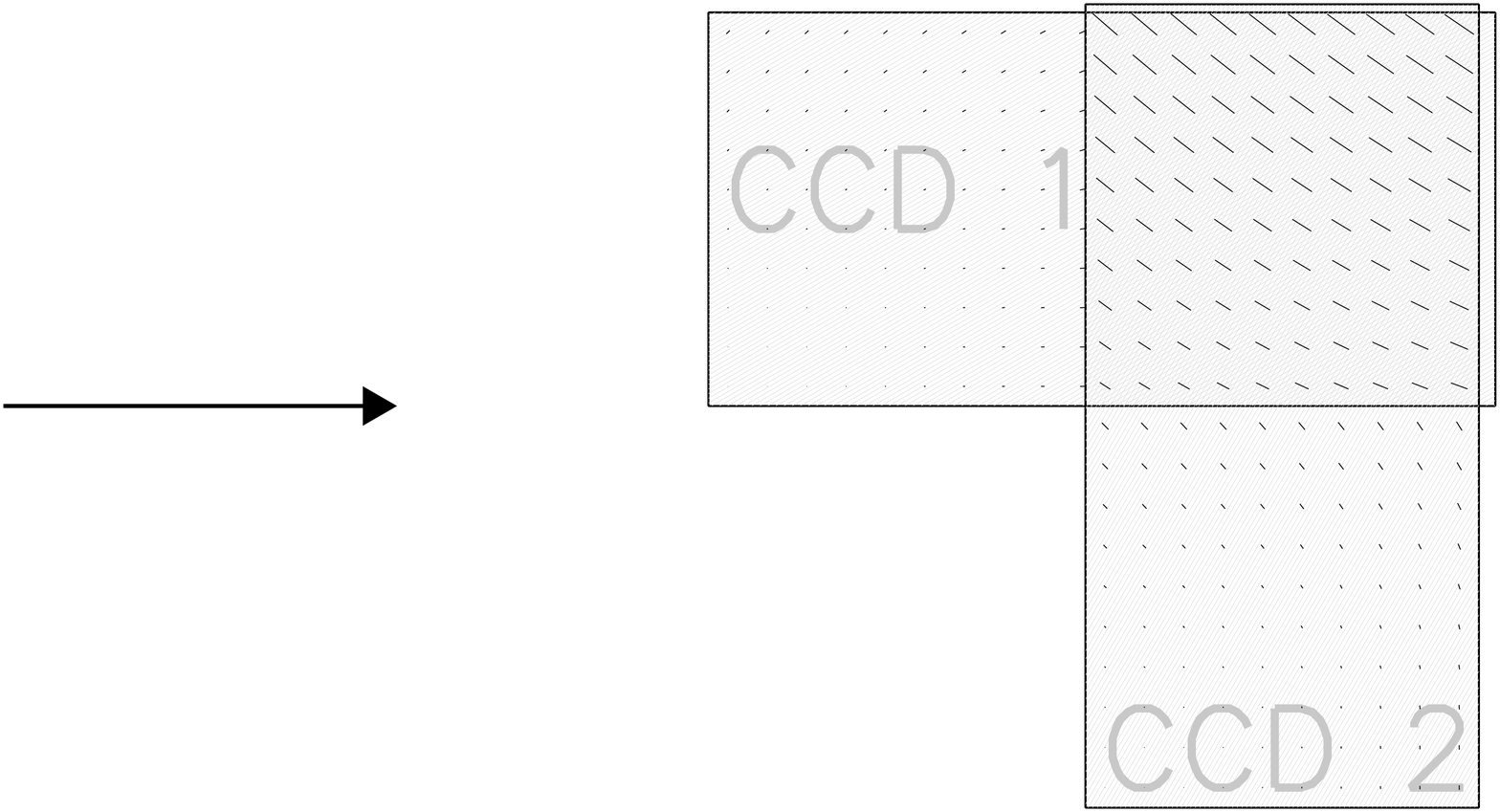}
\caption{Complication in PSF modeling due to image stacking. Weak-lensing analysis is typically performed on a stacked image, which often exhibits sharp PSF discontinuities. 
The figure schematically shows how this complication arises. When we combine the two images in the left panel, the resulting PSF (right) possesses abrupt ellipticity changes
across the boundaries of input frames.
\label{fig_psf_overlap}}
\end{figure*}

As most lensing signals come from distant, faint galaxies, which sometimes are not even detected in single exposures, these source galaxies are commonly examined after multi-epoch data are combined to produce the deep stack image (i.e., single 900s exposure vs.  cumulative 18,000s exposure). Therefore, it is important that the PSF modeling closely mimics the image stacking procedure (e.g., offsets, rotations, geometric distortion corrections, etc.).
Figure~\ref{fig_psf_overlap} schematically illustrates how image stacking complicates the PSF pattern. After stacking is performed, across the image boundaries of input frames we often observe a discontinuous change
of PSF as displayed. This discontinuity prohibits us from interpolating PSFs based on the information obtained only from the final stacked image.  Hence, in our DLS weak lensing analysis, the PSF modeling
is performed with the following two steps to address the issue.
First, we construct a PSF model for each CCD image using a PCA method. Then, the PSF on the final mosaic image is computed by weighted combination of all contributing PSFs from each CCD image. 
Below we provide the details for each step.

\subsubsection{Step 1: PSF Modeling with PCA for each CCD image}
A $2\degr\times2\degr$ mosaic image for each DLS field is created with the {\tt SCAMP/SWARP} software. The {\tt SCAMP} program automatically refines WCS headers of images by first cross-identifying astronomical objects with external standard catalogs and then by ``tweaking" the WCS information of each header in such a way that internal consistency is maximized. Because the astrometric solution is already obtained in the photometric pipeline to the weak-lensing precision, we feed the Pipeline I catalogs into 
{\tt SCAMP} as an external catalog.

The {\tt Swarp} program utilizes the series of these refined WCS headers to define the global WCS for the final mosaic. Then, the input images are resampled and combined to create the final mosaic. We use the Lanczos3 interpolation kernel, which mimics the ideal sinc kernel and is known to suppress the correlation
between pixels. We estimate that the covariance between adjacent pixels is about 7\% of
the variance. This inter-pixel correlation leads
to underestimation of both photometric errors and shape errors. 
The slight shift in shape errors also changes the weight in our
shear correlation computation. In principle,
we can remedy the situation by increasing our rms map to compensate for
this underestimation. However, we conclude that this step is unnecessary because the
resulting change in weight distribution is small and well within the interval of the shear calibration 
marginalization (\textsection\ref{section_shear_calibration}).

What we should potentially be concerned about is the systematics (multiplicative) 
in shear calibration. The inter-pixel correlation somewhat smears the galaxy profile
and on average circularizes the shapes. Fortunately, since we use the same
Lanczos3 kernel in image simulations for our shear calibration (\textsection\ref{section_shear_calibration}), 
the resulting multiplicative factor already includes this effect.

The {\tt Swarp} program provides an option to keep the intermediate resampled images (hereafter {\tt RESAMP} images). We use these {\tt RESAMP} images to identify stars and model
PSFs because they are properly rotated, shifted, and distortion-corrected. Some frames are found to possess rather large ($\gtrsim0.2$ pix)
systematic offsets with respect to the stacked image. In addition, the PSF of some frames are significantly larger than our criterion (FWHM=$1\arcsec$).
About 5\% of the data fall into this group, and we exclude these images for the creation of the final stack.

\begin{figure}
\includegraphics[width=8.8cm]{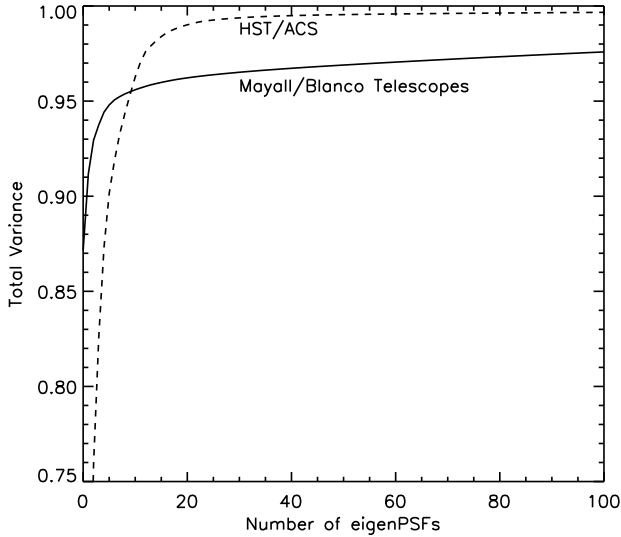}
\caption{Variance vs. number of PSF basis functions (eigenPSFs).  To determine the number of basis functions for a compact description of the PSF, we examine
fractional data variance for different number of basis functions. For $HST$/ACS PSFs, we observe that the growth slows down notably after $\mytilde20$ (Jee et al. 2007).
For the PSF of the 4-m Mayall/Blanco telescopes, this happens at $\mytilde5$. The simpler profile of the ground-based PSF (as opposed to complex, diffraction limited PSF of $HST$) requires fewer basis functions. However, because of the larger FWHM variation (i.e., atmospheric seeing), the total variance
remains slightly lower  than in the case of $HST$/ACS ($\mytilde96$\% vs. $\mytilde99$\% at 20 ). The PSF reconstruction of the Mayall/Blanco Telescopes 
does not show any significant difference in quality as long as the number of eigenPSFs is $\gg5$. In the current study, we choose to keep 20 eigenPSFs.
\label{fig_pca_variance}}
\end{figure}

The $2\degr\times2\degr$ mosaic image for each field consists of more than $\sim1,200$ CCD images. Consequently, we construct and verify an automated procedure to select high S/N isolated stars and apply PCA to them.  Our star-selection algorithm relies on the size versus magnitude relation with some important fail-safe procedures.  The algorithm starts with an initial guess of the half-light radius and magnitude range of the ``good'' stars. Of course, because of the variation in telescope seeing and exposure time, the stellar locus shifts exposure by exposure. Thus, we search the two-parameter space iteratively for the stellar locus in the half-light radius range from 1.4 pixels to 5.5 pixels and the magnitude range whose minimum value (maximum flux) is adjusted depending on the saturation level of the input frame. We discard stars if their {\tt SExtractor} (Bertin \& Arnouts 1996) flags are not zero or if their normalized profiles are significantly different from the median.
The resulting clean stars are used to derive the principal components (eigenPSF), and the coefficients (i.e., amplitude along the eigenPSF) are computed. 
To determine the number of basis functions for a compact description of the PSF, we examine
fractional data variance for different number of basis functions (Figure~\ref{fig_pca_variance}). 
The total variance does not increase rapidly after five, and thus the choice is somewhat arbitrary. We choose to keep 20 eigenPSFs, which accounts for
$\mytilde96$\% of the total variance.

After we obtain these 20 eigenPSFs, the $k^{th}$ star image is decomposed as
\begin{equation}
C_k (i,j)= \sum_{n=0}^{n_{max}} a_{kn} P_n (i,j) + T(i,j),
\end{equation}
\noindent
where $C_k (i,j)$ is the normalized pixel value of the $k^{th}$ star image at the
pixel coordinate $(i,j)$,
$P_n$ is the $n^{th}$ eigenPSF, $a_{kn}$ is
the projection of the $k^{th}$ star in $P_n$, and $T$ is the mean PSF. Because $P_n$'s are orthogonal to
one another, one can determine $a_{kn}$ by multiplying the corresponding eigenPSF to the mean-subtracted star image.

Approximately 50-200 stars are available per CCD per exposure depending on galactic latitude, and we fit  3rd order polynomials to the spatial variation of the coefficients to enable interpolation at any arbitrary position within the CCD. When we experiment with 4th order polynomials instead, the interpolation becomes occasionally
unstable for some frames, where the number of high S/N stars is not sufficient. In addition, we find that the interpolation by 2nd order polynomials
slightly underfit the spatial variation with respect
 to the 3rd order polynomial result, increasing the amplitude of the residual correlation by 10\%-20\%.

The PSF solution on each CCD on each exposure is iteratively refined by comparing the model PSF with the observed star and eliminating significant outliers. This procedure is justified because we expect the spatial variation of the DLS PSF to be continuous across each CCD; in the long-exposure limit the PSF variation is dominated by instrumental aberration. The iteration improves the purity of the stars by removing compact
galaxies accidentally included in the initial size-magnitude-based selection, although the contamination rate is already low because we target
bright objects. Inevitably, some galaxies may remain after the iteration if their surface brightness profiles closely resemble those of stars. However, in practice these objects can be
treated as legitimate point sources and thus help our PSF sampling rather than bias the resulting PSF model. 
As a sanity check, we create a color-color plot ($B-V$ vs. $R-z$) using our star candidates and compare their loci with stellar and galaxy tracks. Less than 1\% of the data points fall outside the main sequence track. Our visual inspection of these off-the-main-sequence objects show that  their morphologies are still indistinguishable from point sources, which reassures that
galaxy contamination is negligible in our PSF sampling. 

\subsubsection{Step 2:  PSF STACKING \label{section_psf_stacking}}

The PSF models for individual {\tt RESAMP} images are inverse-variance weight-averaged to create the PSF on the mosaic image, where the weak-lensing signal is measured.  
The image header of the {\tt RESAMP} file contains the shift information (i.e., integer offsets). For each object, we need to loop over the list of the {\tt RESAMP} files to stack PSFs.
 In order to determine whether or not an object is observed by a given {\tt RESAMP} image and also to find the exact weight value used in co-adding, 
 we utilize the corresponding ``projected'' weight map 
 generated by {\tt Swarp}. If the object
is found to be within the weight map, we compute the PSF at the shifted location (remember that the {\tt RESAMP} images are already rotated to properly align with the final stack) and applied the corresponding weight.

\begin{figure*}
\includegraphics[width=18cm]{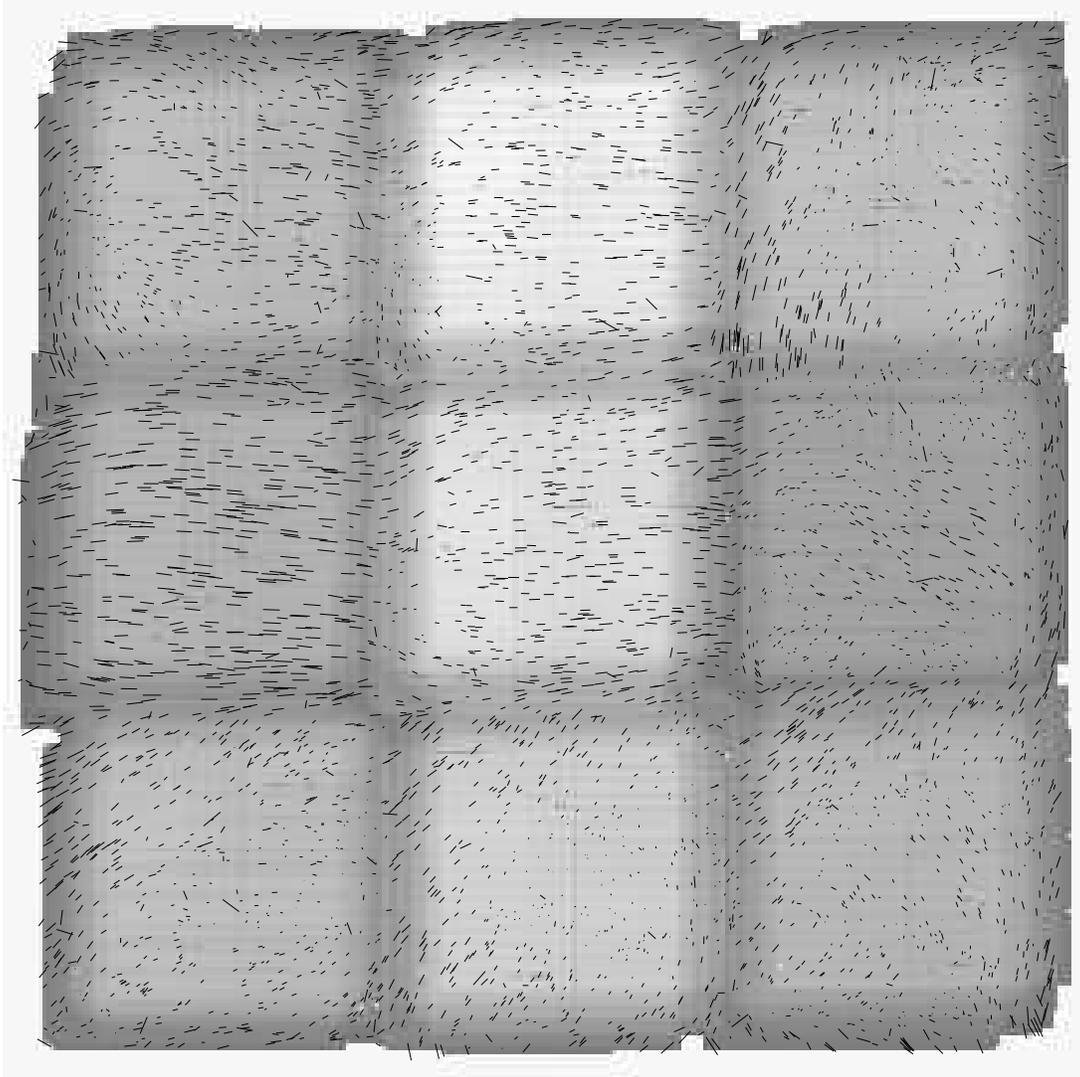}
\caption{Observed PSF ellipticity in the stacked image for F2. The whiskers display the ellipticity distribution of the stars directly measured from the $2\degr\times2\degr$ mosaic image.
The background shade represents the weight map (darker shade indicates lower value) derived from both exposure maps and photon statistics, and illustrates the complexity of the weight distribution. PSF discontinuities occur at exposure boundaries (i.e., at discontinuities in the weight map).
We did not clean up outliers (e.g., cosmic-ray hit stars, binary stars, etc.), and they do not represent real PSFs.
\label{fig_psf_entire}}
\end{figure*}

\begin{figure*}
\includegraphics[width=18cm]{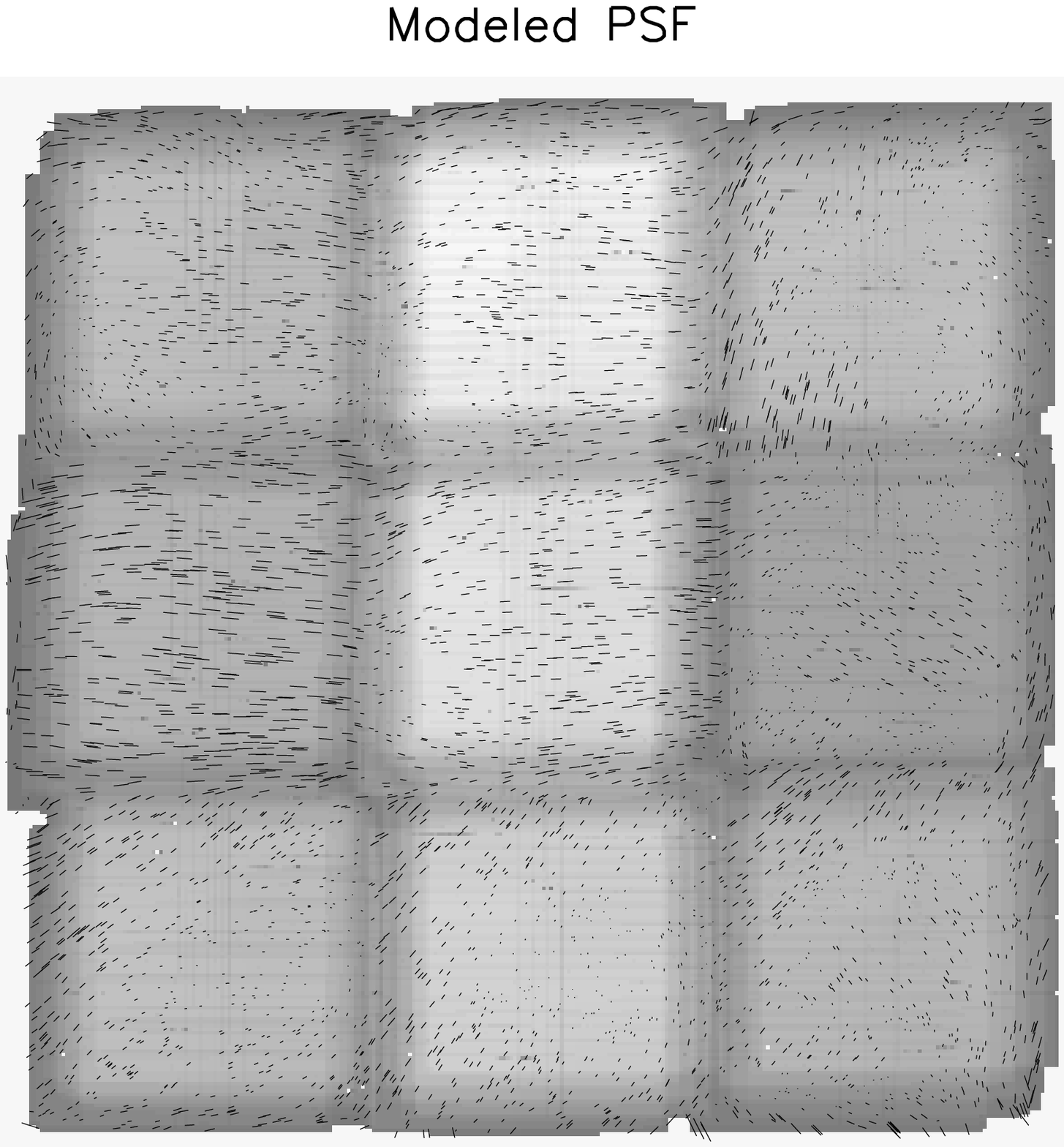}
\caption{PCA PSF reconstruction in the stacked image for F2. The whiskers show the ellipticity of the PSFs evaluated (interpolation + stacking) at the location of the stars in Figure~\ref{fig_psf_entire}. 
The agreement between observation and model in both the size and direction of the whiskers across the entire field is remarkable.
\label{fig_psf_entire_model}}
\end{figure*}

Figure~\ref{fig_psf_entire} and~\ref{fig_psf_entire_model} illustrate that the above PSF stacking scheme closely reproduces the observed ellipticity pattern in F2. In Figure~\ref{fig_psf_entire}, the whiskers show the ellipticity distribution of the stars directly measured from the $2\degr\times2\degr$ mosaic image. The PSF ellipticity change pattern mimics the $3\times3$ pointing pattern of the DLS observation. In addition, it is
easy to see that across the exposure boundaries (where the level of the shade changes) the PSF ellipticity often exhibits a sudden change. The whiskers in Figure~\ref{fig_psf_entire_model} display
the ellipticity of the model PSFs evaluated (interpolation + stacking) at the location of these stars. The similarity in both the size and direction of the whiskers across the entire field is remarkable.

Despite this seemingly nice agreement in the PSF on the stacked image, however, we find that this initial PSF model must be ``tweaked" to remove the PSF-induced anisotropy to our satisfaction.
This tweaking is carried out in two steps, for which we provide the details as follows.

First, the model PSF tends to have systematically lower ellipticity by $\delta e=0.001\sim0.003$ with respect to the data PSF.
This is because the procedure in the PSF sampling from noisy stars slightly circularizes the model PSF. Using this imperfect PSF model for our galaxy shape measurement
leads to non-negligible under-correction. Hence, we compensate for this circularization by increasing the ellipticity (without altering the position angle) of the model PSF by  $\delta e=0.001\sim0.003$. This 
``re-stretching'' is implemented by shearing the PSF image in real space, and the applied shear is a constant (fixed for each DLS field) fraction of
the PSF ellipticity.

\begin{figure*}
\includegraphics[width=6cm,height=7.8cm,trim=0.6cm 0.4cm 0.6cm 0.6cm]{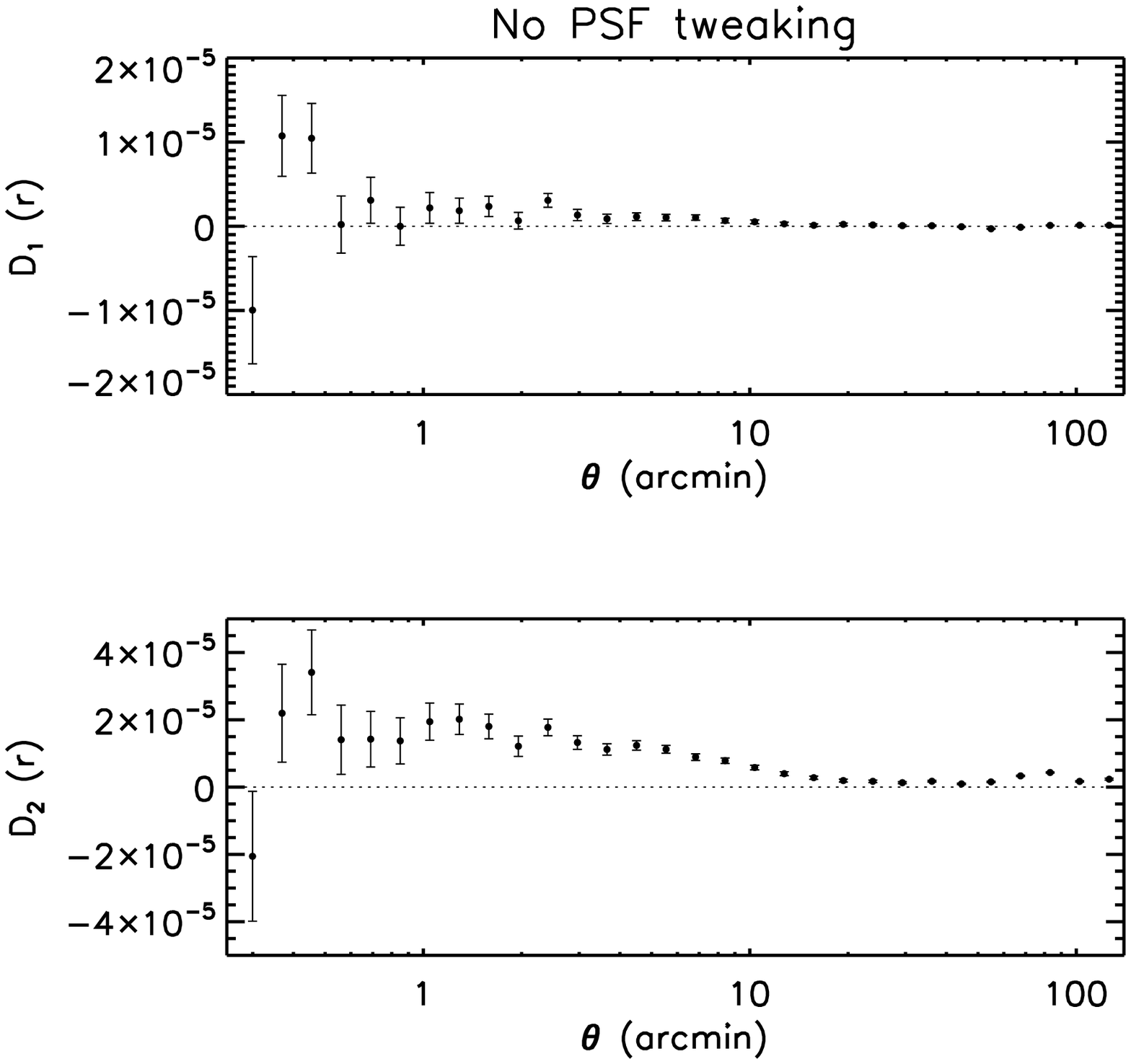}
\includegraphics[width=6cm,height=7.8cm,trim=0.6cm 0.4cm 0.6cm 0.6cm]{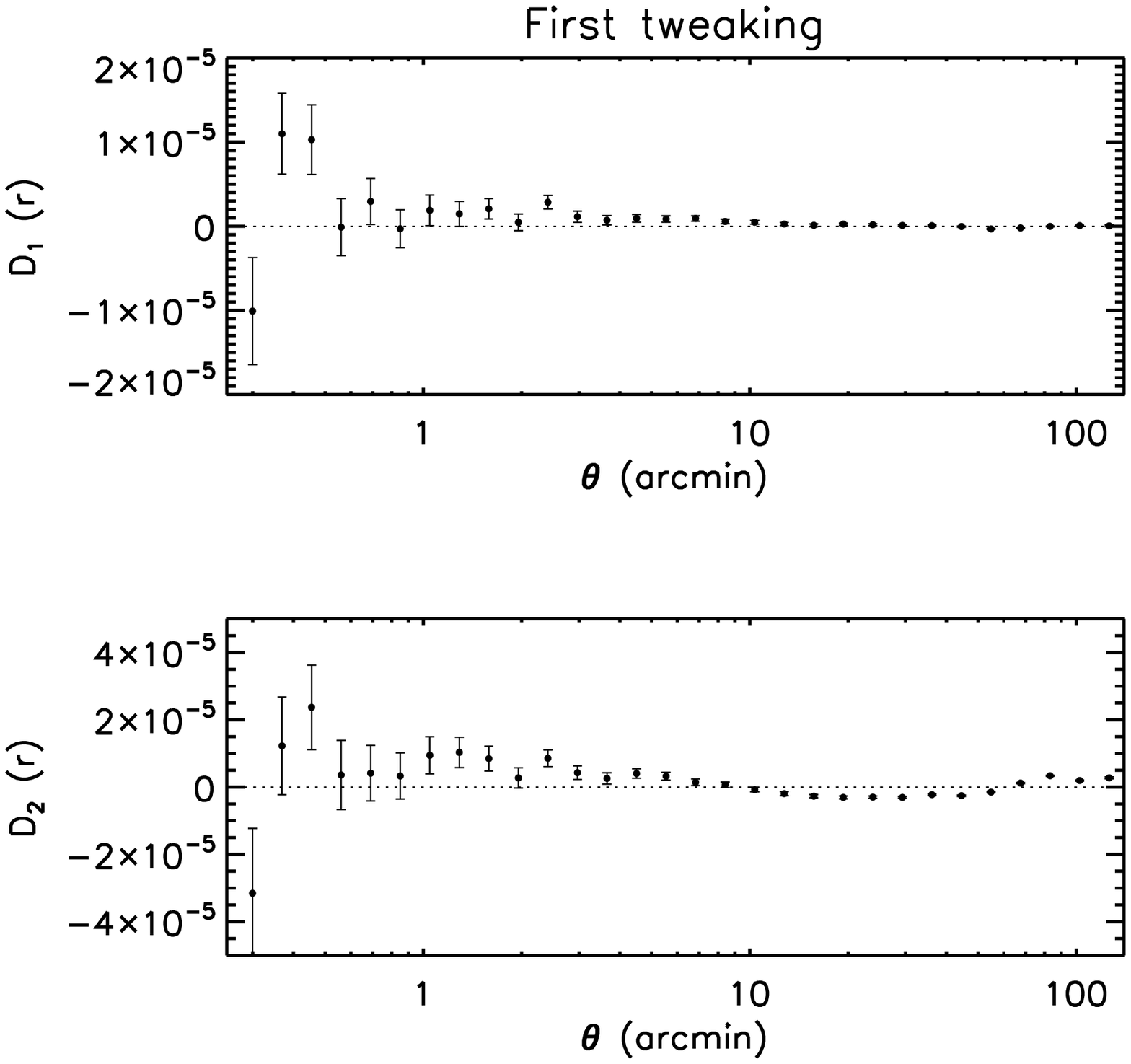}
\includegraphics[width=6cm,height=7.8cm,trim=0.6cm 0.4cm 0.6cm 0.6cm]{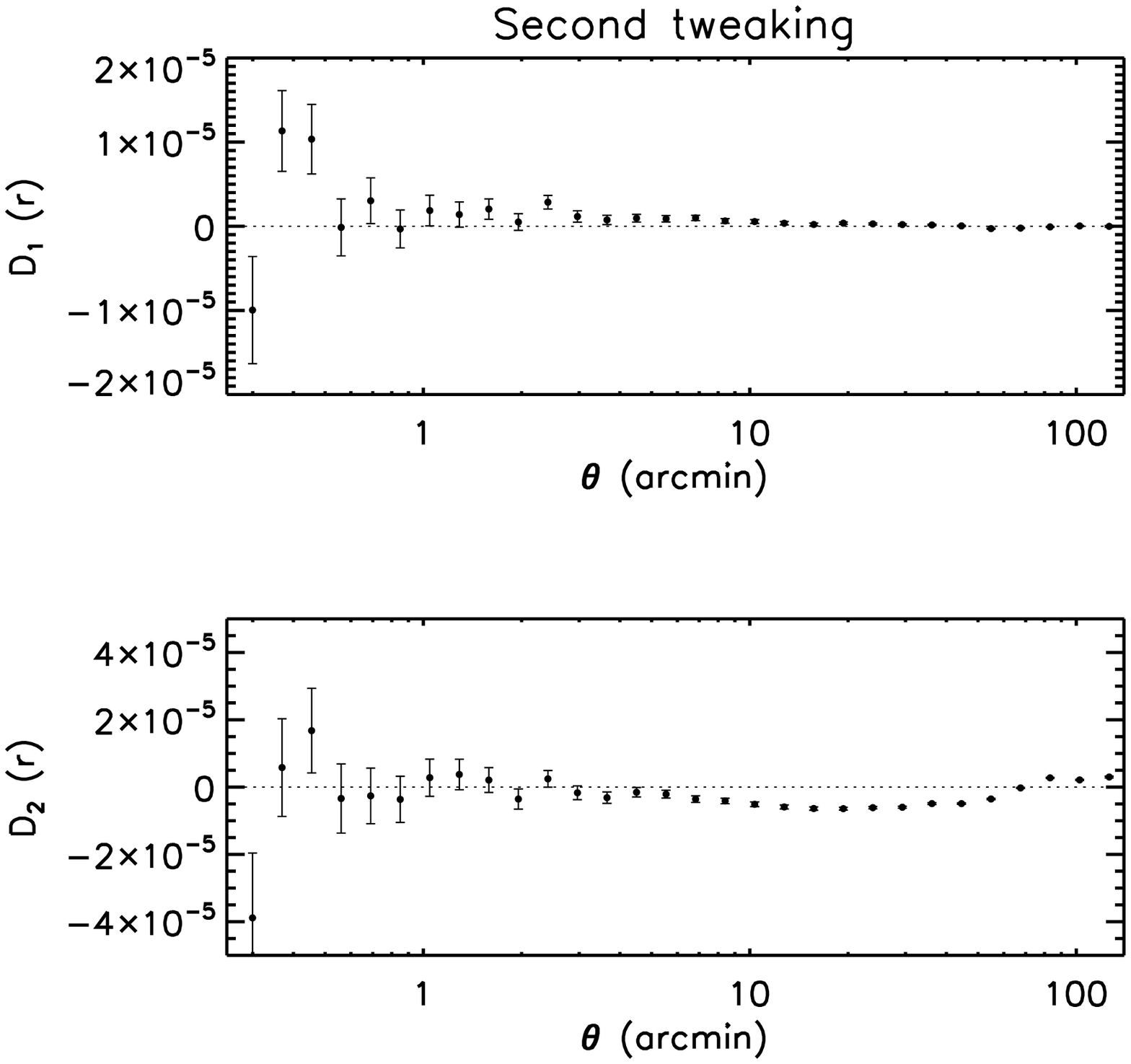}
\caption{Diagnostic ellipticity correlation functions for PSF modeling. $D_1 (r)$ and $D_2 (r)$ are the residual autocorrelation
and the data-residual cross-correlation, respectively (Rowe 2010). Any significant departure from zero indicates that
the model possesses non-negligible systematics. Here we show the case for F2.
The left panel displays the result directly obtained from our $raw$ PSF model whereas the middle and right panels shows the 
results obtained after the application of the first-level and second-level corrections, respectively.
The first tweak is needed to improve the PSF ellipticity agreement between the model and data. 
This is done by increasing the ellipticity of the model PSF by $0.001-0.002$.
However, this first-tweak model does not remove the PSF-induced anisotropy in galaxy images completely due to the centroid bias. We have to further
increase the ellipticity of the first-tweaked PSF by $\mytilde3\times10^{-4}$ to remove this residual bias; note that this makes $D_2(r)$ deviates from zero at $8\arcmin \lesssim \theta \lesssim 70 \arcmin$.
One should remember that $D_2(r)$ is in general more sensitive to systematics than $D_1(r)$ in part because the intrinsic ellipticity of the PSF is much larger
than the residual PSF ellipticity (see Rowe 2010 for extensive discussion on the issue).
\label{fig_psf_correlation}}
\end{figure*}

The exact amount of re-stretching for this first-level tweaking is determined using the following two diagnostic functions proposed by Rowe (2010):
\begin{eqnarray}
D_1(r) & \equiv & \left < (e_{d} - e_{m})^{*} (e_{d}-e_{m}) \right > (r)   \\
D_2(r) & \equiv & \left < e_d^* (e_d - e_m) + (e_d - e_m)^* e_d \right> (r)  
\end{eqnarray}
\noindent
where $e_d$ and $e_m$ are the ellipticity of the data and model PSFs, respectively in complex notation (see \textsection\ref{section_shear}). Consequently, $D_1$ and $D_2$ 
show the residual autocorrelation and the data-residual cross-correlation, respectively.
Rowe (2010) suggests that  a combined use of these two functions provides an insight into systematics of the model.
In Figure~\ref{fig_psf_correlation} we display $D_1(r)$ and $D_2(r)$ for F2. The left panel displays the result directly obtained from our 
PSF stacking whereas the middle panel shows the result
when this PSF model on the left panel is re-stretched to compensate for the PSF circularization. 
The improvement is more noticeable in $D_2(r)$.
When comparing the amplitudes of $D_1$ and $D_2$, one should remember that $D_2(r)$ is in general more sensitive to the presence of systematics
than $D_1(r)$ in part because
$D_2(r)$ is a sum of two data-residual ellipticity correlation functions (in order to cancel the imaginary part), and in part because the ellipticity of
the PSF is higher than that of the residual. For other possible reasons, we refer readers to Rowe (2010).

The small residual correlation functions in the middle panel of Figure~\ref{fig_psf_correlation} suggests that the above re-stretched PSF model
is an excellent description of the data. However, we notice that the shapes of galaxies obtained with this re-stretched PSF (middle) tends to be
still under-corrected. In other words, collectively speaking, galaxy shapes are still biased toward the initial anisotropy of the PSF.
We suspect that this phenomenon is in part related to the so-called centroid bias mentioned by Bernstein \& Jarvis (2002) and Kaiser (2010), 
where it is argued that 
even a perfect PSF model will not remove the PSF bias completely because the centroid of the object is
more uncertain along the elongation of the PSF.  This bias does not go away even if we treat the centroid as free parameters because
the PSF-induced pixel correlation still makes the resulting centroid distribution anisotropic\footnote{Although ``centroid bias" might not be
the most adequate term to describe the phenomenon, we refer to it as such for the lack of better term.}.
 
\begin{figure*}
\includegraphics[width=6.cm,height=4.5cm]{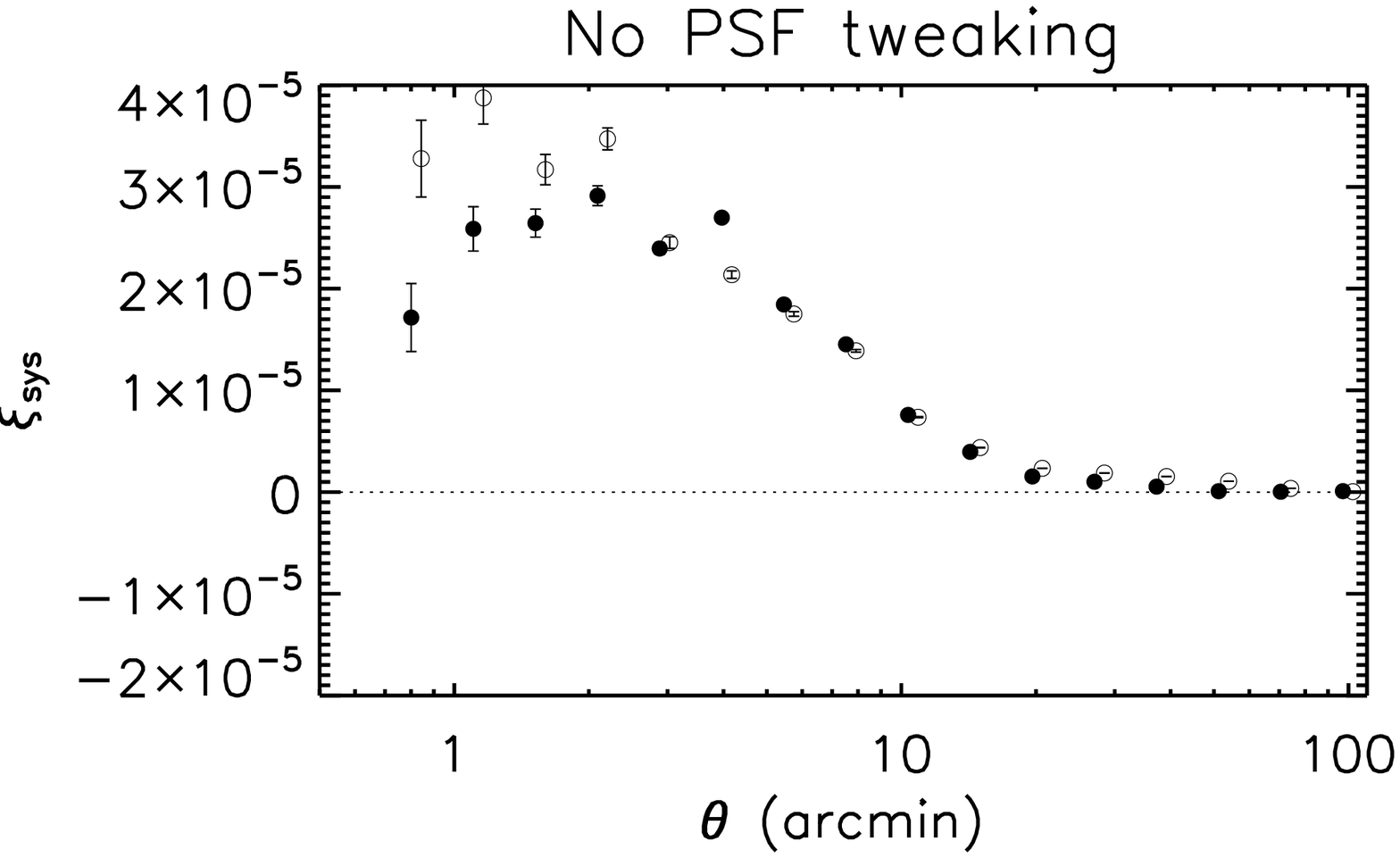}
\includegraphics[width=6.cm,height=4.5cm]{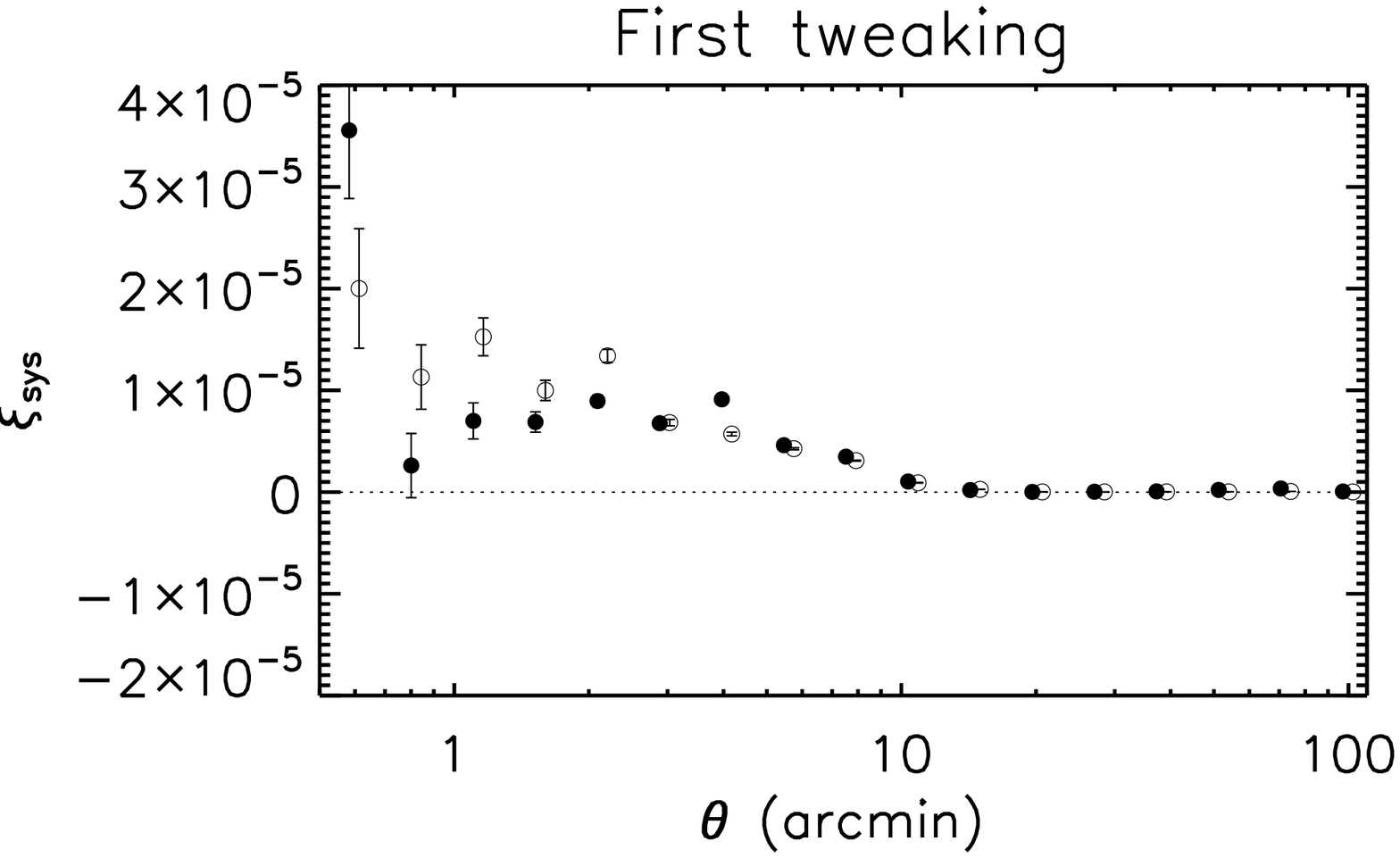}
\includegraphics[width=6.cm,height=4.5cm]{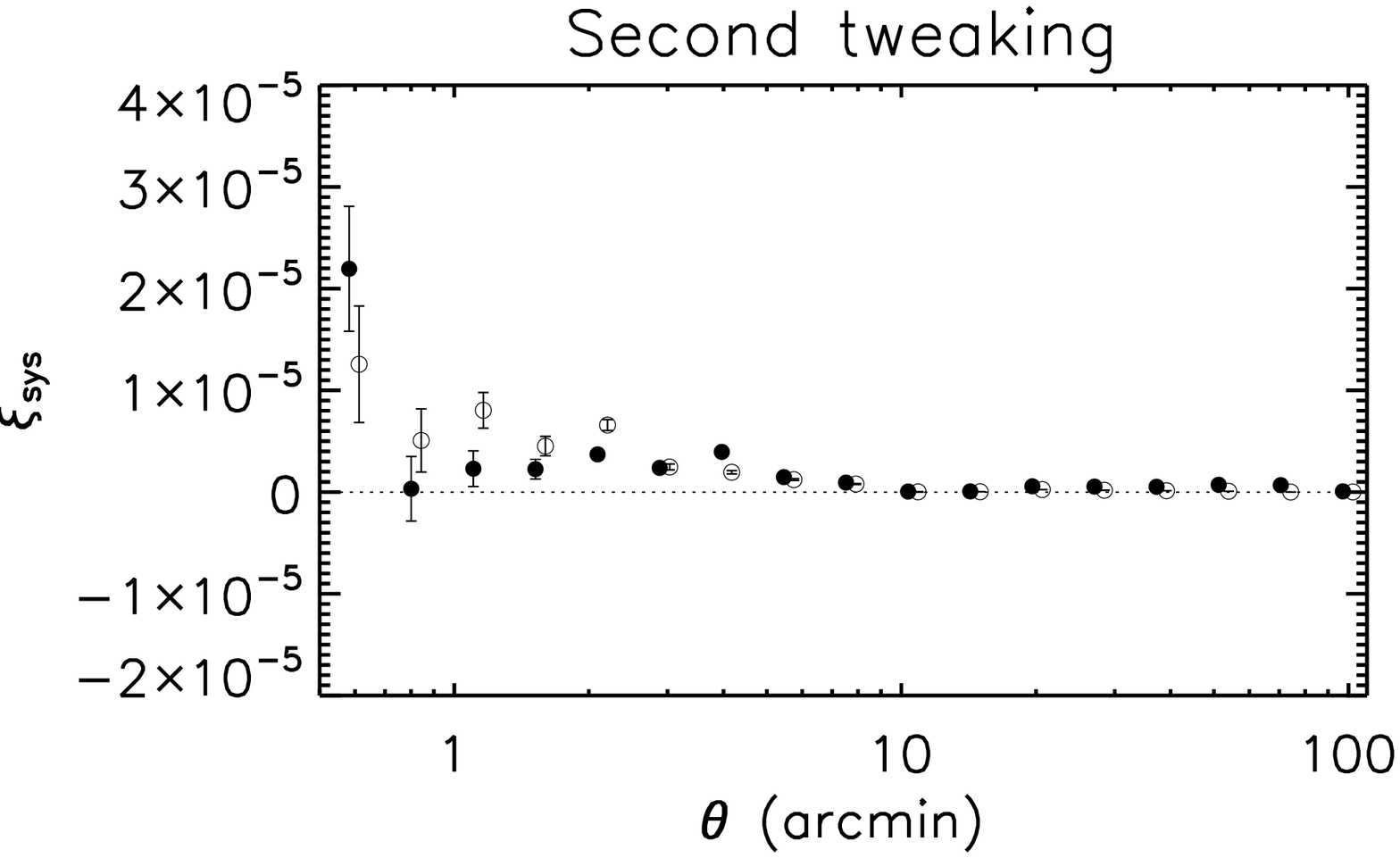}
\caption{Star-galaxy correlation as diagnostics of PSF model. We show the case for F2. The left, middle, and right panels correspond to the PSF models
shown in the left, middle, and right panels in Figure~\ref{fig_psf_correlation}. Filled and open circles correspond to the
``$tt$'' and ``$\times \times$'' (see \textsection\ref{section_equations} for the definition)  correlations, respectively.
\label{fig_star_galaxy_tweak}}
\end{figure*}

\begin{figure*}
\includegraphics[width=18.cm]{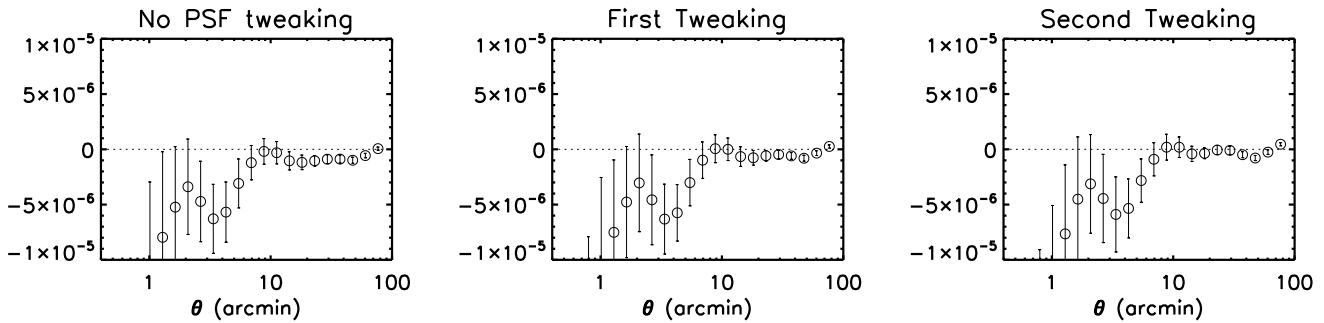}
\caption{B-mode signal in aperture mass dispersion. We show the case for F2. The left, middle, and right panels correspond to the PSF models
shown in the left, middle, and right panels in Figure~\ref{fig_psf_correlation}.
The negative B-mode signals at $\theta\lesssim10\arcmin$ represent not residual systematics, but artifacts arising from the missing data on small scales (see \textsection\ref{section_b_mode}).
\label{fig_b_mode_tweak}}
\end{figure*}

This is the reason that we need a second-level tweaking mentioned above.
We address this issue by further stretching the model PSF so that the ellipticity increases by additional $\left< \delta e \right> \sim 3\times10^{-4}$.
The amount of this additional stretching is also a fixed  (for each DLS field) fraction of the PSF ellipticity, and the first-order value is
determined mainly utilizing our image simulations, where galaxies are randomly oriented (i.e., no shear is present).
We adjust the stretching factor until the PSF-induced residual shear signal vanishes. 
Then, we refine this factor by  making sure that 
 the amplitude of star-galaxy correlations (\textsection\ref{section_star_galaxy}) and B-mode signals (\textsection\ref{section_b_mode})
 also decreases simultaneously. The right panel of Figure~\ref{fig_psf_correlation} shows the resulting $D_1(r)$ and $D_2(r)$
 diagnostic functions when this second-tweak is applied to the first-tweak PSF model shown in the middle panel.
 Note that this increases the deviation of $D_2(r)$ from zero at $8\arcmin \lesssim \theta \lesssim 70 \arcmin$, 
 although this final PSF removes the PSF-induced anisotropy from galaxy images most satisfactorily among the three cases shown here.
Figure~\ref{fig_star_galaxy_tweak} displays the star-galaxy correlation functions (see \textsection\ref{section_star_galaxy} for the definitions)
for the three cases shown in Figure~\ref{fig_psf_correlation}. It is obvious that the PSF model that we obtain from the second tweak gives
the smallest amplitude for star-galaxy correlations, although the amplitude of the diagnostic function (especially $D_2$) of this PSF is
not the smallest. 
Finally, we show the impacts of this PSF tweaking on B-mode signals in Figure~\ref{fig_b_mode_tweak}. Although the difference is somewhat
small compared to the test results carried out with the residual PSF and star-galaxy correlation, we observe that the B-mode
signals are closest to zero when we use the second-tweak PSF model. Here we display the B-mode signals in aperture mass statistics, and
thus the negative B-model signals at $\theta\lesssim10\arcmin$ represent not residual systematics, but artifacts arising from the missing data
on small scales (see \textsection\ref{section_equations} and \textsection\ref{section_b_mode} for the definition of
the aperture mass statistic and the discussion of aliasing artifacts, respectively).

One should not be misled into thinking that our PSF-tweaking removes any arbitrary B-mode signal. Systematics arising
from non-centroid bias cannot be made to disappear by simply increasing the ellipticity of every model PSF uniformly by a constant factor.
In addition, the above PSF-tweaking cannot arbitrarily get rid of intrinsic alignment signals.

\subsection{Galaxy Ellipticity Measurement}
There exist a number of algorithms for galaxy shape measurement in the context of weak lensing. Depending on the approaches to removing PSF effect, 
we can classify the existing algorithms into moments-based methods and profile-fitting methods.
The former methods measure second-moments for both galaxies and PSFs and use them to estimate the pre-seeing ellipticity.  This
approach was pioneered by Kaiser, Squires, \& Broadhurst (1995; KSB hereafter) and Fischer \& Tyson (1997), and many variations exist.
The latter algorithms approximate the surface profile of galaxies with some analytic profiles. These analytic profiles are convolved with PSF models before being fit to the images rather than
fit to a deconvolved image. 
While the classic, moments-based methods continue to be popular and updated, 
cosmic shear studies are relying more on the second, profile-fitting approach to overcome the potential limitations (Kaiser 2000) of the moments-based approach.

Our shape measurement algorithm belongs to the second category. We fit a PSF-convolved elliptical Gaussian to a galaxy image.
Of course, an elliptical Gaussian profile is not the best approximation of galaxy profiles. This sub-optimal fitting is
termed ``underfitting" (Bernstein 2011) and has been shown to cause some bias in shear estimation. However,
we find that this bias is only multiplicative and thus can be calibrated out with careful image simulations (discussed in \textsection\ref{section_shear_calibration}).
Our experiments with S\'ersic profiles show that although this multiplicative factor is reduced, the measurement uncertainties increase. This increase in
ellipticity uncertainty is attributed to the following two facts. First, S\'ersic profile fitting takes into account more pixels farther from the object center, introducing larger noise.
Second, S\'ersic profile fitting involves more free parameters to marginalize over.
We want to include as many faint galaxies as possible for shear measurement as long as the net noise (quadratic sum of systematic and statistical noise) goes down, 
and we find that using Gaussian over other more sophisticated profiles increases the overall S/N of our cosmic shear signal.

Formally, a description of a galaxy image with an elliptical Gaussian requires the following seven free parameters: normalization, semi-major and semi-minor
axes, position angle, background level, and two parameters for the centroid. We fix the centroid\footnote{When we free the object centroid, 
the number of usable galaxies decrease by 
$\mytilde8$\% and the multiplicative shear calibration factor increases by $3-6$\%.} 
and the background using the SExtractor's {\tt xwin\_image}, 
{\tt ywin\_image}, and {\tt background} so that the total number of free parameters is only four, which further stabilizes the minimization and reduces the ellipticity uncertainty.
The initial guesses for these four parameters are computed utilizing SExtractor measurements.

For each object, square postage stamp images are extracted from the final stack and rms map.
We choose the size of this postage stamp image to be $(8a + 20)$ pixels on a side, where $a$ is the semi-major
axis initially determined by SExtractor. In most cases, the image contains pixels belonging to other objects and we need to
mask them out. This is implemented by replacing the rms values of these pixels with very large numbers, thus masking them out in further processing. The identification
of these pixels is based on the information in the segmentation map output by SExtractor.
The shape measurement code is written in IDL, and the {\tt MPFIT}\footnote{available at  http://www.physics.wisc.edu/\mytilde craigm/idl/.} module was
employed as a minimizer. {\tt MPFIT} estimates parameter uncertainties from a Hessian matrix. We convert these errors
to ellipticity uncertainties by error propagation.

\subsection{Shear Estimation}

\subsubsection{Shear Estimator \label{section_shear}}
Gravitational lensing transforms the shape in the source plane to the image plane according to the following matrix:
\begin{equation}
\textbf{A}=(1-\kappa) \left (\begin{array} {c c} 1 - g _1 & -g _2 \\
                      -g_2 & 1+g _1
          \end{array}  \right ), \label{eqn_lens_trans}
\end{equation}
\noindent
where $\kappa$ is the projected mass density in units of the critical lensing density and $g$ is the reduced shear $g = \gamma/(1-\kappa)$.
In the weak-lensing regime, $\kappa$ is small and thus the $\gamma \simeq g$ assumption is often made.
The $(1-\kappa$) factor affects the overall magnification, which is observable through the measurement of bias in object number density or size distributions.
The transformation matrix shears a circle into an ellipse with an ellipticity $g=(g_1^2+g_2^2)^{1/2} = (1-r)/(1+r)$, where $r$ is the ratio of the semi-minor axis to
the semi-major axis (i.e., $b/a$). The position angle of the ellipse is given by $1/2~ \mbox{tan}^{-1}(g_2/g_1)$.

Using complex notation $\textbf{g} = g_1 + \textbf{i} g_2$, we can also express the ellipticity transformation when an object has an initial
ellipticity $\textbf{e} = e_1 + \textbf{i} e_2$ as
\begin{equation}
\textbf{e}^{\prime} = \frac{\textbf{g} + \textbf{e}}  {1 + \textbf{g}^{*} \textbf{e}}, 
\end{equation}
\noindent
where the asterisk represents complex conjugation and $\textbf{e}^{\prime}$ is the measured ellipticity.
If we assume that the distribution of $\textbf{e}$ is isotropic, we can derive $\textbf{g}$ from averaging over a population of galaxies using
\begin{equation}
\textbf{g} = \frac{1}{\mathcal{R}}  \frac{ \sum \mu_i  \textbf{e}^{\prime}_i }{ \sum \mu_i}, \label{eqn_shear_estimator}
\end{equation}
where $\mu_i$ is a weight for each galaxy $i$.  In the current paper, we use the following inverse variance as weight:
\begin{equation}
\mu_i  = \frac {1} { \sigma_{SN}^2 + (\delta e_i)^2} \label{eqn_shear_weight}, 
\end{equation}
\noindent
where $\sigma_{SN}$ is a shape noise of the population per component ($\mytilde0.25$) and $\delta e_i$ is the ellipticity measurement error per component.
In equation~\ref{eqn_shear_estimator}, $\mathcal{R}$ is called the shear responsivity, which is a calibration factor necessary to reconcile the difference between the average ellipticity and the shear.
It is easy to show that $\mathcal{R}\approx1$ if no measurement noise is present and galaxy morphology can be described by a simple elliptical isophote.
However, because neither is true in the real world, one must estimate $\mathcal{R}$ with  care, and this is one of the most critical issues in future large
lensing surveys since the result will not be limited by statistical uncertainties.

Ideally, it is desirable to estimate $\mathcal{R}$ analytically from first principles and use image simulations only to verify the accuracy. Bernstein \& Jarvis (2002) provided an important 
contribution and their prescription has been used in quite a few studies (e.g.,  Jarvis et al. 2006, Hirata et al. 2004).  Nevertheless, it relies on some assumptions which are
not strictly true of real data or highly realistic simulations. For the current DLS cosmic shear analysis, we find that the shear responsivity derived with the Bernstein \& Jarvis (2002) method
agrees reasonably well with the value obtained from our weak-lensing image simulations for bright ($R < 22$) galaxies, but gradually underestimates the shear dilution effect as the S/N of the objects decreases. Our DLS shear calibration hereafter is purely based on our image simulation studies, which are described in detail below.

\subsubsection{Image Simulation}  \label{section_shear_calibration}

The translation of the measured ellipticity to the applied shear is not straightforward. First, a response to a shear depends on galaxy populations. 
This is because the change in the second moments under a given shear $\gamma$ depends not only on the second moments themselves, but also on the
higher moments (Mandelbaum et al. 2012). This makes the effects of morphological features such as radial profiles, bulge-to-disk ratios, spiral arms, etc. non-negligible. As we model a galaxy light distribution with an elliptical Gaussian in the current study, we should understand how much the lack of details in the model biases the lensing signal. Second, ellipticity measurement is a noisy process. As most lensing signals come from faint galaxies, this measurement noise significantly dilutes the signal. Third, a nontrivial fraction of galaxies are affected by catastrophic shape measurement errors. The sources of these catastrophic shape errors include substructures of galaxies (e.g., HII regions), crowding,  ``bleed'' trails, clipped objects, galactic cirrus, 
spurious detection around bright objects, etc.
As the ellipticity measurement from these sources does not contain any lensing signal, the direction of the bias will always be toward underestimation.
In the current paper, instead of quantifying the effect of each factor separately, we choose to derive a global value for shear responsivity $\mathcal{R}$. Although it is worth investigating the effect of
each factor in isolation, marginalizing over other parameters increases the required number of simulated image sets considerably, which is beyond the scope of the current study.

We utilize a modified version of the Large Synoptic Survey Telescope (LSST) image simulator presented in Jee \& Tyson (2011). The simulator samples galaxy images from the {\it Hubble Space Telescope (HST) } / Ultra Deep Field 
(UDF; Beckwith et al. 2003) images and convolves them with the PSFs computed from the atmospheric turbulence model and the telescope optics. 
The purpose of this modified image simulator is to calibrate the conversion of ellipticity to shear. Given the same galaxy profile, the size and intrinsic ellipticity of the PSF are the most important factors affecting this calibration parameter. The main difference in the PSF between LSST and the two 4-m telescopes comes from 1) different f-ratios (f/1.24 and f/2.7 for LSST and Mayall, respectively), 2) exposure time (15 s vs. 900 s), and 3) atmospheric seeing ($0\farcs65$ versus $0\farcs85$).
We address 2) and 3) by changing the atmospheric parameters (e.g., Fried parameter and outer scale) in such a way that the resulting seeing distribution is close to the observation. We cannot address 1) directly without replacing the current LSST optical design model with the most up-to-date Mayall/Blanco telescope models. However, it is possible to approximate the effect by degrading the focus (and optical alignment) so that when the diffraction limited PSF is convolved with the atmospheric PSF, it matches the DLS pattern. Without this adjustment, the delivered DLS PSF is severely circularized by atmosphere (longer exposure and large atmospheric PSF). After this modification, we obtain a distribution of PSF ellipticity ranging from 2\% to 7\%, matching the DLS data. 
Another important question might be whether or not the resulting spatial variation within a single DLS CCD is realistic. If our simulated PSF lacks a small scale variation compared to that of the data, the PSF model in the simulation may be easier to describe than in real situations. We find that the residual PSF correlation  
for both simulation and data shows a similar residual amplitude, which suggests that the spatial variation of the PSF on both simulated images and
DLS data possess a similar level of complexity.

Most lensing signals come from $z\sim1$ galaxies, the median of the $n(z)$ counts,  which are dominated by the faint blue galaxy (FBG) population. Hence, it is important to test shear measurement
from these UDF galaxies rather than from synthetic galaxies with analytic profiles. We randomize both the orientation and the position of the HST galaxies so that the net shear vanishes.  
We refer to these images as zero shear sky (ZSS). In a strict sense, the ZSS images are already convolved by the PSF of the Advanced Camera for Surveys, and thus are not the ``true'' sky images in the absence of the instrument seeing. However, because the size of the ACS PSF is a factor of eight smaller, the effect in the creation of the final DLS images is limited to a scale far smaller than the DLS pixel ($\sim0.257\arcsec$). 
We apply a gravitational shear using bi-cubic interpolation to the ZSS images. As we have not down-sampled the UDF images yet, any interpolation
artifacts and their propagation are expected to be insignificant on the final image. 
Then, we convolve these sheared sky (SS) images with spatially varying DLS PSFs. Readers are referred to Jee \& Tyson (2011) for the details of the algorithm involved in this step.
Finally, we down-sample the convolved sheared (CS) images and add noise to the result in order to match the pixel scale and depth of our DLS images.

\begin{figure}
\includegraphics[width=8.8cm]{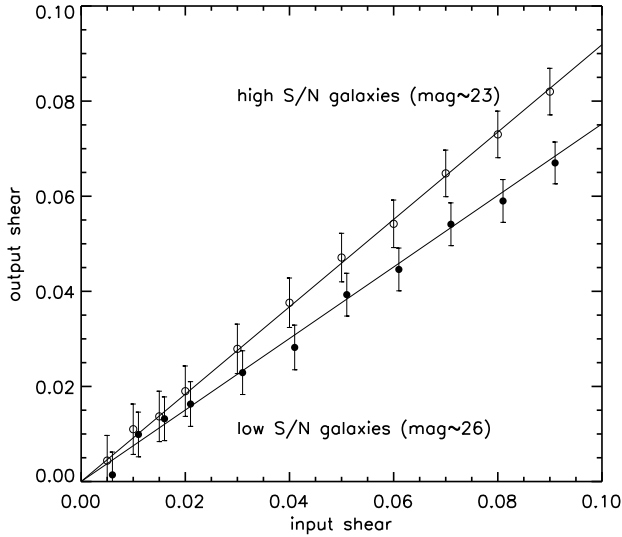}
\caption{Shear recovery test. The calibration factor (inverse slope of the solid lines) depends on the S/N of the source galaxies.
The high S/N galaxies (median magnitude $\sim23$) require a calibration factor of $\mytilde1.05$ whereas the calibration factor of
the low S/N galaxies (median magnitude $\sim26$)  is as high as $\sim1.28$.
\label{fig_shear_recovery}}
\end{figure}

\begin{figure}
\includegraphics[width=9.8cm]{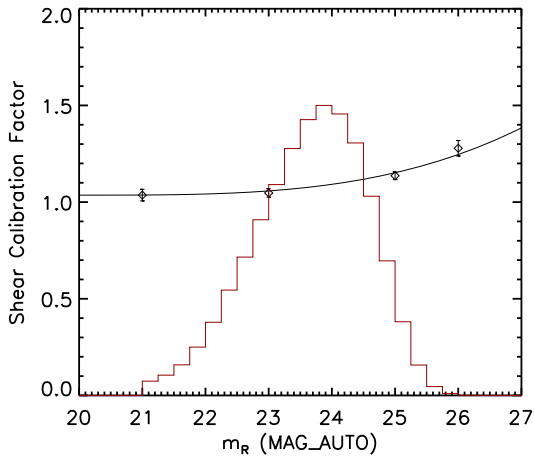}
\vspace{-28pt}
\caption{$R$-band magnitude versus shear calibration factor. The data points ( {\it diamond} ) are measured from image simulations.
The black solid line represents our parameterization of the shear calibration factor.
The mean shear calibration factor derived by multiplying this curve and the weighted magnitude
distribution of source galaxies (red) is $\mytilde1.08$.
\label{fig_shear_calibration}}
\end{figure}

Our goal is to determine the relation between input shears and weighted sum of the ellipticities as a function of magnitude. Figure~\ref{fig_shear_recovery}
shows the results for two of these simulated populations where the mean apparent magnitudes are approximately 23 (open) and 26 (filled). We omit the results from intermediate magnitude objects to avoid clutter.
The slope of the line is the shear responsivity in equation~\ref{eqn_shear_estimator}. For the bright population, we obtain $\mathcal{R}\sim0.95$. This is similar to
the value $\mathcal{R}\sim0.93$ that we obtain using eqn.~5.33 of Bernstein \& Jarvis (2002).  However, for the faint population the shear responsivity $\mathcal{R}\sim0.78$ determined from our image simulation
is lower than the analytic estimate $\mathcal{R}\sim0.89$.
Figure~\ref{fig_shear_calibration} summarizes the results when we combine the results
of this shear recovery test for four magnitude bins.
We parameterize the dependence of the multiplicative factor $m_{\gamma}=1/\mathcal{R}$ on the $r-$band  magnitude ($m_R$) with
the following form:
\begin{equation}
m_{\gamma}=6\times10^{-4} (m_R - 20 )^{3.26} + 1.036,
\end{equation}
\noindent
where $m_R$ is SExtractor's {\tt MAG\_AUTO}. In deriving an average multiplicative factor $\left <m_{\gamma} \right >$, we
need to consider both the magnitude and weight distributions of the source population; the source selection criteria are discussed in detail in \textsection\ref{section_source_selection}.
The magnitude distribution of our source galaxies peaks at $m_R\sim24$ and then precipitously decreases (virtually no galaxies beyond $m_R\sim26$).
In addition, smaller weights are given to ellipticities of faint galaxies (eqn. \ref{eqn_shear_estimator} and \ref{eqn_shear_weight}). 
The red histogram displays this weighted magnitude distribution.
The resulting mean multiplicative factor is  estimated to be $\left<m_\gamma \right>=1.08\pm0.01$.

As this multiplicative factor is determined from the galaxies in the UDF, it is possible that the above shear calibration may need to be refined further
for the DLS data. Indeed, our experience suggests that galaxy morphology and size distributions are non-negligible factors in shear calibration. 
Currently, the UDF images are the only available space-based images that contain virtually noiseless galaxy images down to the limiting magnitude of the DLS.
Therefore, to assess the effect of the sample variance, we perform another suite of lensing image simulations, but this time with an analytic description of galaxy
profiles. We utilize two publicly available software packages {\tt Stuff} and {\tt SkyMaker}, which create
astronomical catalogs and images, respectively \footnote{http://www.astromatic.net/software}.  The default generation of the ellipticity distribution in the {\tt Stuff} catalog
is rather unrealistic, and thus we modify the output in such a way that the ellipticity distribution per component matches the one
in the UDF data. Because the exact ellipticity correlation between bulge and disk is unknown, we choose to align
bulge and disk with an identical axis ratio as a conservative measure. 
Gravitational lensing shear is applied at the catalog level by altering the object ellipticity. This allows
us to minimize the dilution of the lensing signal from the interpolation noise. We first create space-based images and then convolve the results with the DLS-like
PSF. The rest of the simulation follows the steps in our UDF-based analysis.
From this second set of simulations, we determine the mean multiplicative factor to be $\left <m_\gamma \right>=1.05\pm0.01$.  
As we observe that galaxies with analytic profiles tend to require smaller calibration factors, the $\mytilde3$\% decrease in $\left<m_\gamma \right>$
in the latter experiment is consistent with our expectation.
Because it is unlikely that most of the galaxies in the DLS can be decomposed into bulge and disk as are done here,
we can assume that the difference in
the galaxy population between the first and second sets of simulations represents an extreme case\footnote{When 
we relax the bulge-disk alignment constraint, the mean multiplicative factor becomes $\left < m_{\gamma} \right >\sim1.06$, moving closer to the UDF case
$\left < m_{\gamma} \right >\sim1.08$.}. 
Therefore, we adopt the difference in $\left <m_\gamma \right >$ as the maximum deviation due to the sample variance, and we
marginalize over $1.05 < m_{\gamma} < 1.11$ with a flat prior in our cosmological parameter estimation. 

We did not participate directly in previous community shear calibration efforts such as  STEP and GREAT, although it is worth mentioning here that our elliptical Gaussian fitting method is similar to the Bernstein \& Jarvis (2002) method where the authors propose to shear a galaxy image iteratively until it  matches a circular Gaussian. These shear calibration programs provided important contributions by raising the public awareness regarding the key issues of future weak-lensing surveys.  However,  our independent simulations include the following real-world features that both STEP and GREAT have not fully addressed yet. 

First, our training set data include a spatially varying PSF and its estimation through noisy stars. No spatially varying PSF was addressed in STEP and GREAT08 (Bridle et al. 2009). GREAT10 (Kitching et al. 2012) addressed the issue but only in a limited way. In the Galaxy Challenge where the participants are asked to measure pre-seeing ellipticity, the spatially varying PSF was given as a known function. In the separate challenge called  the Star Challenge, the PSF at star positions was also provided as a known function. The only challenge in the latter is to interpolate/extrapolate the PSF to the galaxy location using the known PSFs at star positions. In real-world weak-lensing analysis the PSF must be estimated from a finite number of noisy stars, and this imperfect model PSF is then applied when measuring pre-seeing galaxy shapes. 

Second, we address the effect of galaxy morphology on shear measurement by using both real galaxy images and analytic profiles. 
As mentioned above, our simulation finds that in extreme cases the difference in the multiplicative factor is $\mytilde3$\% (i.e., 1.08 vs. 1.05).
Since the DLS is deep, it is important to include many faint ($\gtrsim24$ ABmag) galaxies, whose high S/N proxy images are only available in the UDF data.

Third, we simulate the effect of object blending.  Both GREAT and STEP have assumed that a galaxy is isolated from the rest of the objects. 
However, when a survey goes deep as the DLS, a significant fraction of the objects overlap with one another.  Obviously, the details of how one treats this blending of objects in source detection and shape measurement affect shear calibration. Since we use the same source detection and shape measurement algorithm for both DLS and training-set data, our shear calibration bias from blended objects is not an issue.

\subsection{Photometric Redshift Estimation  \label{section_photo_z}}

Since the strength of gravitational lensing signal depends on the distance ratios between the observer, lens, and source, one's ability to characterize the amount of bias in photometric redshift estimation is 
as critical as the ability to control shear systematics for any precision cosmic shear analysis.
Because the lensing kernel is broad, individual galaxy photo-$z$ estimation errors
are of less concern than a skewed probability distribution.
On the other hand,  substantial effort should be made to address catastrophic errors, which can
bias knowledge of the overall redshift distribution.
The photometric redshift catastrophic errors
often arise when there are multiple peaks in the estimated probability distribution $p(z)$. In many
cases, these catastrophic errors are inevitable because of the inherent degeneracy between galaxy colors and redshift.
In order to properly interpret the cosmic shear signal amplitude, it is important to understand the direction of the bias and quantify the fraction of catastrophic outliers.

We thus stack probability distributions of individual galaxies (instead of 
single-point, best-fit values) to reconstruct the final redshift histogram of our source population using the following equation:
\begin{equation}
P(z) = A \sum_i p_i(z) \mu_i
\end{equation}
\noindent
where $p_i(z)$ is the redshift probability distribution of an individual galaxy, $\mu_i$ is the weight used for our shear estimation (see \textsection\ref{section_shear}), and $A$ is the normalization constant.
As argued by Wittman (2009), stacking $p(z)$ provides a way to fairly represent the population with multimodal distribution in their $p(z)$. In addition, 
even for galaxies with unimodal redshift distribution, their $p(z)$'s are asymmetric in many cases because of the nonlinear mapping of color space into redshift. 

Detailed description of the DLS photometric redshift estimation is presented by Schmidt \& Thorman (2012), and here we provide a brief summary.
We use the publicly available Bayesian Photometric Redshift code (BPZ; Benitez 2000). 
The six CWW+SB SED templates (Coleman et al. 1980; Kinney et al. 1996) enclosed with
the BPZ code are tweaked so that we improve the agreement between
best-fit values and known spectroscopic redshifts.
We utilize the Smithsonian HEctospec Lensing Survey (SHELS; Geller et al. 2005) spectroscopic redshift data (complete down to $m_R\sim20.7$)
for this template tweaking.

The advantage of the BPZ code is the use of magnitude priors to partially break the color-redshift degeneracy.
We use the data from the VIMOS-VLT Deep Survey (VVDS; Le Fevre et al. 2005) to obtain magnitude- and type-dependent priors for our DLS photometric
redshift estimation:
\begin{equation}
p(z|m,T)=p(T|m)\times p (z|T,m),
\end{equation}
\noindent
where the type dependence $p(T|m)$ is parameterized for three types (E, Sp, and Im/SB) as
\begin{equation}
p(T|m)=f_t \exp{\left [ -k_t (m-20) \right ]} \label{eqn_type}
\end{equation}
\noindent
and the redshift dependence $p (z|T,m)$ is parameterized as
\begin{equation}
p(z|m,T)=z^{\alpha} \exp{ \left [ - \frac{z}{ z_m}  \right ]^{\alpha} }  \label{eqn_redshift}.
\end{equation}
\noindent
In equation~\ref{eqn_type}, $f_t$ and $k_t$ are the type-dependent constants. $z_m$ in equation~\ref{eqn_redshift} is the median
redshift for the magnitude $m$.
Note that the faint tails of the priors are constrained strongly by the above functional forms, not by
the small number of faint galaxies in the VVDS sample.
The comparison of our priors with those from the Hubble Deep Field (HDF) shows that the two sets of priors are
very similar to each other. If we switched to HDF priors, this would
shift the mean redshift of our DLS source galaxies by $\mytilde3$\%.

\begin{figure*}
\includegraphics[width=8.8cm]{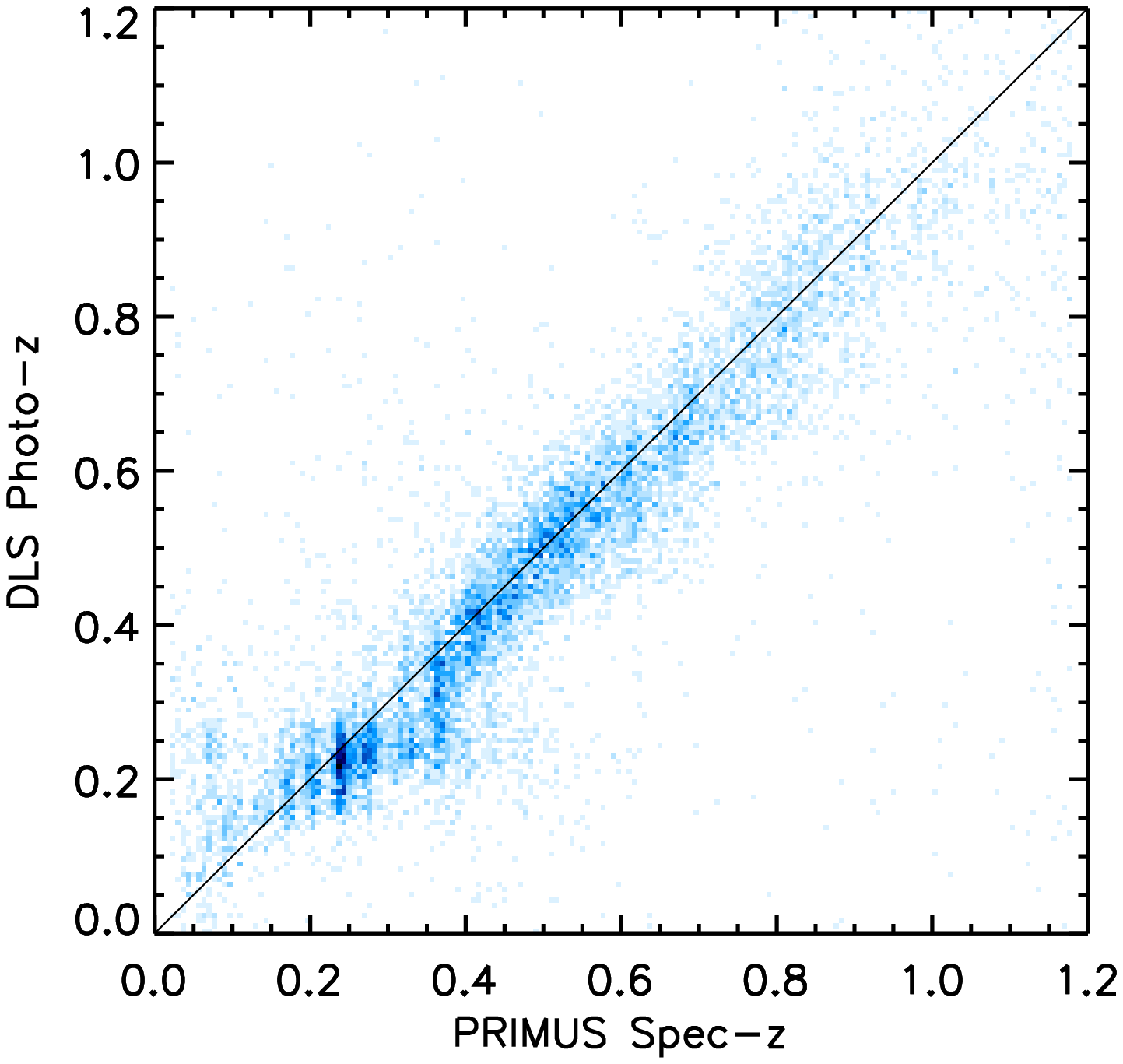}
\includegraphics[width=8.8cm]{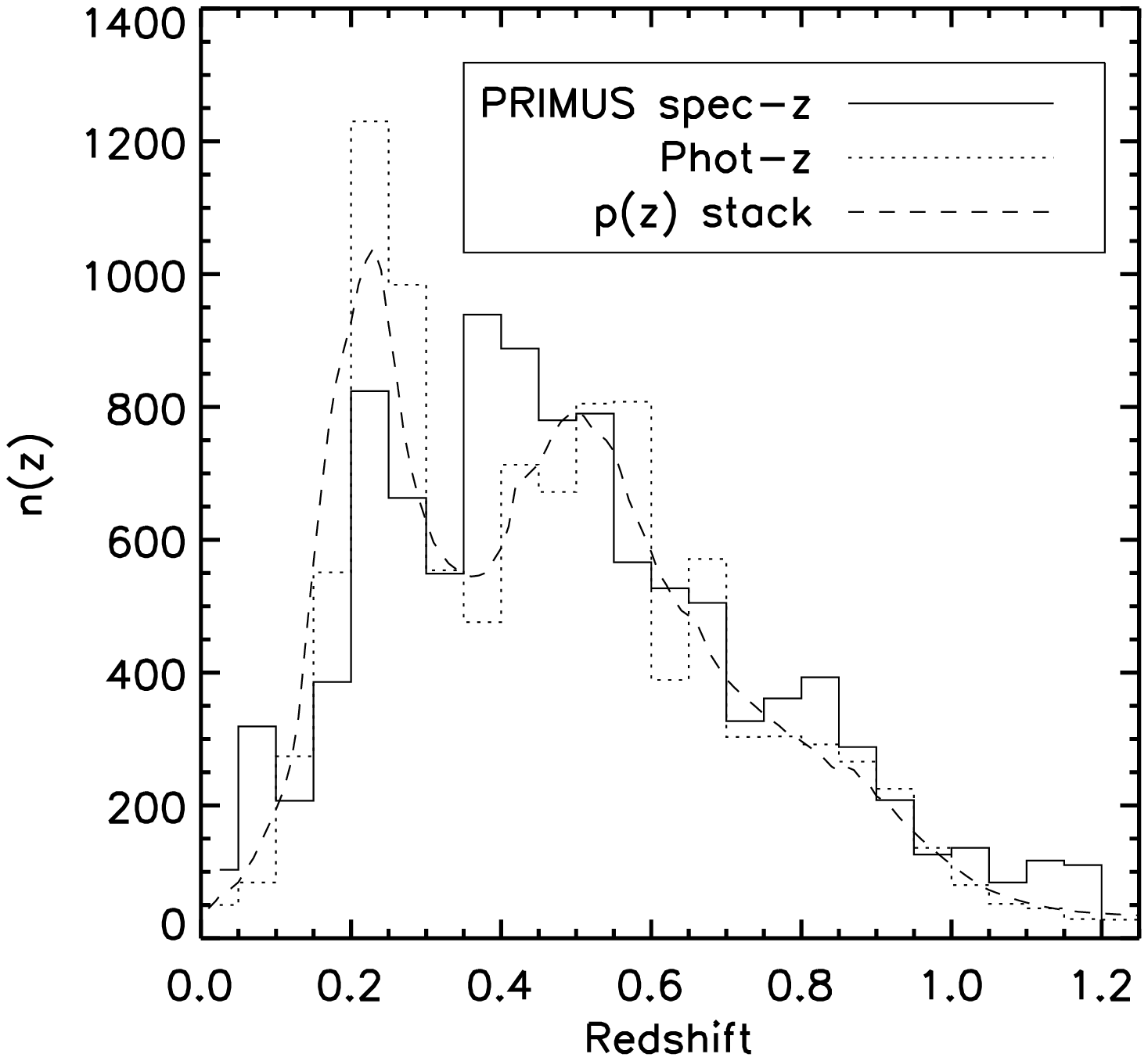}
\caption{Comparison of spectroscopic redshift with photometric redshift in DLS. In this plot, we display 10,231 objects in F5, whose spectroscopic redshifts were measured
by the PRIMUS project and quality flags ({\tt ZCONF}) are greater than 2. The plot on the left panel compares the PRIMUS spec-$z$'s with DLS photo-$z$'s directly whereas
the plot on the right compares the distribution of the redshifts. The $p(z)$ curve was obtained by stacking the photometric redshift probability distribution of individual galaxies.
 \label{fig_spec_z_vs_phot_z}} 
\end{figure*}

\begin{figure}
\includegraphics[width=8.8cm]{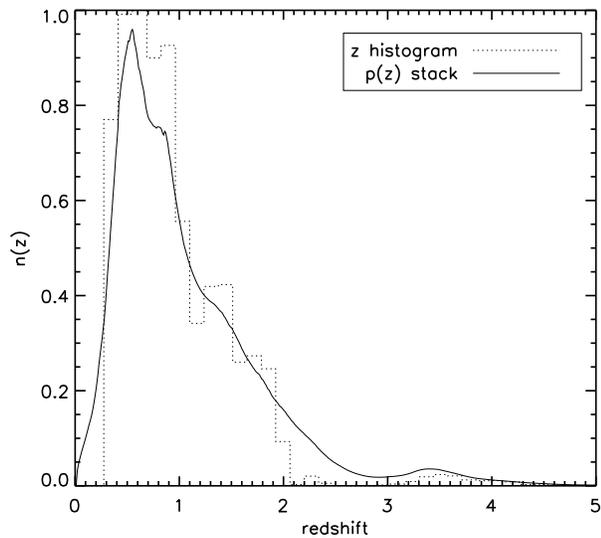}
\caption{Redshift distribution of source galaxies. The histogram constructed from best-fit photo-{\it z} values is compared with the $p(z)$ curve obtained by stacking individual 
probability distributions. We discard the sources with $z_{phot}<0.3$ because our systematics in photo-$z$ estimation
might be slightly biased at $z_{phot} \lesssim 0.2$, and because throwing away these low redshift galaxies removes a large fraction of luminous red galaxies (\textsection\ref{section_source_selection}).
\label{fig_p_of_z}}
\end{figure}

As a spectroscopic test for fainter galaxies, we use an independent survey.
The comparison between our DLS photo-$z$ and the PRIsm MUlti-object Survey (PRIMUS; Coil et al. 2011) spec-$z$ is shown in Figure~\ref{fig_spec_z_vs_phot_z}
for one field. The SHELS galaxies are $\mytilde2$ mag brighter than the PRIMUS galaxies. Hence, this PRIMUS spec-z vs. our DLS photo-$z$ comparison is a fair method to evaluate the performance of the above 
template tweaking carried out with the SHELS data.
We compare 10,231 objects in F5, whose spectroscopic redshifts were measured
by the PRIMUS project and quality flags ({\tt ZCONF}) are greater than 2.
Although  some systematic errors are indicated at the low ($z_{spec}\lesssim0.15$) redshift end, the overall agreement is excellent between
$z_{spec}$ and $z_{phot}$; Schmidt \& Thorman (2012) show that the agreement is even better using stacked $p(z)$.
The PRIMUS data is approximately complete down to $m_{r}\sim23$ and about $\mytilde1,100$ redshifts are available
for the objects in the $23\lesssim m_{r}\lesssim24$ range. Hence, it is reasonable to assume that the DLS photo-$z$'s are of good quality for the
$m_R\lesssim24$ galaxies, which accounts for $\mytilde57$\% of our source population. We estimate that
the aforementioned systematics at $z_{spec}\lesssim0.15$ would affect the amplitude of the predicted shear signal
by less than $\sim2$\% because the lensing kernel for our source population is broad.
Beyond $m_R\sim24$, the fraction of catastrophic errors is expected to increase more rapidly with magnitude, and thus we argue that stacking $p(z)$
should give a more realistic representation of the redshift distribution of the source population in this regime, too.
We display the stacked $p(z)$ of the entire source population in Figure~\ref{fig_p_of_z}. Also plotted is the $n(z)$ histogram computed from
single-point photometric redshifts.  The single-point photo-$z$ histogram virtually truncates at $z\sim2$, and this feature is rather unphysical.
Typically, cosmic shear studies parameterize the redshift distribution with an analytic form (e.g., $ p(z) \propto (z/z0)^{\alpha} \exp [ - (z/z_0)^\beta ]$).
Applying the method would smooth the distribution and recover the high-redshift tail as shown by the $p(z)$ curve.

The role of the magnitude prior becomes more  important progressively with magnitude as the SED constraint weakens.
At the faint end, we expect that the redshift probability $p(z)$ of a nontrivial fraction of galaxies may default to the prior. If we had taken a prior from small-field results
such as the HUDF studies (Coe et al. 2006), the sample variance would have been a  dominant source of bias for the $m_R \gtrsim 24$ population.
However, because the VVDS prior used here is obtained from a relatively large survey ($\mytilde$2,000 times larger than the HDFN in area), the impact of the sample variance on our photometric redshift estimation in DLS should be small. 

Therefore, it is fair to argue that the accuracy of the current $p(z)$ for the $m_R\gtrsim24$ sources is limited by the systematics in the VVDS prior itself.
Unfortunately, there is currently no solid method available for testing the systematics of the prior. Possible sources of systematics include
the effects of the galaxy SED evolution, inclination-dependent reddening, non-Gaussian photometric redshift errors, etc.
For our cosmological parameter estimation, we assume a $\mytilde 3$\% systematic error in the source mean redshift and marginalize 
over this interval; we implemented this by running  a Monte Carlo Markov Chain (MCMC) while compressing/stretching the $p(z)$ curve horizontally by
a random factor drawn from the [0.97,1.03] interval for each chain.
This 3\% interval is the difference of the mean redshift when we compare the results from the VVDS priors and the HDF priors as mentioned above.
In fact, the SED constraint is strong for a significant fraction of our source galaxies. Therefore, the above $\mytilde 3$\% systematic error is
a conservative value. 

\subsection{Source Galaxy Selection} \label{section_source_selection}

\begin{deluxetable*}{ll}
\tabletypesize{\scriptsize}
\tablecaption{Parameters Used for Source Selection}
\tablenum{2}
\tablehead{\colhead{Name} &  \colhead{Condition} \\} 
\tablewidth{0pt}
\startdata
$R$-band magnitude                             &   $21<m_R<27$  \\
ellipticity measurement error               &   $\delta e < 0.25$  \\
photometric redshift                               &  $z_{phot} > 0.3$  \\
semi-minor axis                                      &  $b > 0.4$ pixels  \\
elliptical gaussian fitting convergence  &  {\tt STATUS = 1}  \\
source extraction flag                             & {\tt FLAG $\neq$ 8, 16, 32} \\
masking                                                   &  excluded if near or within the region\\
\enddata
\end{deluxetable*}

\begin{figure*}
\includegraphics[width=18cm]{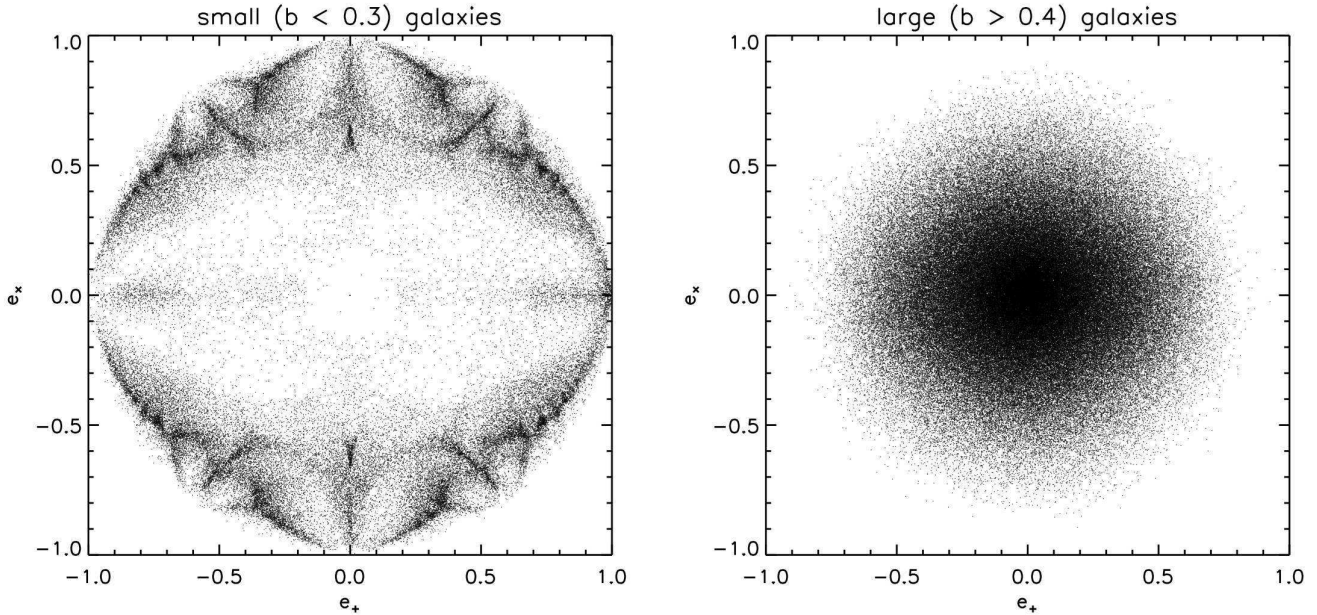}
\caption{Effect of object size on ellipticity measurement. Extremely small objects show
distinctive patterns regardless of their S/N
in their ellipticity distribution mainly due to the pixellation, which disappears for relatively large objects.
We threshold the semi-minor axis $b$ measured from our elliptical Gaussian fitting to be greater than 0.4 pixels in order
to make the pixellation effect negligible. The objects in both panels have a ``reported'' ellipticity measurement error less than 0.25.
\label{fig_small_galaxy_ellipticity}}
\end{figure*}

\begin{figure}
\includegraphics[width=8.8cm]{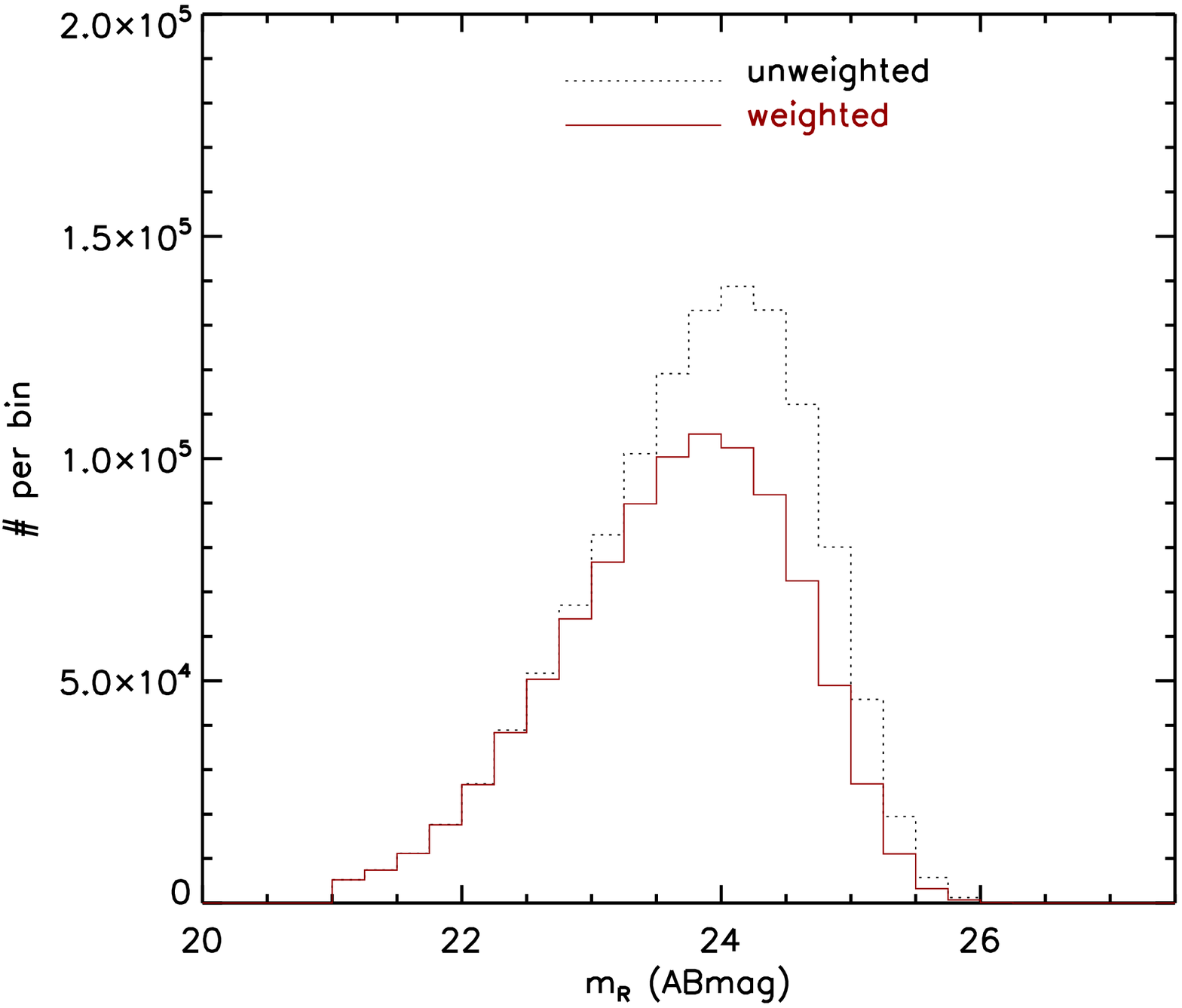}
\caption{$R$-band magnitude distribution of source galaxies. We select a total of $\mytilde$1.2 million galaxies from the five DLS fields. When we convert galaxy ellipticities into shears,  the measured ellipticity is weighted with the inverse variance, which is discussed in the text. The red histogram shows the magnitude distribution when this weighting is considered. The black dotted histogram displays the raw magnitude distribution without this weighting. The mean shear calibration factor should be derived using the red histogram.
\label{fig_source_mag_dist}}
\end{figure}

Our source selection is based on various parameters. Table 2 summarizes the names of these parameters and the values used as threshold.  
The lower magnitude cut is needed to prevent an accidental inclusion of very low redshift galaxies and saturated objects. The upper limits for
the magnitude and ellipticity error disallow spurious detections to be included.
We set the lower bound of the photometric redshift to be $z_{phot} > 0.3$ because there is some indication that our systematics in photo-$z$ estimation
might be slightly biased at $z_{phot} \lesssim 0.2$, and because throwing away these low redshift galaxies minimizes a potential contamination
of our cosmic shear from intrinsic alignments.
The lower limit in galaxy size is required to avoid stars and also very small galaxies whose shape measurements
are noisy and biased because of the pixellation effect.
In many cases, despite their misleadingly small ellipticity measurement errors, the ellipticities of these small galaxies show
distinctive patterns as shown in Figure~\ref{fig_small_galaxy_ellipticity}.
We threshold the semi-minor axis $b$ measured from our elliptical gaussian fitting to be greater than $\mytilde 0.3$ pixels in order
to make the pixellation effect negligible.
The {\tt MPFIT} minimization engine reports the status of the convergence. Although the author comments that all values greater than zero in the keyword {\tt STATUS}
can represent success, we select the objects only with {\tt STATUS $=1$}, which is the most conservative indicator of the convergence.
The SExtractor software saves the history of the source extraction in the binary switch format in the {\tt FLAG} parameter.
We discard an object if any of the third (8), fourth (16), or fifth (32) bits is turned on.  This helps us to exclude the objects on or close to
image borders. We mask out the regions affected by bright stars (PSF wings and bleeding streaks).
The resulting total number of source galaxies in our $\mytilde20$ sq. degree area is $\mytilde1.2$ million, which gives a mean number
density of $n_{source}\sim17$ galaxies per sq. arcmin. Without the $z_{phot} > 0.3$ threshold, this number density would increase to $n_{source}\sim21$ galaxies per sq. arcmin.
The field-to-field variation of $n_{bg}$ is listed in Table 1.
The magnitude distribution of the source population is shown in Figure~\ref{fig_source_mag_dist}.

\subsection{Intrinsic Alignment and Luminous Red Galaxies}

A potentially important systematic in cosmic shear of astrophysical origin is intrinsic alignment of source galaxies (e.g., Hirata \& Seljak 2004).
If a large-scale tidal gravitational field significantly affects intrinsic alignments, our interpretation of cosmic shear measurements
must include these intrinsic ellipticity correlations. Ellipticity correlations between galaxies subject to a common large
scale gravitational field are often called the intrinsic-intrinsic (II) signal whereas the correlation between
foreground galaxy ellipticity (tidal) and background galaxy shear (gravitational lensing) is called  the gravitational-intrinsic (GI) signal.

In tomographic studies,  the auto-correlation
function within a narrow redshift shell may be severely influenced by the II signal. However, in the current  non-tomographic
study, the II signal is expected to be negligible because galaxy pairs within a close
redshift interval are substantially outnumbered by those with very different redshifts. In addition, because the II signal increases
for decreasing angular scale, we can mitigate this potentially small contribution further by discarding close pairs.

However, the GI signal may be non-negligible even in non-tomographic studies because the fraction of foreground-background
pairs is certainly overwhelming. Mandelbaum et al. (2006) report a significant detection of the GI signal from the Sloan
Digital Sky Survey (SDSS). They conclude that luminous red galaxies (LRGs) are the main sources of the intrinsic
alignment and project that cosmic shear surveys at $z\sim1$ may underestimate the linear amplitude of fluctuations
as much as 20\%.

We address removing LRGs from our source catalog by identifying the population
with our photometric redshift catalog. We classify galaxies as LRGs
whose SED template is consistent with the elliptical galaxy template and absolute
magnitude is brighter than $M_R < -20$. We limit the maximum redshift to be $z_{phot} < 0.9$
because LRGs at higher redshift should provide negligible contribution to the GI signal.
From this procedure, we discard about 5\% ($\sim61,200$ objects) of our original source galaxies.

A comparison of the cosmic shear signals with and without the LRGs shows that the
shift in signal amplitude is much smaller than the statistical errors and thus does not affect
our cosmological parameter estimation. This non-detection of the GI effect is rather unexpected.
However, given the depth of the DLS and thus the fact that 
most lensing signals come from high redshift, it is possible that the GI contribution may become relatively insignificant.
In addition, because we do not use galaxies at $z_{phot}<0.3$,  we suspect that the GI
effect is already suppressed even before we remove LRGs.

No significant detection of the GI effect was  claimed in some previous studies.
Fu et al. (2008) report that their marginalized likelihood analysis with the CFHT data
does not show any negative GI signal as predicted by the theory. Schrabback et al. (2010)
state that their redshift scaling analysis is not affected by the presence of LRGs.
The early CFHT study by Fu et al. (2008) may have suffered from non-negligible weak-lensing systematics
(Heymans et al. 2012, Kilbinger et al. 2009). Also, the redshift scaling test by Schrabback et al. (2010)
is not very sensitive to the GI signal. Hence, it is premature to draw any firm conclusion from these results.
For the DLS, we have yet to perform a full analysis of the intrinsic alignment, including source galaxies at $z_{phot}<0.3$.
We thus drop these galaxies from our sample.

\section{COSMIC SHEAR MEASUREMENT} \label{section_measurements}
\subsection{Theoretical Background } \label{section_equations}

We carry out cosmic shear analysis with two-point shear-shear correlation functions. These second-order statistics and their derived quantities 
have been widely used, and the mathematical tools and the algorithms
are in a mature stage. 
The computation of correlation functions is time-consuming if brute-force algorithms are employed.
Therefore, we developed a fast tree-code, which closely approximates the result from the exact brute-force method.
Our tree-code is cross-checked against some publicly available codes \footnote{e.g., http://www2.iap.fr/users/kilbinge/athena/,
http://code.google.com/p/mjarvis/, etc.}.  We note that although we used our fast tree-code for intermediate steps, 
the final results presented in this paper are obtained through our brute-force two-point correlation estimation code. 

The shear-shear correlations are evaluated as
\begin{equation}
\xi_{tt} (\theta) = \frac{\Sigma_{i,j} w_i w_j e_{t,i} e_{t,j}} {\Sigma w_i w_j} \label{eqn_xi_tt}
\end{equation}
and
\begin{equation}
\xi_{\times\times} (\theta) = \frac{\Sigma_{i,j} w_i w_j e_{\times,i} e_{\times,j}} {\Sigma w_i w_j}. \label{eqn_xi_xx}
\end{equation}
\noindent
where the summation is carried out over every possible combination of $i^{th}$ and $j^{th}$ galaxies ($i< j$).
The two subscripts
$t$ and $\times$ refer to the two projections of the ellipticity along the tangential and 45$\degr$ angle with respect to the line connecting
the galaxy pair, respectively.   $w_{i (j)}$ is the weight associated with the ellipticity of the $i(j)^{th}$ galaxy.
The angle between the galaxy pair is $\theta$.

The two following linear combinations of $\xi_{tt}$ and $\xi_{\times\times}$ are useful to express
the derived shear statistics that we discuss hereafter:
\begin{equation}
\xi_+ = \xi_{tt} + \xi_{\times\times}
\end{equation}
and
\begin{equation}
\xi_- = \xi_{tt} - \xi_{\times\times}.
\end{equation}

Derived shear statistics commonly used include shear variance $\left <\gamma^2_{E/B} \right >$, aperture mass $\left <M^2_{E/B} \right >$, and correlation function $\xi_{E/B}$, where
the subscripts E and B denote the contributions from the so-called E- and B-mode signals, respectively. The E/B signals are analogous to electric versus magnetic fields in the electromagnetic theory, where the former is the gradient of a scalar field and the latter
is the curl of a vector field. Because gravitational lensing only produces E-mode (curl free) signals, in principle the B-mode signal must be consistent with zero.
As image distortions due to systematics (e.g., inaccurate PSF correction) are not always curl-free, the B-mode measurement is frequently used as a diagnostic of the residual systematic
errors.

The top-hat shear variance is given by
\begin{equation}
\scriptsize
\left <\gamma^2_{E/B} \right > (\theta) = \frac{1}{2\theta^2} \int_0^{\infty}  \vartheta d \vartheta \left [ S_+ \left (\frac{\vartheta}{\theta} \right ) \xi_+(\vartheta) \pm S_- \left (\frac{\vartheta}{\theta} \right ) \xi_-(\vartheta)  \right ]. \label{eqn_tophat}
\end{equation}
\noindent
The aperture mass statistic is
\begin{equation}
\scriptsize
\left <M^2_{E/B} \right > (\theta) = \frac{1}{2\theta^2} \int_0^{2\theta}  \vartheta d \vartheta \left [ T_+ \left (\frac{\vartheta}{\theta} \right ) \xi_+(\vartheta) \pm T_- \left (\frac{\vartheta}{\theta} \right ) \xi_-(\vartheta)  \right ]. \label{eqn_ap_mass}
\end{equation}
\noindent

The difference between equations~\ref{eqn_tophat} and~\ref{eqn_ap_mass} is the use of different filter functions ($S_{+,-}$ and $T_{+,-}$), which sample different angular parts of $\xi_{+,-}$.
The filter functions $S_{+,-}$ and $T_{+,-}$ are defined in Schneider et al. (2002).
The E- and B-mode decomposition for the correlation function $\xi$ requires the definition of the following quantity $\xi^{\prime}$:
\begin{equation}
\scriptsize
\xi^{\prime} (\theta) = \xi_- (\theta) + 4 \int_{\theta}^{\infty} d \vartheta \frac{\xi_- (\vartheta)}{\vartheta}   -  12 \theta^2 \int_{\theta}^{\infty} d \vartheta \frac{\xi_- (\vartheta)}{\vartheta^3},
\end{equation}
which is also the result obtained by filtering $\xi_{-}$. Using $\xi^{\prime}$, we can define $\xi_{E/B}$ as follows:
\begin{equation}
\xi_{E/B} (\theta) = \frac{ \xi_+(\theta) \pm \xi^{\prime} (\theta) } {2}.  \label{eqn_xi_enb}
\end{equation}

The above three lensing statistics assume that we can measure the $\xi_{+/-}$ data on arbitrarily small and large scales. Therefore, when
these statistics are evaluated using real data with a cutoff at both ends, the resulting E/B-decomposition values deviate from the theoretical
ones (e.g., Kilbinger et al. 2006). This limitation motivates some authors to suggest new E/B-mode statistics that do not
suffer from the finite-interval problem (e.g., Schneider \& Kilbinger 2007; Eifler et al. 2010; Fu \& Kilbinger 2010; Schneider et al. 2010; Becker 2012). Among these, we consider the ring statistics (Schneider \& Kilbinger 2007) using a scale-dependent integration limit $\eta$ suggested by Eifler et al. (2010). The parameter $\eta$ refers to the ratio of the smallest separation to the largest separation $\eta=\theta_{min}/\theta$.
The ring statistics are evaluated in a similar fashion as above except that the integration limit is over a finite interval $[\eta\theta,\theta]$ as the following:
\begin{equation}
\scriptsize
\left < RR_{E/B} \right > (\theta) = \frac{1}{2} \int_{\eta\theta}^{\theta} \frac{\mbox{d}\vartheta}{\vartheta} \left [ Z_+ (\vartheta,\eta) \xi_+ (\vartheta)
 \pm Z_- (\vartheta,\eta) \xi_- (\vartheta) \right ], 
\end{equation}
where the filter function $Z_{+,-}$ is defined in Schneider \& Kilbinger (2007).

Now, in order to compare the above statistics obtained from our DLS data with the prediction for a given cosmology, we need a method to predict the signal.
We begin with the following cosmic convergence/shear power spectrum:
\begin{equation}
\scriptsize
P_\kappa^{kl} (\ell) = \frac{9}{4} \Omega_M^2 \left ( \frac{H_0}{c} \right )^4 \int_0^{\chi_{max}} d \chi \frac{ g_k (\chi) g_l (\chi) }{a^2 ( \chi ) } P_{\delta}  \left ( \frac{\ell}{f_K (\chi)}, \chi \right ), 
\end{equation}
\noindent
where $H_0$ is the Hubble parameter, $\Omega_M$ is today's matter density, $a (\chi) $ is the scale factor at a redshift corresponding to a comoving distance $\chi$, $f_K$ is the
comoving angular diameter distance, $P_{\delta}$ is the linear power spectrum,
and $g_k$ is the lensing efficiency factor:
\begin{equation}
g_k(\chi) = \int_{\chi}^{\chi_{max}} d \chi^{\prime} p_k (\chi^{\prime}) \frac{ f_K (\chi^{\prime} - \chi) }{f_K (\chi^{\prime} ) }
\end{equation}
for the $k^{th}$ redshift shell with a redshift distribution $p_k(\chi)$. Note that only the redshift probability $p_k(\chi)$ connects the survey to the shear power spectrum.
Also keep in mind that in the current non-tomographic study only a single broad redshift shell is used.

With this convergence  power spectrum $P^{kl}_{\kappa}$, we can obtain predicted cosmic shear statistics for any of the above.
For example, the basic correlation function $\xi_{+,-}$ is the convolution of the shear power spectrum with a Bessel function $J_{0,4}$ (Crittenden et al. 2002).
That is,
\begin{equation}
\xi_{+,-}^{k l} (\theta) = \frac{1}{2 \pi} \int_0^{\infty} d \ell \ell J_{0,4} (\ell \theta) P_{\kappa}^{k l} (\ell). \label{eqn_xi}
\end{equation}

In principle, the aperture mass statistics $\left <M^2_{E/B} \right >$ and $ \left <\gamma^2_{E/B} \right>$ can be evaluated by substituting these $\xi_{+,-}$ functions into Equations~\ref{eqn_tophat} and~\ref{eqn_ap_mass}, respectively. However, it is more convenient to obtain them by directly convolving the shear power spectrum $P^{kl}_{\kappa}$ with corresponding kernels in a similar way to Equation~\ref{eqn_xi}.

\subsection{Weak-lensing Systematics in DLS}

One, if not the most,  critical source of systematics in weak lensing is the inaccurate removal of the PSF effects from shear measurements.
The problem arises either because the PSF model is in error or because the removal procedure is suboptimal despite the correct model.
In any case, the effects of this PSF correction error can be classified into two kinds: shear calibration and residual anisotropy.
The degree of the shear calibration and anisotropy systematics are often parameterized by the following
multiplicative $m_{\gamma}$ and additive factors $c_{\gamma}$, respectively:
\begin{equation}
{\gamma}= m_{\gamma} \hat{\gamma} + c_{\gamma},
\end{equation}
\noindent
where $\hat{\gamma}$ and $\gamma$ are the weight-averaged ellipticity and true shear, respectively.
The multiplicative factor $m_{\gamma}$ can also be affected by the degree of complexity in galaxy shape modeling, but
we do not distinguish the effect of PSF from that of the galaxy shape modeling in this study. Instead, we determine the
global value of $m_{\gamma}$ from the image simulations described in \textsection\ref{section_shear_calibration}.

\subsubsection{Star-Galaxy Correlation} \label{section_star_galaxy}

The additive error is mostly caused by imperfect PSF-induced anisotropy removal.
A useful diagnostic of the additive error $c_{\gamma}$ is the following star-galaxy correlation (Bacon et al. 2003):
\begin{equation}
\xi_{\mbox{sys}}  = \frac{ \left < e^* \gamma \right >^2 }  {\left < e^* e^* \right >}, \label{eqn_star_galaxy}
\end{equation}
\noindent
where $e^*$ is the ellipticity of uncorrected stars. The intrinsic size of the PSF ellipticity depends on many conditions and varies
widely. Equation~\ref{eqn_star_galaxy} makes sure that the amplitude of  the star-galaxy correlation $\left < e^* \gamma \right >$ is
normalized by this intrinsic size of the PSF ellipticity to enable a fair comparison with other observations.
However, occasionally  the intrinsic PSF correlation $\left < e^* e^* \right >$ crosses zero  where the transitions between
positive and negative correlations occur. In these cases, $\xi_{\mbox{sys}}$ can rise abruptly because of the small denominator if $\left < e^* \gamma \right >$
is indeed uncorrelated with $\left < e^* e^* \right >$.  We use a mean amplitude of $\left < e^* e^* \right >$  in the $10\arcmin < \theta <100\arcmin$ range
to avoid this artifact.

\begin{figure*}
\includegraphics[width=6cm]{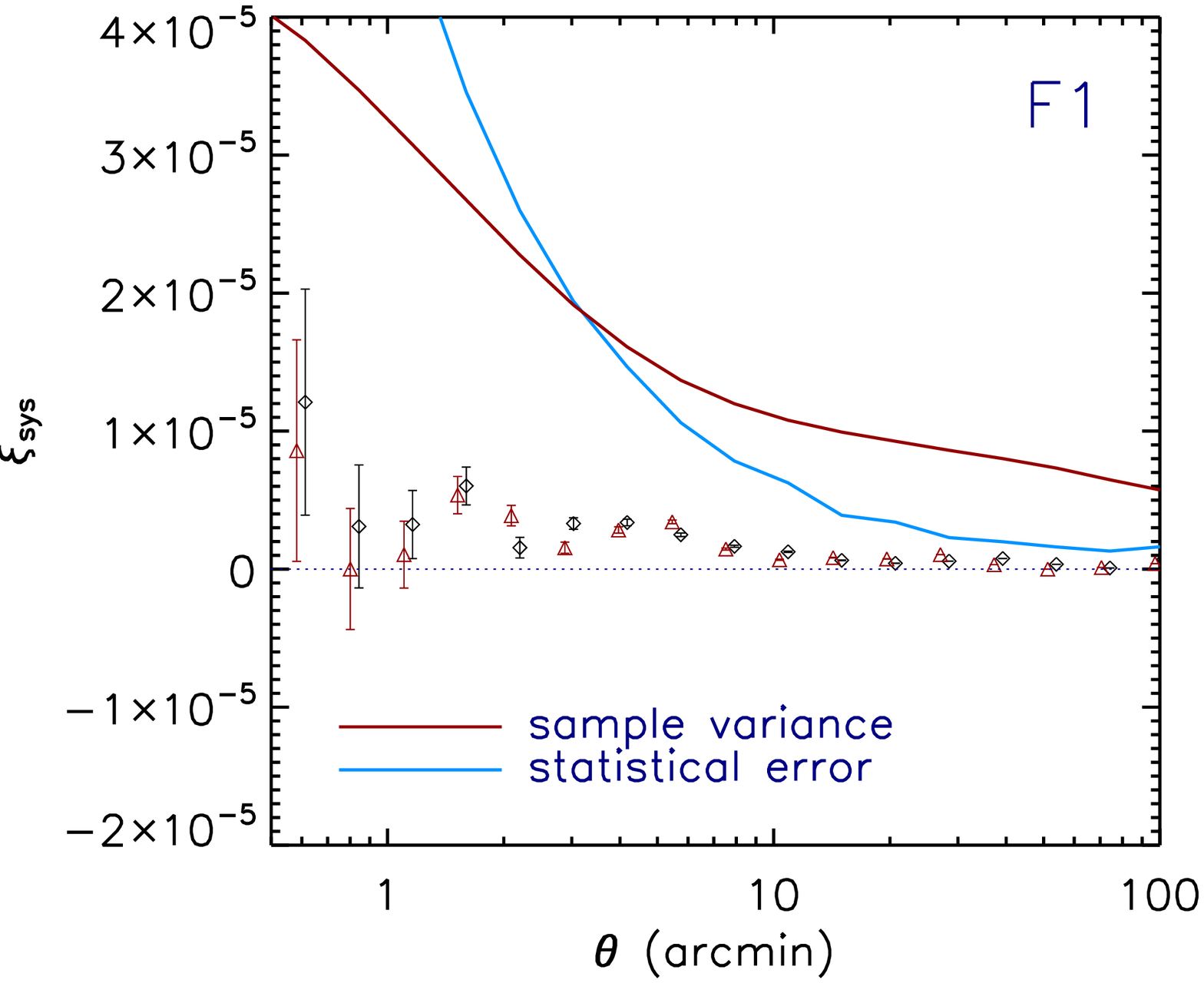}
\includegraphics[width=6cm]{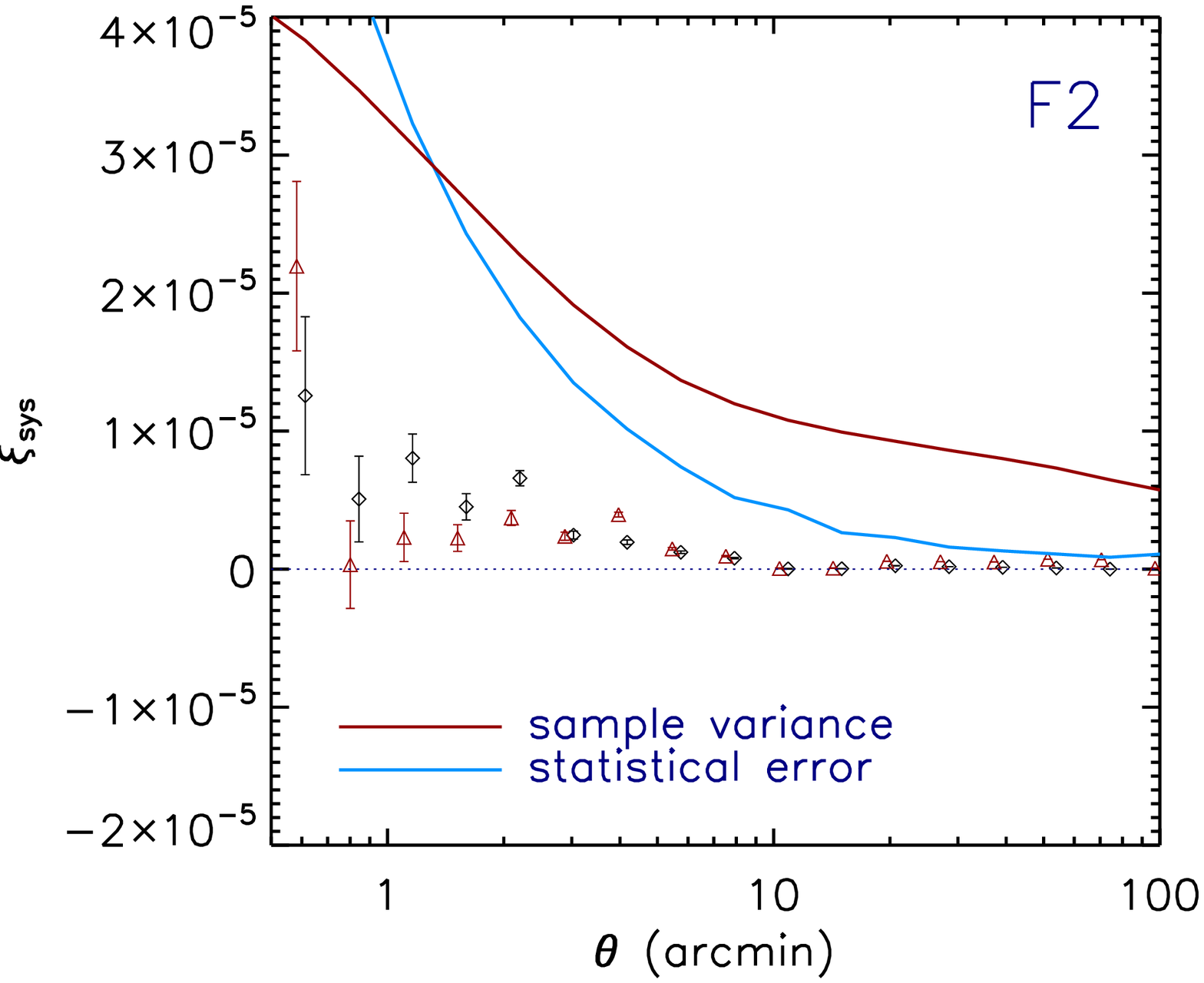}
\includegraphics[width=6cm]{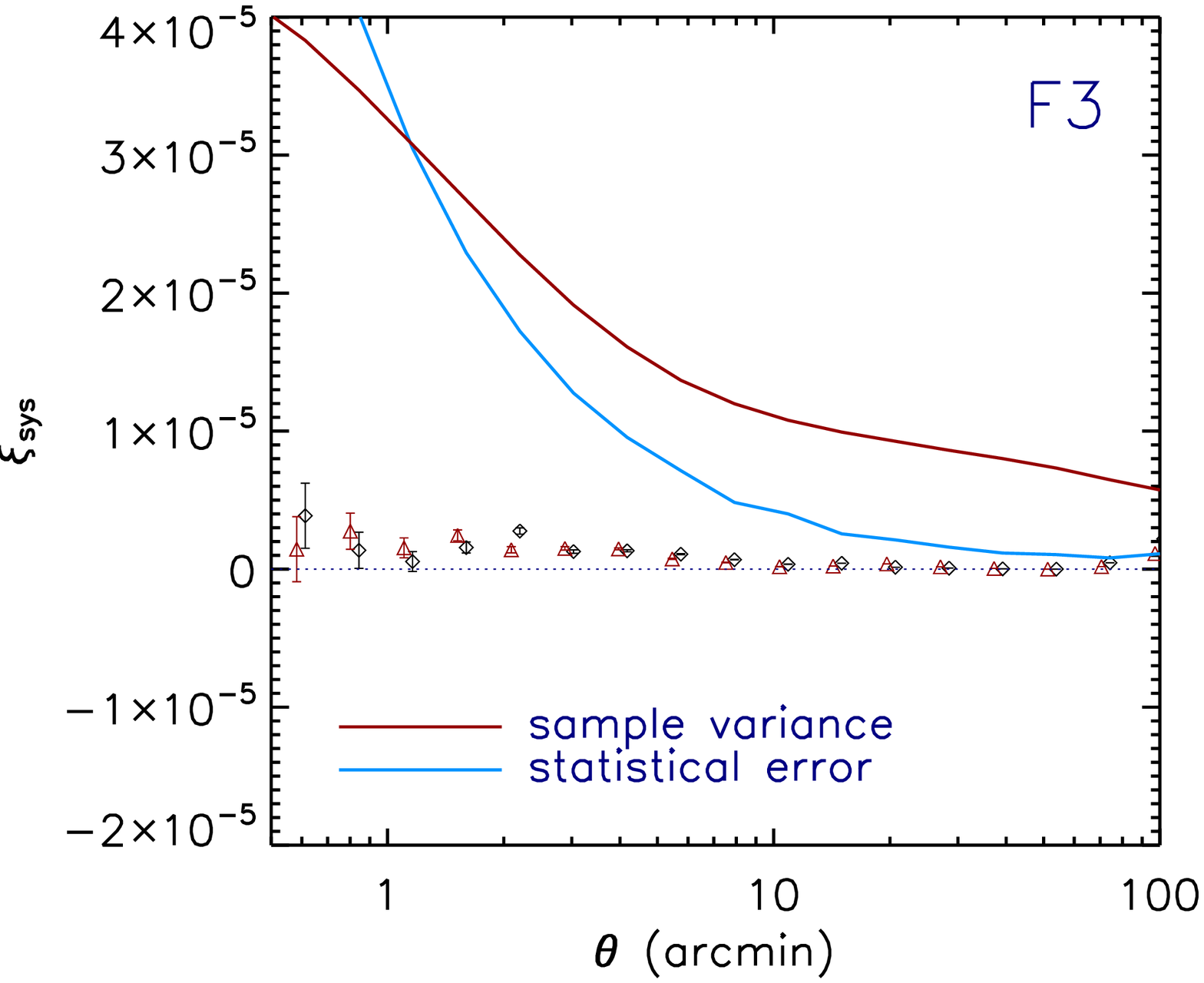}
\includegraphics[width=6cm]{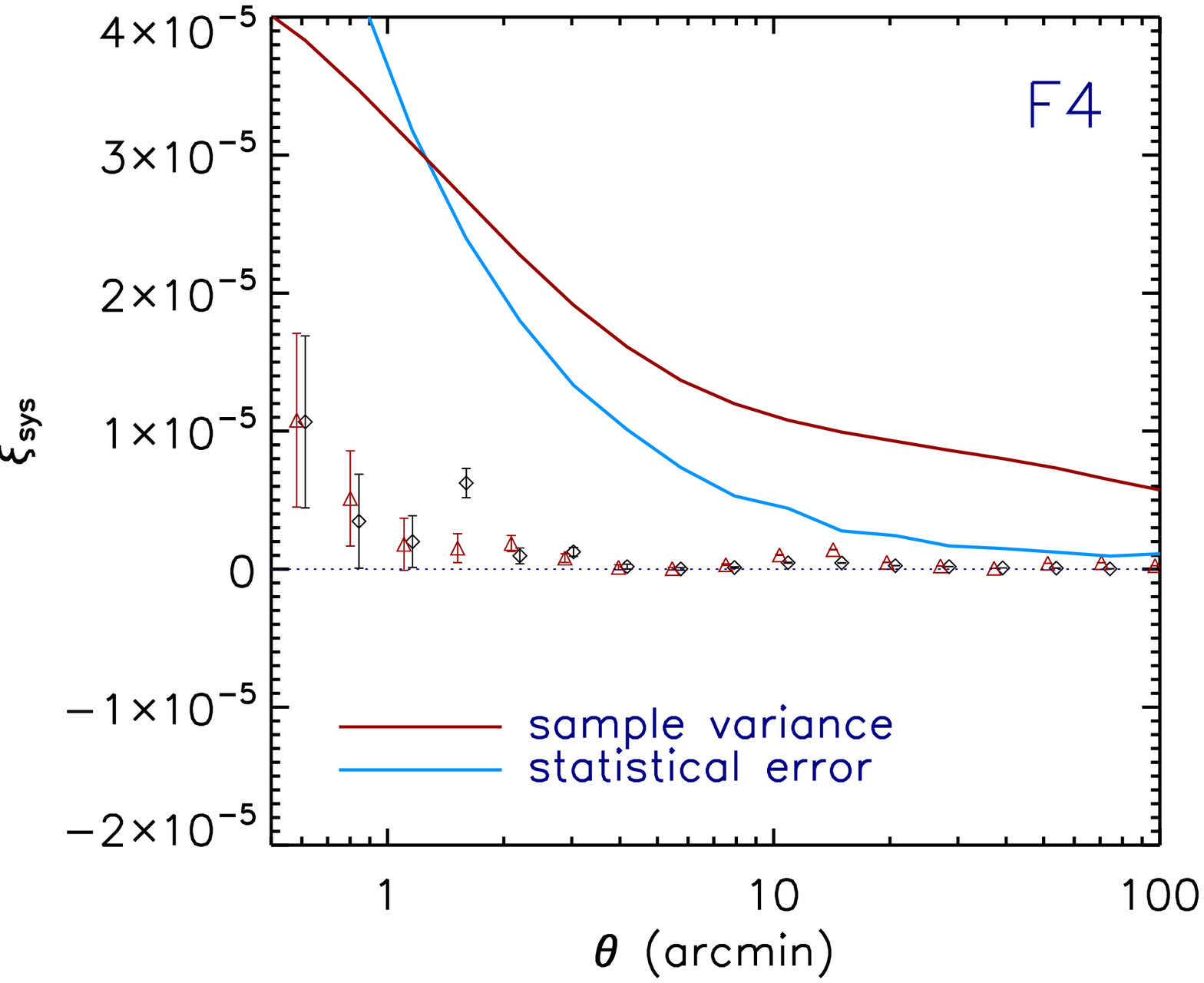}
\includegraphics[width=6cm]{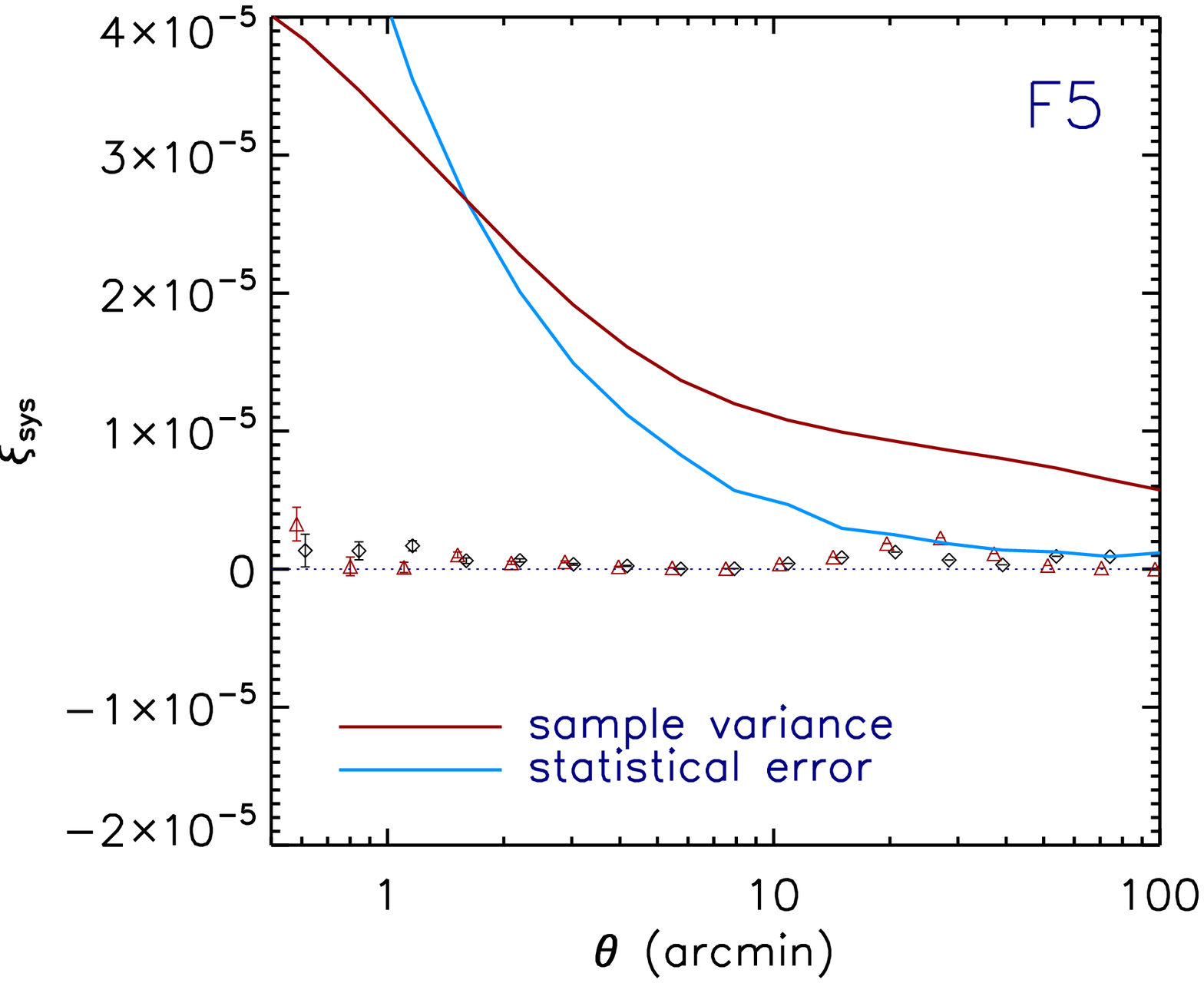}
\caption{Star-galaxy correlation. 
The diamond and triangle symbols represent the tangential ($\left < e_{+}^* \gamma_{+} \right > $) and cross ($\left < e_{\times}^* \gamma_{\times} \right > $)  components  of star-galaxy pairs
divided by auto-correlation functions of stellar ellipticity. Overall, the amplitude of the star-galaxy correlation is far smaller than the cosmic shear uncertainties evaluated for the field.
The sample variance is estimated using the Sato et al. (2011) method for 4 sq. deg. The statistical error is computed by counting the number of pairs in each bin. \label{fig_star_galaxy}}
\end{figure*}

\begin{figure*}
\includegraphics[width=6cm,height=5cm]{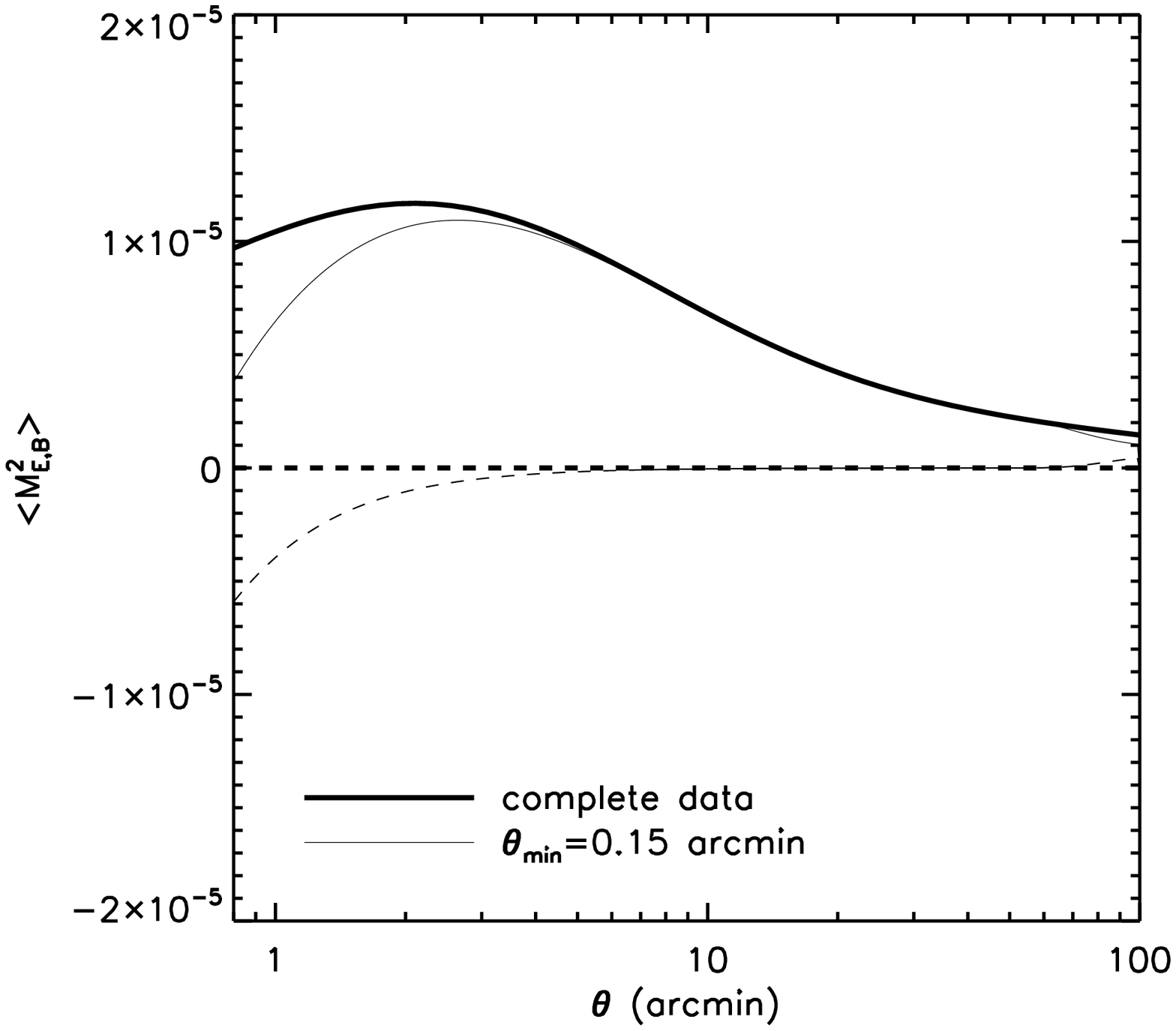}
\includegraphics[width=6cm,height=5cm]{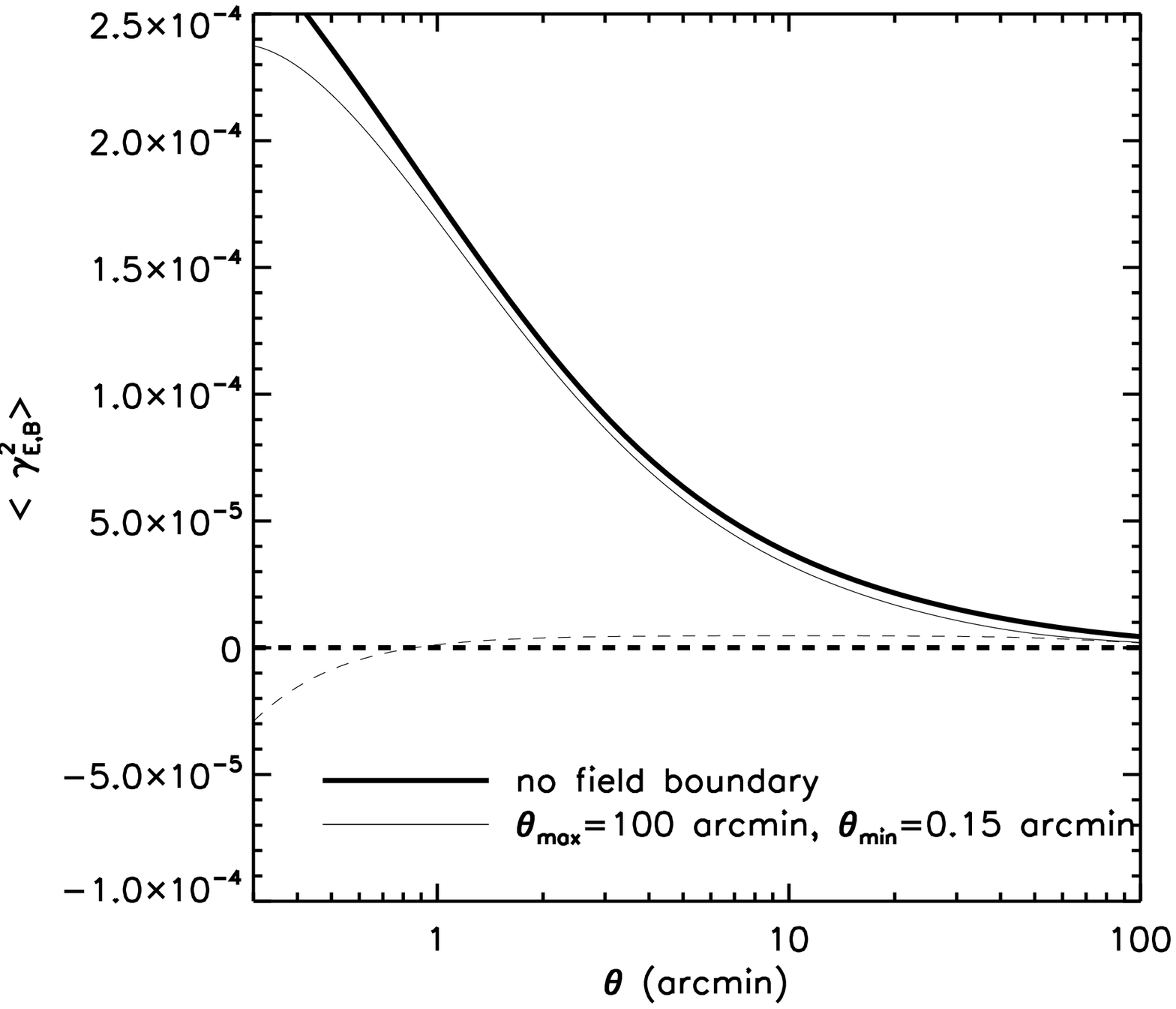}
\includegraphics[width=6cm,height=5cm]{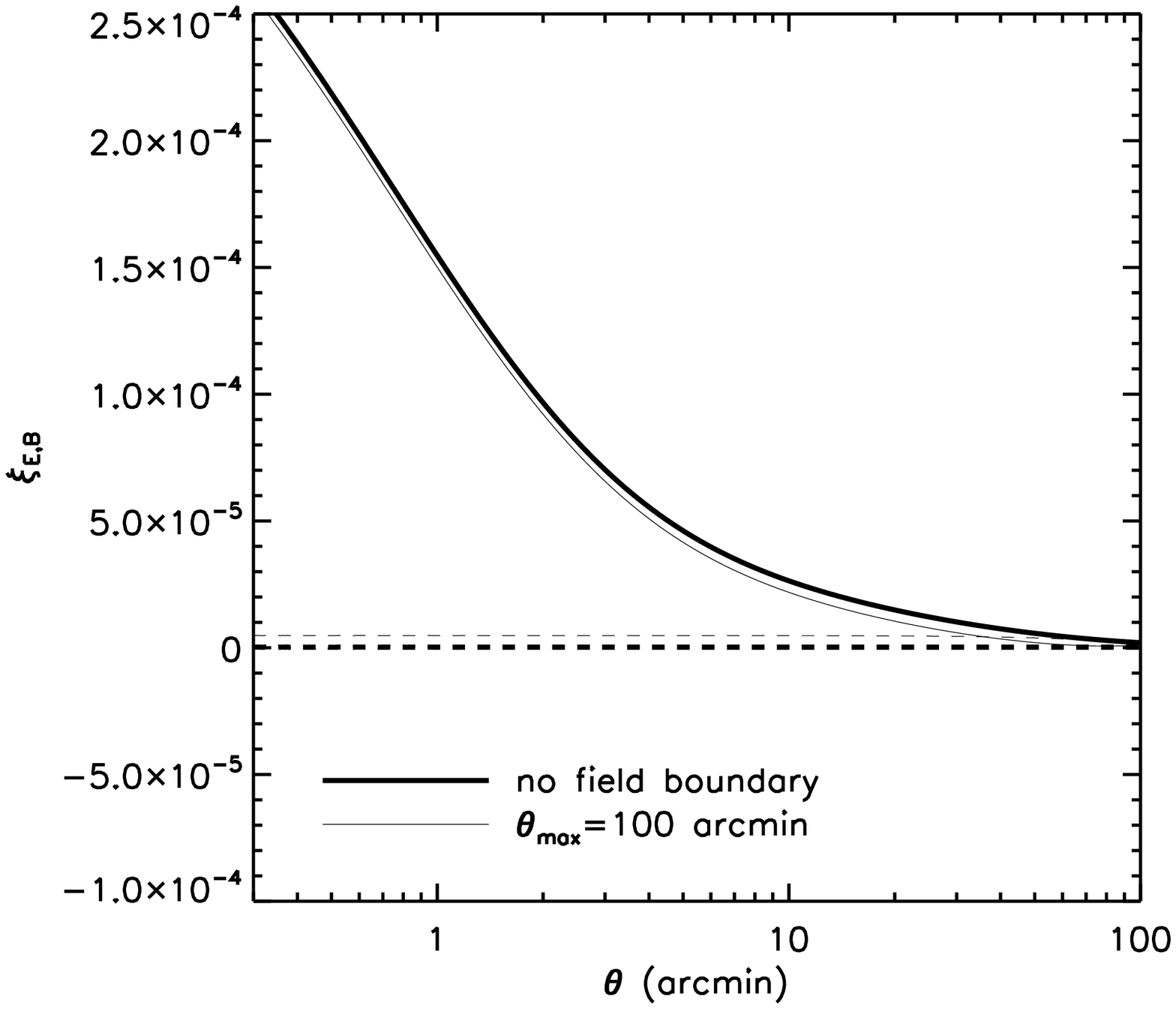}
\caption{Effects of missing data on the $\xi_{\mbox{ \tt E/B}}$, $\left <\gamma^2_{E/B}\right >$ and $\left <M^2_{E/B}\right >$ statistics. We assume
the WMAP7 cosmology and the source redshift distribution shown in Figure~\ref{fig_p_of_z}.
The left plot illustrates the effect on $\left <M^2_{E/B} \right >$ due to the missing data on small scales ($\theta<0\farcm15$) whereas
the right plot shows the effect of the missing data on large scales ($\theta>100\arcmin$) on the $\xi_{\mbox{ \tt E/B}}$ statistics.
The tophat shear variance $\left <\gamma^2_{E/B}\right >$ (middle) suffers from the missing data on both ends.
 The solid and dashed lines
represent the E- and B-mode signals, respectively. \label{fig_effect_of_missing_data}}
\end{figure*}

In Figure~\ref{fig_star_galaxy} we display the star-galaxy correlation for all five DLS fields. Overall, the amplitude of the star-galaxy correlation is far smaller than the cosmic shear uncertainties evaluated for the field.  Note that at $\theta \sim 30\arcmin$, the star-galaxy correlation amplitude in F5  is comparable to the statistical errors. However, 
the sample variance error is still much larger even in this rare case. It is important to remember that
the galaxy shapes used here are measured with the second-tweak PSF discussed in \textsection\ref{section_psf_stacking}.
These small star-galaxy correlations strongly suggest that the residual systematics after PSF correction will be a highly subdominant source of uncertainty in this DLS cosmic shear study.

\subsubsection{E- and B-mode Decomposition} \label{section_b_mode}
As mentioned in \textsection\ref{section_equations}, the B-mode statistics provide useful insights into the residual systematics in gravitational lensing. 
However, the evaluation of equations~\ref{eqn_tophat}, \ref{eqn_ap_mass}, and \ref{eqn_xi_enb}  formally requires the measurement of $\xi_{+,-}$ 
on large and/or small scales inaccessible to the DLS data. This is a well-known problem in cosmic shear studies. According to Kilbinger et al.  (2006), the missing data on large scales make  the $\xi_{E/B}$ statistics valid only up to an offset over the entire range
whereas those on small scales suppress the $\left <M^2_{E/B}\right >$ statistics on small scales. The exact amount of deviation with respect
to the result from the ideal case depends on the cutoff angle, redshift distribution, and cosmology. Using the parameters specific to our DLS study, we examine these effects quantitatively. 
The missing data on small scales is relevant\footnote{Strictly speaking, the upper limit of the integration is $2\theta$, and this
causes a slight deviation near the maximum angular separation because of the missing data between $\theta$ and $2\theta$.} to $\left <M^2_{E/B}\right >$
 and displayed in the left panel of
Figure~\ref{fig_effect_of_missing_data}. Although our cutoff angle happens at $\theta_{min}=0\farcm15$, the suppression 
effect is visible out to $\theta\sim4\arcmin$. This implies that our (observed) E/B-decomposition using this statistic is
meaningful only at $\theta\gtrsim4\arcmin$. On the other hand, the $\xi_{E/B}$ statistics are affected by the missing data
on large scales ($\theta\gtrsim100\arcsec$) and the effect is manifested as a constant offset over the entire scale 
(right panel of Figure~\ref{fig_effect_of_missing_data}).
This predicted offset in $\xi_B$ is $\mytilde4.4\times10^{-6}$ for the WMAP7 cosmology, which
is in excellent agreement with the observed  value $\mytilde4.5\times10^{-6}$ 
when we carry out the E-/B-decomposition with the finite-field DLS data.
Although the exact value of the offset depends on the assumed cosmology, the variation within $\sigma_8=0.7-0.9$ is comparable to the sample variance of the DLS. Therefore, in our presentation of the $\xi_{ E/B}$ statistics hereafter we show the results obtained by
supplementing $\xi_{-}$ at $\theta>100\arcmin$ with theoretical predictions for the given redshift distribution. The addition of the synthetic
data at $\theta>100\arcmin$  precisely cancels the $\mytilde4.4\times10^{-6}$ offset.
This is justified because
the result is virtually cosmology-independent.
However, when it comes to $\left <M^2_{E/B}\right >$, since the predicted  $\xi_{+,-}$ signal at $\theta< 0\farcm15$ is sensitive to cosmology,  we  present
the results without filling in the missing data on this small scale. Consequently, our DLS $\left <M^2_{E/B} \right>$ values are suppressed
at $\theta  \lesssim 4\arcmin$. The tophat shear variance $\left <\gamma^2_{E/B}\right >$ (middle panel of Figure~\ref{fig_effect_of_missing_data} ) suffers from the missing data on both ends.
For the evaluation of $\left <\gamma^2_{E/B}\right >$, we only supplement $\xi_{+,-}$ at $\theta>100\arcmin$ with theoretical values.
We note that the amount of the observed signal suppression on small scales for $\left <M^2_{E/B}\right >$ and $\left <\gamma^2_{E/B}\right >$ is
consistent with the theoretical prediction.

\begin{figure*}
\includegraphics[width=18cm,height=4.3cm]{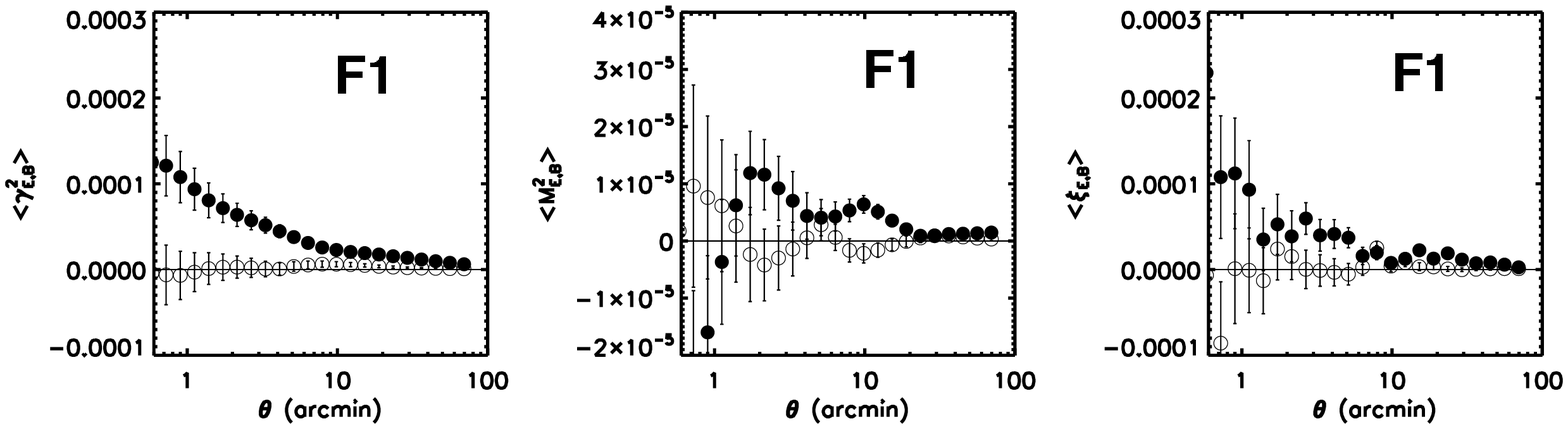}
\includegraphics[width=18cm,height=4.3cm]{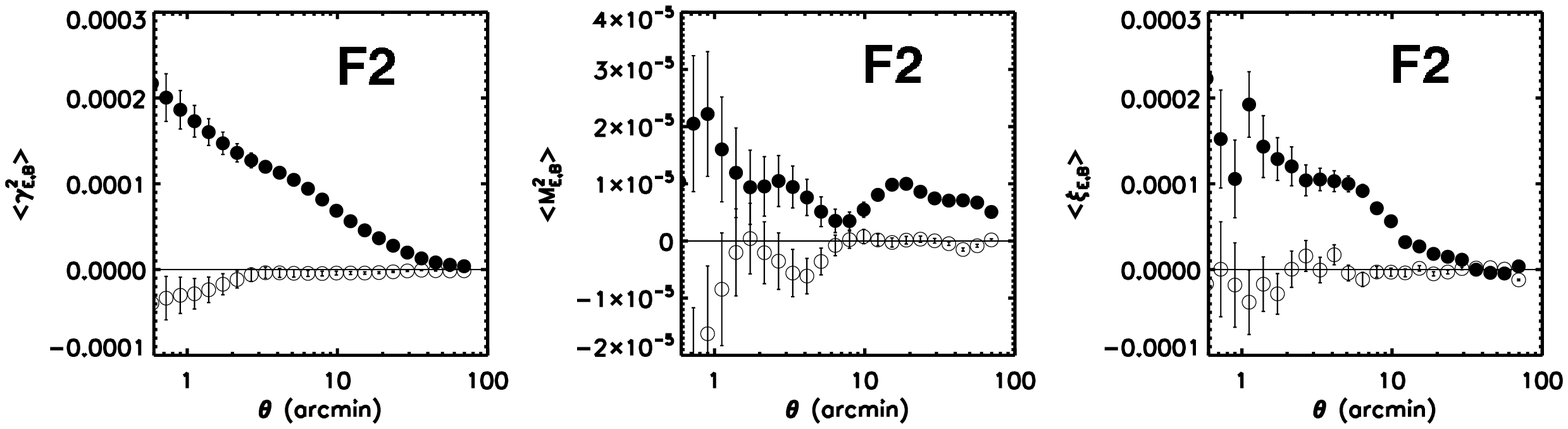}
\includegraphics[width=18cm,height=4.3cm]{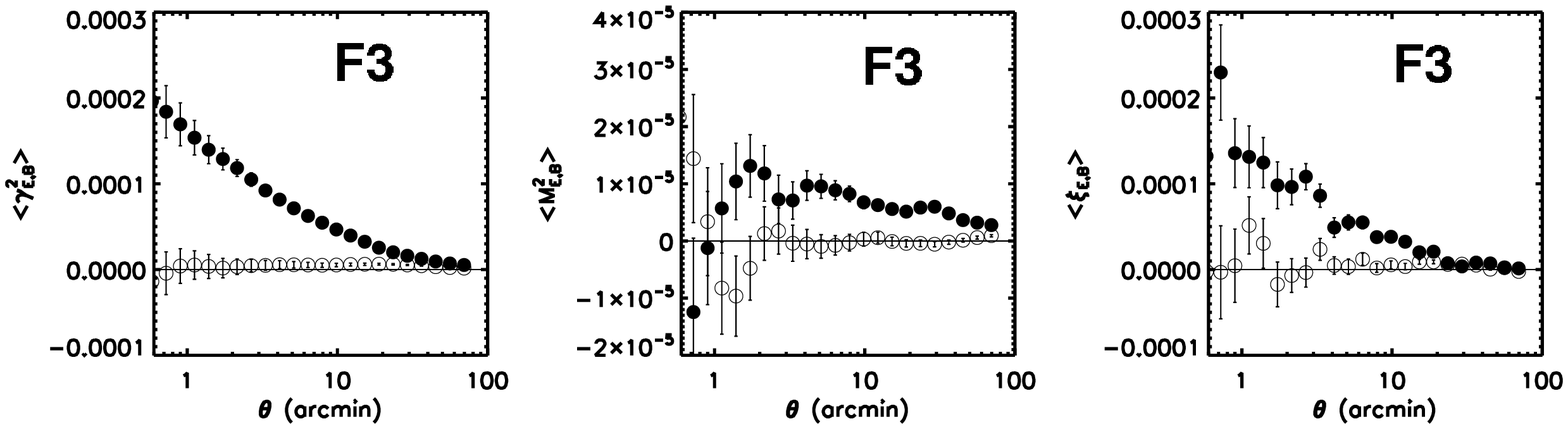}
\includegraphics[width=18cm,height=4.3cm]{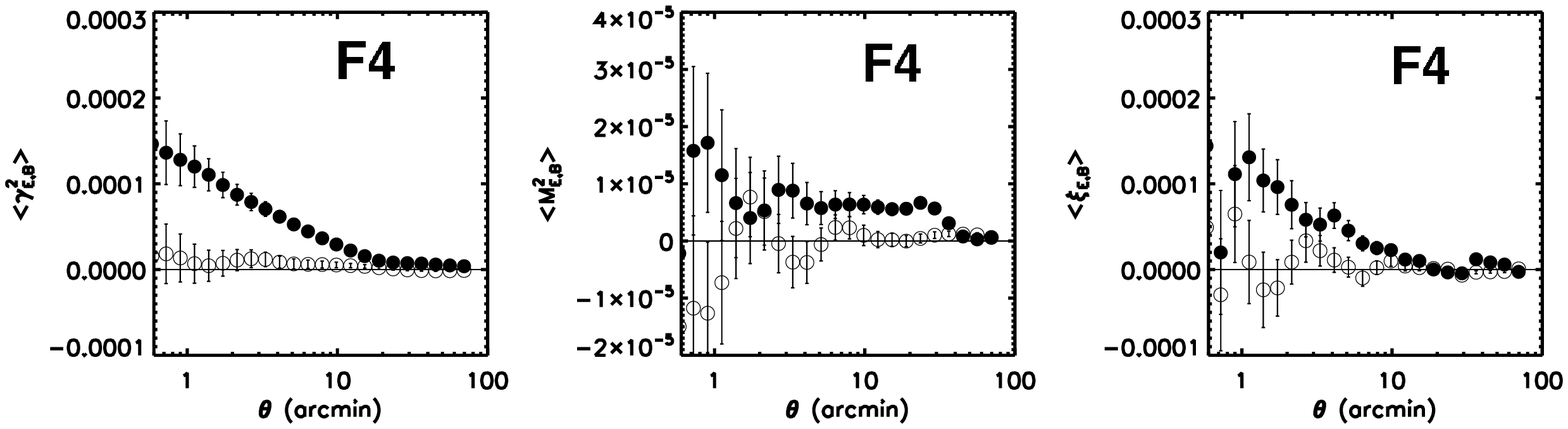}
\includegraphics[width=18cm,height=4.3cm]{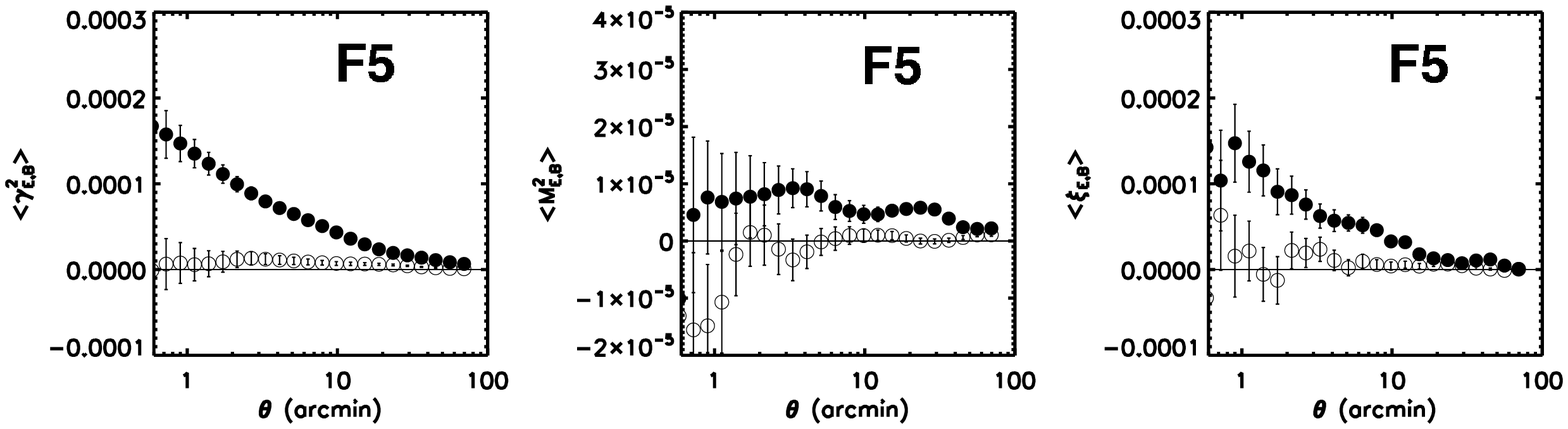}
\caption{Cosmic shear statistics for each DLS field. The plots in the first, second, and third columns show shear variance, aperture mass, and correlation functions, respectively. Filled and open circles represent E- and B-mode signals, respectively. The error bars show only shot noise. No shear calibration has been applied yet.
\label{fig_dls_cs}}
\end{figure*}

\begin{figure}
\includegraphics[width=8.8cm]{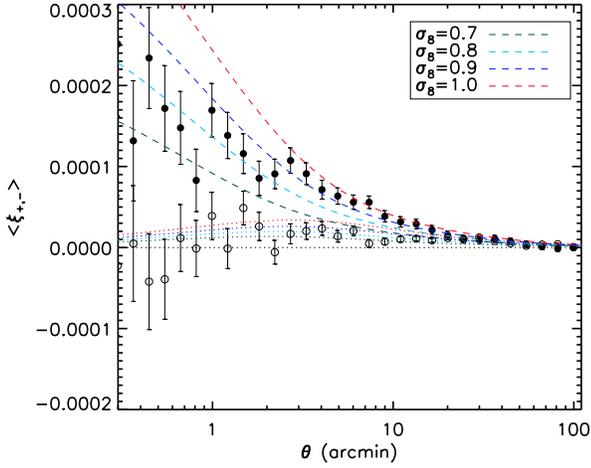}
\caption{$\xi_{+,-}$  correlation functions from the combined (F1-F5) cosmic shear data. Filled and open circles represent $\xi_+$ and $\xi_-$ signals, respectively. Both 
$\xi_+$ and $\xi_-$ error bars include the shot noise and the sample variance.
Dashed (dotted) lines represent predicted $\xi_+$ ($\xi_-$) signal for different $\sigma_8$ values while we assume the WMAP7 cosmology for the rest of the parameters. Shear calibration has been applied.}
\label{fig_dls_xi}
\end{figure}

\begin{figure}
\includegraphics[width=8.8cm]{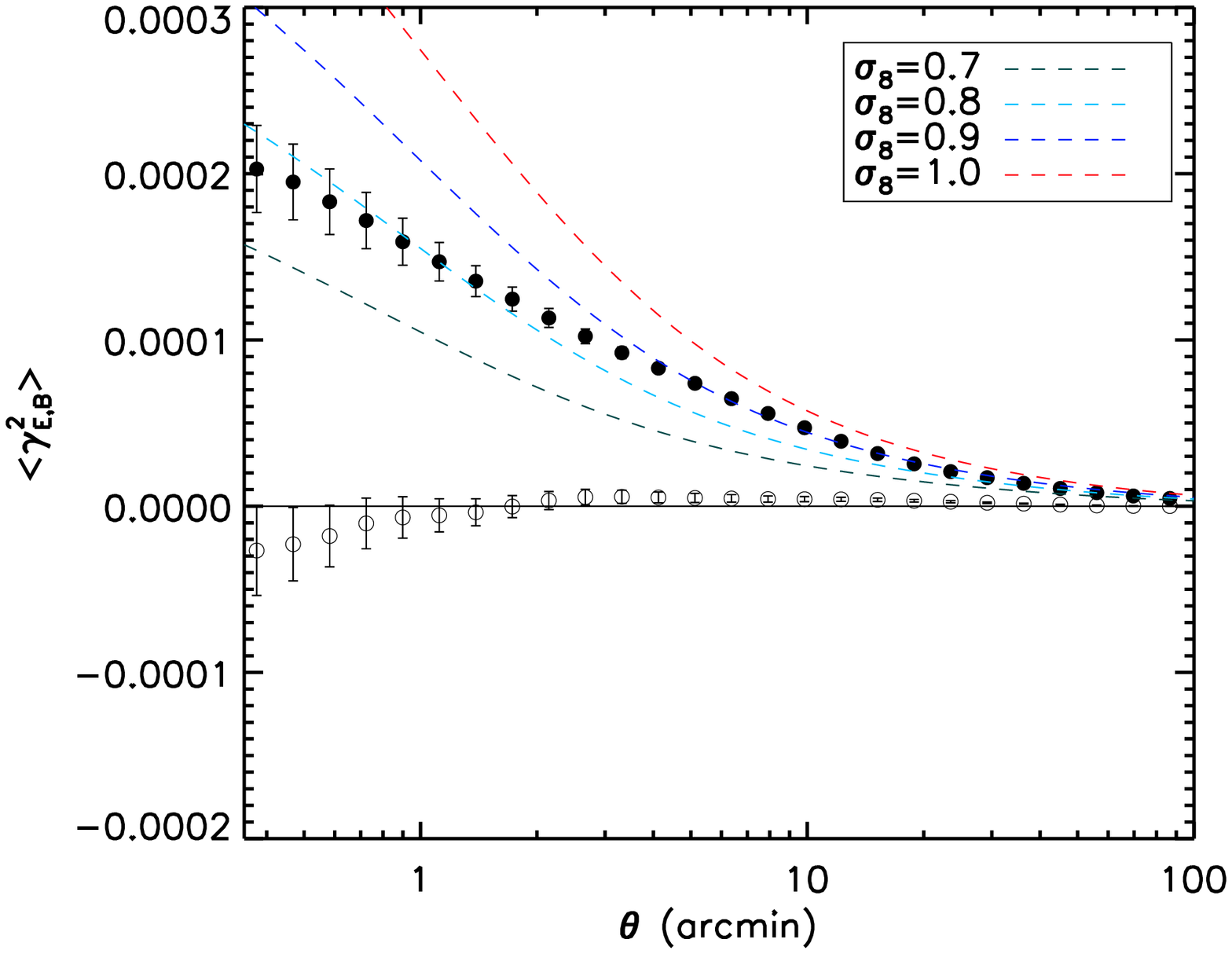}
\caption{Top-hat shear variance from the combined (F1-F5) cosmic shear data. Filled and open circles represent E- and B-mode signals, respectively. The error bars show only galaxy shot noise.
Dashed lines represent predicted signal for different $\sigma_8$ values while we assume the WMAP7 cosmology for the rest of the parameters. Shear calibration has been applied.}
\label{fig_dls_tophat}
\end{figure}

\begin{figure}
\includegraphics[width=8.8cm]{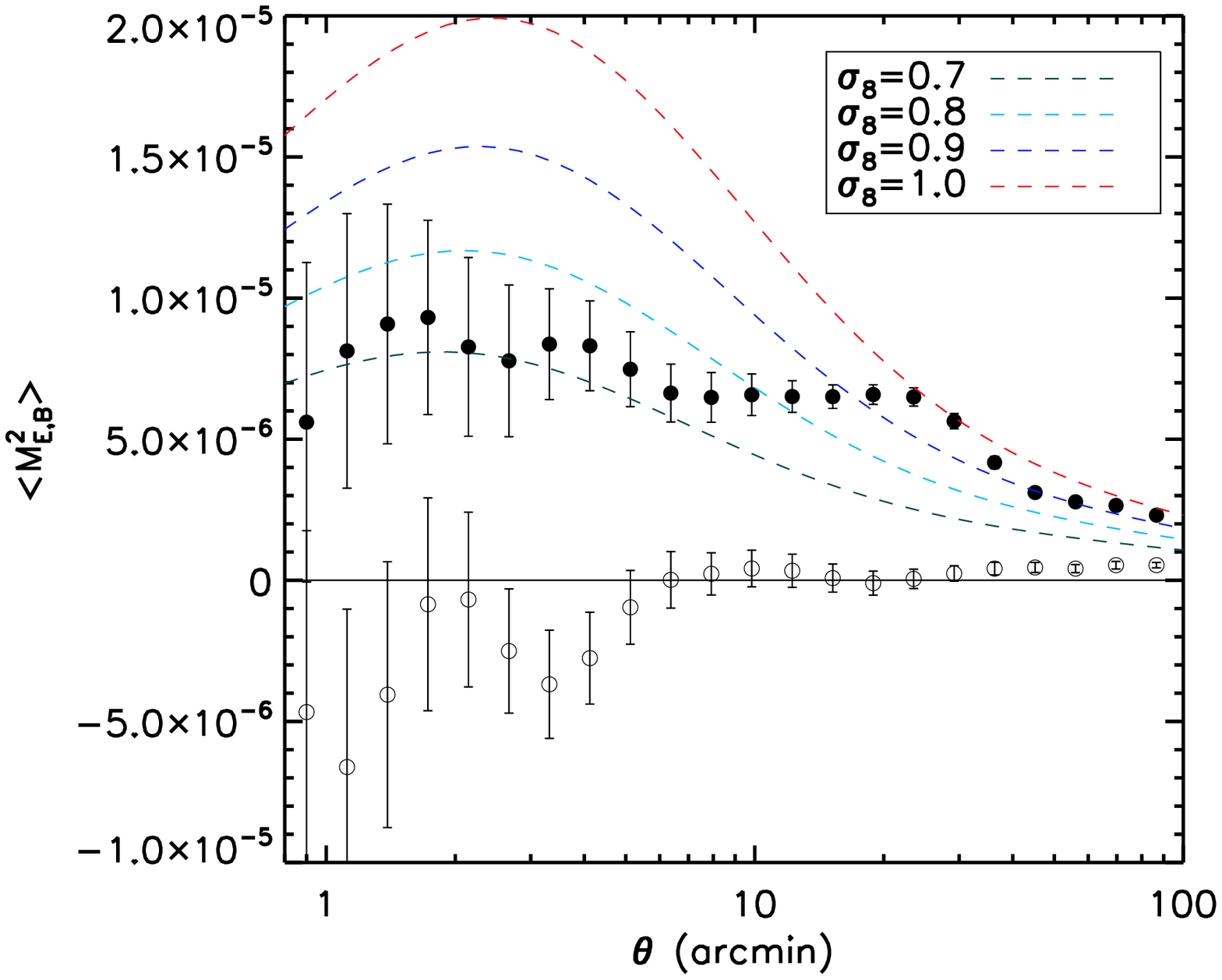}
\caption{Aperture mass variance from the combined (F1-F5) cosmic shear data. Filled and open circles represent E- and B-mode signals, respectively. The error bars show only shot noise.
Dashed lines represent predicted signal for different $\sigma_8$ values while we assume the WMAP7 cosmology for the rest of the parameters. Shear calibration has been applied.}
\label{fig_dls_map}
\end{figure}

\begin{figure}
\includegraphics[width=8.8cm]{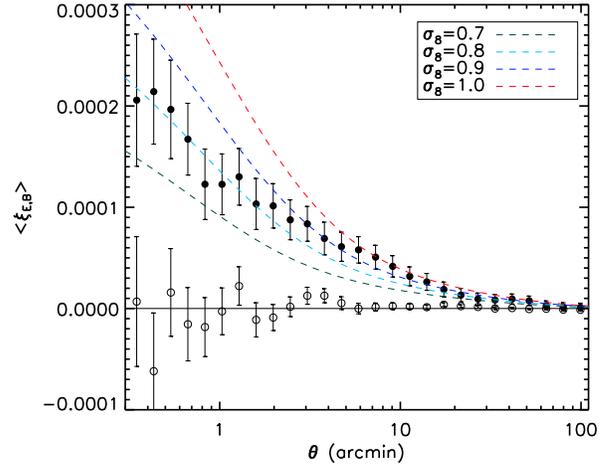}
\caption{$\xi_{E/B}$ statistics from the combined (F1-F5) cosmic shear data. Filled and open circles represent E- and B-mode signals, respectively. The error bars include both the shot noise and the sample variance.
Dashed lines represent predicted signal for different $\sigma_8$ values while we assume the WMAP7 cosmology for the rest of the parameters. Shear calibration has been applied.}
\label{fig_dls_xi_enb}
\end{figure}

\begin{figure}
\includegraphics[width=8.8cm]{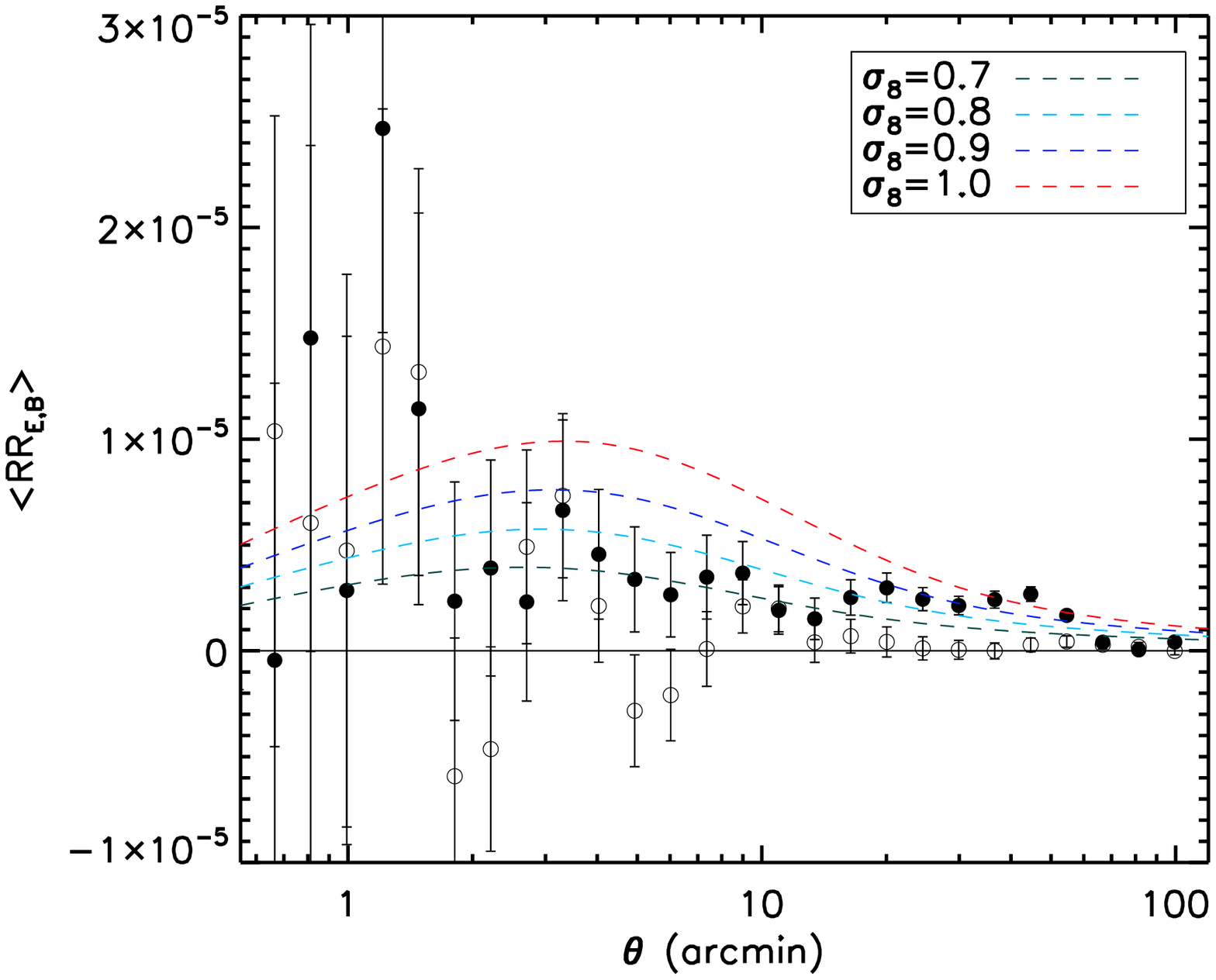}
\caption{$\left < RR_{E/B} \right > $ statistics from the combined (F1-F5) cosmic shear data. Filled and open circles represent E- and B-mode signals, respectively. The error bars here include only shot noise.
Dashed lines represent predicted signal for different $\sigma_8$ values while we assume the WMAP7 cosmology for the rest of the parameters. Shear calibration has been applied.}
\label{fig_dls_rr}
\end{figure}

We present these cosmic shear statistics in Figure~\ref{fig_dls_cs}, where the DLS field number (F1-F5) runs from top to bottom.  The displayed error bars represent only shear shot noise.
The cosmic shear signal is clearly seen in all three ($\left <\gamma^2_E \right >$, $\left <M^2_E \right >$, $\xi_E$) statistics whereas the corresponding B-modes are all close to zero.
The shear variance ($\left <\gamma^2_E \right >$) measures the mean dispersion within an aperture and hence the data points are highly correlated. The signal shape from
all five fields  is similar and the amplitude is roughly proportional to the amount of large scale structure in each field. For example, $\left <\gamma^2_E \right >$ in F2
is nearly a factor of two higher at $\theta\sim10\arcmin$ than the signal in the other fields, and this is consistent with the rather unusually large number of
structures in F2 seen by both red sequence distribution and convergence map (Kubo et al. 2009).  These structures are known to be at $z\sim0.3$ and $0.6$, and their angular scales are
consistent with the $\mytilde10\arcmin$ scale of excess shear correlation. Because we do not exclude the data in F2 in our cosmological parameter estimation,
it is possible that the results are slightly shifted toward high normalization, although the DLS fields were randomly chosen.
The signal from $\xi_{E}$ statistics
is similar to $\left <\gamma^2_E \right >$ except that the data points are less correlated. As expected, the field to field variation in signal is the largest in aperture mass $\left <M^2_{E} \right>$, and
the correlation between the data points is the least among the three statistics shown here. Therefore, although the S/N of $\left <M^2_E \right >$ for each data point is lower than the other two
statistics, the information content as a whole is in fact comparable.

In order to maximize the constraining power of our DLS data on cosmology, we need to combine the measurements $\xi_+$ and $\xi_-$ from all five fields
with a careful choice of weighting scheme. The errors are dominated by statistical errors on small scales whereas
the sample variance is more important on large scales. For our DLS data, the transition of this error dominance occurs at $\theta\sim2\arcmin$ (see Figure~\ref{fig_star_galaxy}).
We weight the measurements in each field with the following inverse variance:
\begin{equation}
w = \frac{1}{\sigma_{SV}^2 + \sigma_{shot}^2 + \sigma_{sys}^2},
\end{equation}
\noindent
where $\sigma_{SV}$ is the sample variance, $\sigma_{shot}$ is the shot noise, and $\sigma_{sys}$ is the residual systematic error estimate.
The sample variance $\sigma_{SV}$ is a function of both cosmology and angle, and we discuss the issue in detail in \textsection\ref{section_covariance}.
The shot noise $\sigma_{shot}$  is given by
$\sigma_{SN}^2 N^{-0.5}$, where $N$ is the number of all galaxy pairs used per angular bin.
The residual systematic error is estimated from the analysis of B-mode signals and star-galaxy correlation. Although our residual systematics are negligible compared to statistical errors and the sample
variance, the star-galaxy correlation functions (see Figure~\ref{fig_star_galaxy}) indicate that the star-galaxy correlation might be non-zero at small angular scales. We model the trend with:
\begin{equation}
\sigma_{sys} (\theta) = 3\times10^{-6}  \left( \theta/1 \arcmin   \right)^{-0.2}. \label{eqn_sigma_sys}
\end{equation}

In Figure~\ref{fig_dls_xi} we show the combined $\xi_+$ and $\xi_-$ correlation functions from all five fields. We assume the WMAP7 cosmology for the estimation of the sample variance $\sigma_{SV}$ here, which however is allowed to vary in our cosmological parameter estimation (\textsection\ref{section_cosmology}). 
Figures~\ref{fig_dls_tophat},~\ref{fig_dls_map}, and~\ref{fig_dls_xi_enb} display the three derived statistics $\left <\gamma^2_{E/B} \right >$, $\left <M^2_{E/B} \right >$, and $\xi_{E/B}$, respectively.
As mentioned above, the missing data on a small scale ($\theta<0\farcm15$) suppresses $\left <\gamma^2_{E/B} \right >$ and $\left <M^2_{E/B} \right >$ on small scales, consistent with the prediction (Figure~\ref{fig_effect_of_missing_data}). In addition, no constant offset is observed
in  $\left <\gamma^2_{E/B} \right >$ and $\xi_{E/B}$ because we supplement the DLS data with the synthetic data at $\theta > 100\arcmin$.
Finally, we display the ring statistic $\left < RR_{E/B} \right >$ in Figure~\ref{fig_dls_rr}. The plot confirms that the B-mode signal is also consistent
with zero in this statistic that does not require any synthetic data. Together with the star-galaxy correlation presented in \textsection\ref{section_star_galaxy}, the absence of the B-mode signals here supports
our success controlling weak-lensing systematics.

Note that in Figures~\ref{fig_dls_xi}, \ref{fig_dls_tophat},~\ref{fig_dls_map}, ~\ref{fig_dls_xi_enb}, and ~\ref{fig_dls_rr},  we scale the cosmic shear signal using the
shear calibration result (\textsection\ref{section_shear_calibration}). 
The various lines represent the predicted signal for different $\sigma_8$ values while we assume the WMAP7 cosmology for the rest of the parameters. The comparison of these predicted values with the data points indicates that the DLS data may favor a slightly higher normalization than the WMAP7 prediction (i.e., $\sigma_8\simeq0.81$). In particular,  this tendency is conspicuous for the $\xi_+$ data points in the $3\arcmin \lesssim \theta \lesssim 20\arcmin$ range (Figure~\ref{fig_dls_xi}). We do not find any evidence that this is caused by residual systematics. Rather, we believe that
the rich structures in F2 (see Figure~\ref{fig_dls_cs}) are responsible for this signal excess. 

In deriving cosmological parameters from the DLS data, we use only the two $\xi_+$ and $\xi_-$ correlation functions, which are directly measured from our shape catalog. Although it is
possible to obtain our parameter constraints based on pure E-mode statistics such as $\left <\gamma^2_E \right>$, $\left <M^2_E \right>$,  $\xi_E$, and $\left < RR_E \right >$,
most theoretical studies so far have focused on the correct evaluation of the covariance for $\xi_+$ and $\xi_-$ rather than for those derived statistics.
In principle, a constant shear field can lead to a result, where $\xi_+=const$ and $\xi_-=0$ while those E/B-mode statistics are unaffected (Schneider et al. 2010). However, we are unable to find
any evidence indicating that this constant shear field from residual systematics might be present in our DLS data (e.g., the star-galaxy
ellipticity correlation should reveal this type of systematics).
Hence, we believe that as long as the covariance matrix is constructed from robust error propagation, our
parameter constraints should yield virtually indistinguishable results regardless of the choice in lensing statistics.

\section{COSMOLOGY WITH DLS COSMIC SHEAR} \label{section_cosmology}

\subsection{Shear Power Spectrum Model}
Cosmological parameter estimation is performed by comparing observed signals with those predicted by full simulation of the survey in different cosmologies. 
As discussed in \textsection\ref{section_equations}, the key component in model
predictions is the shear power spectrum $P_\gamma$, which is obtained by integrating the matter power spectrum $P_{\delta}$ along the line of sight weighted by
lensing efficiency. Using a nonlinear  matter power spectrum $P_{\delta}$ is critical, and also it is well known that the details in the method for the computation of the nonlinear power spectrum
produce non-negligible differences in the parameter estimation. For example, the use of Peacock and Dodds (1996) shifts the value of $\sigma_8$ high by $\mytilde5$\% 
with respect to the case when the Smith et al. (2003) method is used.
In this paper, we use the modified transfer function of Eisenstein \& Hu (1998) that includes Baryonic Acoustic Oscillations (BAO) features and the Smith et al. (2003)
``HaloFit" nonlinear power spectrum. 

Note that according to
Hilbert et al. (2009) and Heitmann et al. (2010),
HaloFit (Smith et al. 2003) underpredicts the matter power on small scales, and
for COSMOS cosmic shear data, using HaloFit prediction leads to an overestimation of $\sigma_8$ by 
about 5\% (Schrabback et al. 2009). Another unresolved issue in our theoretical cosmic shear signal modeling is the effect of baryons
on the matter power spectrum, which affects the power spectrum on small scales ($l\gtrsim1000$)  (e.g., Jing et al. 2006, Rudd et al. 2008;
Semboloni et al. 2011).

The shear power spectrum obtained in this way assumes that the reduced shear $g$ is equal to the shear $\gamma$ (\textsection\ref{section_shear}), which
leads to underestimation of the amplitude on small scales.
Kilbinger (2010) provides fitting formulae, which prescribe the amount of correction necessary to improve
the accuracy of the shear power spectrum. As the fitting formulae are valid for
the cosmological parameter set near the fiducial flat $\Lambda$CDM WMAP7-like values, we cannot apply the correction to our model
for the full range of parameters.
However, the correction is very small (e.g., $\sim1$\% at $\theta=0\farcm5$), and we find that this correction has virtually no effect on our cosmological
parameter estimation on the scale of the noise.

We fix the baryon density and spectral index to be $\Omega_b=0.046$ and $n_s=0.96$, respectively (Komatsu et al. 2011). The Hubble parameter $h$ is
marginalized over the $0.64<h<0.80$ range with a flat prior, consistent with the {\it Hubble Space Telescope} Key Project (Freedman et al. 2001).
We only consider flat ($\Omega_{M}$ + $\Omega_{\Lambda}=1$) universes.

\subsection{Data Vector and Covariance Estimation} \label{section_covariance}

Our two-point correlation data vector $\hat{\symvec{\xi}}$ consists of two parts: $\hat{\symvec{\xi_+}}$ and $\hat{\symvec{\xi_-}}$. The likelihood
function is given by
\begin{equation}
\scriptsize
\mathcal{L} =  \frac{1}   { (2 \pi)^{n/2} | \textbf{C} |^{1/2} }   \exp{  \left [ - 0.5  \left (\hat{\symvec{\xi}} - \symvec{\xi} \right )^{\mbox{\tt T}}  \textbf{C}^{-1}     \left ( \hat{\symvec{\xi}} - \symvec{\xi} \right )
        \right ] }, \label{eqn_likelihood}
\end{equation}
\noindent
where  \symvec{\xi} is the model prediction for a given set of cosmological parameters, \textbf{C} is the covariance matrix, and $n$ is the number of elements in the data vector $\hat{\symvec{\xi}}$. 
The determinant $| \textbf{C} |^{1/2}$ in the normalization should not be ignored in studies such as the current one, where we take into account the cosmology
dependence.

The structure of \textbf{C} is
\begin{equation}
\textbf{C}= \left (\begin{array} {c c} \textbf{C}_{++}  &  \textbf{C}_{+-} \\
                      \textbf{C}_{+-}^{\mbox{\tt T}} &  \textbf{C}_{--} ,  
          \end{array}  \right ), \label{eqn_cov_struc}
\end{equation}
\noindent
where the sub-matrices $\textbf{C}_{++}$ and $\textbf{C}_{--}$ are the covariance matrices of $\hat{\symvec{\xi_+}}$ and $\hat{\symvec{\xi_-}}$, respectively, and
the off-diagonal sub-matrix $\textbf{C}_{+-}$ is the covariance between $\hat{\symvec{\xi_+}}$ and $\hat{\symvec{\xi_-}}$.
Because we use the cosmic shear signals in the range $0.3\arcmin < \theta < 100\arcmin$ comprised of 30 data bins, the dimension of \textbf{C} is $60\times60$ in our study.

The covariance $\textbf{C}$ is decomposed as 
$\textbf{C} = \textbf{C}_n + \textbf{C}_s + \textbf{C}_B + \textbf{C}_{\epsilon}$, where $\textbf{C}_n$ is the statistical noise, $\textbf{C}_B$
is the residual systematic error,  $\textbf{C}_s$ is the sample variance, and $\textbf{C}_{\epsilon}$ is a
cross-term between shape noise and shear correlations (Joachimi et al.  2008).
$\textbf{C}_n$ is directly measured from our DLS  shape catalog ($\sigma_{SN}^4/N$). We assume that $\textbf{C}_B$ 
is a diagonal matrix whose elements are given by Equation~\ref{eqn_sigma_sys}.
The contribution from $\textbf{C}_{\epsilon}$ might become important in some cases. However, for our DLS study
we find that including these cross terms only negligibly affects our cosmological parameter constraints.  This
can be understood because $\textbf{C}_{\epsilon}$ mostly contributes to  the $\textbf{C}_{+-}$ covariance between
large-scale $\xi_-$ and small-scale $\xi_+$ values. Since the constraining power of $\xi_-$ is insignificant in our
DLS case, our cosmological contours virtually remain the same, regardless of the presence of these terms.

\begin{figure}
\includegraphics[width=8.8cm]{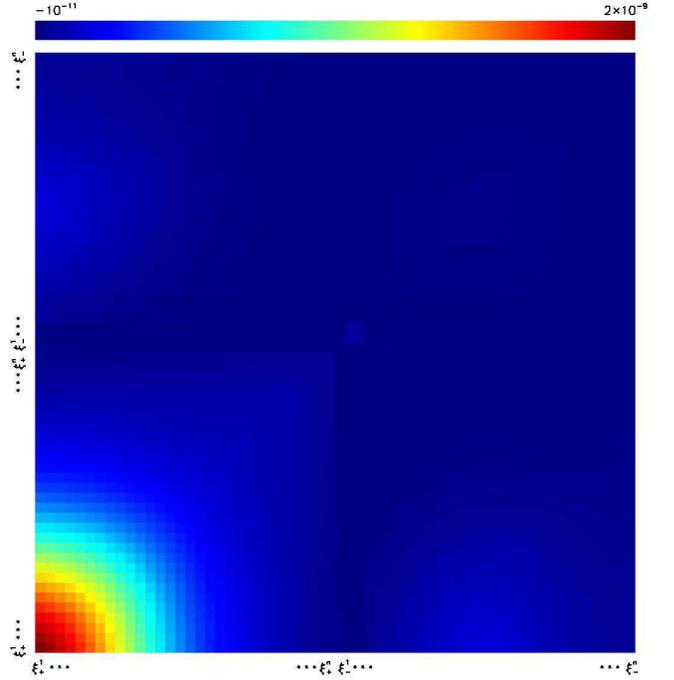}
\caption{Covariance matrix of data vector. We represent the matrix element of Equation~\ref{eqn_cov_struc} using the linear scale color scheme shown on the top.
Only the contribution from the sample variance is displayed for the fiducial WMAP7 cosmology. 
The angular bins are logarithmically spaced with $i=1$ and $i=n$ being $\mytilde0\farcm3$ and $\mytilde100\arcmin$, respectively.
The sample variance is dominated by the nonlinear part of $\xi_+^i$ on small scales ($\theta<10\arcmin$). Also, note that $\mbox{cov}[\xi_{-}^i,\xi_{-}^j]$ and $\mbox{cov}[\xi_{+}^i,\xi_{-}^j]$
are smaller than  $\mbox{cov}[\xi_{+}^i,\xi_{+}^j]$ by an order of magnitude.
}
\label{fig_cov}
\end{figure}

The Gaussian components of $\textbf{C}_s$ can be easily derived from the shear power spectrum. For example, the
Gaussian covariance of $\xi_+ (\theta)$ and $\xi_+ (\theta^{\prime})$ is given by (Joachimi et al. 2008)
\begin{equation}
\scriptsize
\mbox{Cov}[\xi_+ (\theta), \xi_+ (\theta^{\prime}) ] = \frac{1}{\pi A} \int l~ \mbox{d} l~ J_0 (l\theta)J_0 (l \theta^{\prime}) P_{\kappa} (l)^2,  \label{eqn_gauss_cov}
\end{equation}
\noindent

where $A$ is the sky area of the survey. Because the available Fourier modes are limited by the area, equation~\ref{eqn_gauss_cov} in fact leads to
overestimation. According to Sato et al (2011), the discrepancy is roughly a factor of two for $A\sim25~\mbox{deg}^2$. Sato et al. (2011) refer to this bias
as the ``finite field effect". 

The estimation of the non-Gaussian component of $\textbf{C}$ is non-trivial and must be derived from $N$-body simulations with careful ray-tracing. 
Because the computation is expensive, this covariance is commonly assumed to be cosmology-independent in cosmic shear studies.
However, Eifler et al. (2009) found that covariances depend significantly on the cosmology, which impacts the likelihood analysis in cosmological parameter estimation.
Currently, no ray-tracing data for such a wide range of cosmological parameters are available.

Therefore, a practical method to implement this cosmology-dependent covariance (CDC) into one's parameter estimation is to derive ratios between the non-Gaussian
and Gaussian contributions from a particular $N$-body data set and to assume that at least the ratios hold for other cosmological parameters. Since it is
relatively inexpensive to compute the cosmology-dependent Gaussian covariance with Eqn.~\ref{eqn_gauss_cov}, this semi-cosmology dependent covariance (SCDC) estimation is a useful
alternative.

Semboloni et al. (2007) first provided fitting formulae to enable this SCDC estimation. Sato et al. (2011; hereafter S11) performed a similar study but with
a much larger data set. Hilbert et al. (2011; hereafter H11) suggested log-normal approximation as a solution to the problem and provided detailed comparisons of their results with
those from Sato et al. (2011) and Semboloni et al. (2007).

For our parameter estimation, we mainly use the fitting formulae of Sato et al. (2011) to evaluate our SCDC.
The fitting formulae of Sato et al. (2011) do not provide covariance estimation for $\xi_-$ because the shapes are inherently  complicated and cannot be accurately approximated by
simple fitting formulae. The authors kindly provided a table containing these covariances for a fiducial cosmology.
Consequently, in our likelihood analysis
only $\textbf{C}_{++}$ is cosmology-dependent. 
Cosmology-dependence is much more significant in $\textbf{C}_{++}$ than $\textbf{C}_{--}$/$\textbf{C}_{+-}$. 
In addition, the S/N of $\xi_+$ is much higher than that of $\xi_-$. Therefore, 
modeling cosmology independent sample variance for $\xi_-$ (and the cross-covariance) is  a reasonable approximation in the current study.
This argument is supported by an experiment, where we perform parameter constraints
with only $\xi_+$ data. The resulting parameter contours are highly consistent with the current case, where we use both $\xi_+$ and $\xi_-$.

Hilbert et al. (2011) claimed that the covariance of S11 does not meet the positive semi-definiteness criterion, and they
had to perform an eigenvalue decomposition of the S11 matrix and replace the negative eigenvalues with zeros.
However, we do not encounter this problem in our parameter estimation. We suspect that if the claim is true,
perhaps our adding shot noise and systematic noise to the diagonal terms resolves the negative eigenvalue problem.
As the DLS consists of the five separate 4 sq. deg fields, we should not evaluate the sample variance for the $\mytilde20$ sq. deg area. Instead, we compute
the sample variance for a 4 sq. deg area and divide the result by five. 
We show the matrix elements of the Sato et al. (2011) covariance in Figure~\ref{fig_cov} for 4 sq. deg area measured from the $N$-body data simulated for the best-fit parameters of the WMAP7 study.

\subsection{Cosmological Parameter Estimation}

\subsubsection{Constraints on $\Omega_{M}$ and $\sigma_8$}

\begin{figure}
\includegraphics[width=8.8cm]{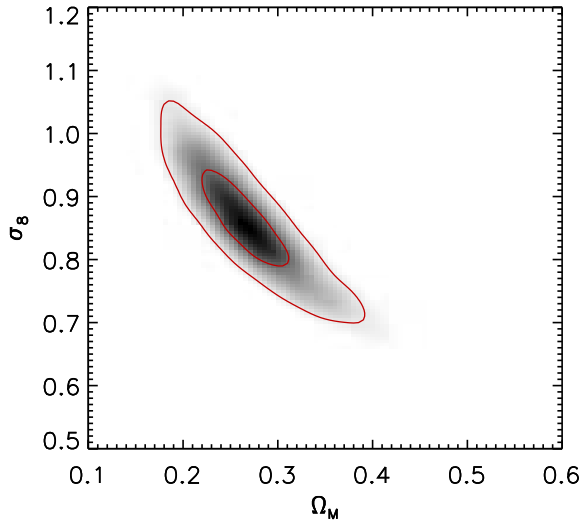}
\caption{Constraints on $\Omega_M$ and $\sigma_8$ from DLS alone. The contours are obtained from $\sim20,000$ sets of parameters explored by our MCMC realizations. The outer and inner contours
show the $2\sigma$ and $1\sigma$ boundaries, respectively. We marginalize over the uncertainties in the Hubble constant, photometric redshift estimation error, and shear calibration.
\label{fig_omega_m_vs_sigma_8}}
\end{figure}

\begin{figure}
\includegraphics[width=8.8cm]{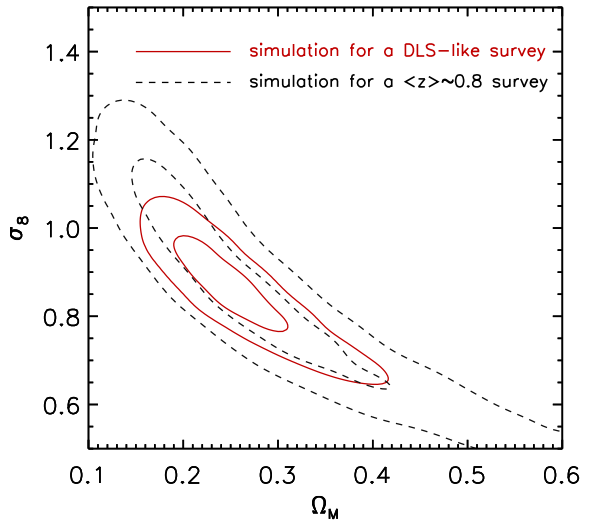}
\caption{Survey depth vs. constraining power. We generate two sets of a``mock'' $\xi_{+,-}$ data vector for a 20 sq. deg field and run MCMC simulations.
For a DLS-like survey, we assume $n_{source}=17$ galaxies per sq. arcmin and $\left <z \right > =1.09$.  For the shallower survey,
we use the parameters in the initial CFHT data studied by Hoekstra et al. (2006), who report $n_{source}\sim12$ galaxies per sq. arcmin
and $\left <z \right >=0.8$. The increase in the survey depth significantly improves the constraining power in three ways.
First, the higher mean redshift boosts the amplitude of the cosmic shear
signal.  Second, the higher source density reduces the shot noise. Third,  the redshift-dependence helps to mitigate the $\Omega_M$-$\sigma_8$ degeneracy.
The outer and inner contours
show the $2\sigma$ and $1\sigma$ boundaries, respectively.
\label{fig_mock_mcmc}}
\end{figure}

\begin{figure*}
\includegraphics[width=9.cm]{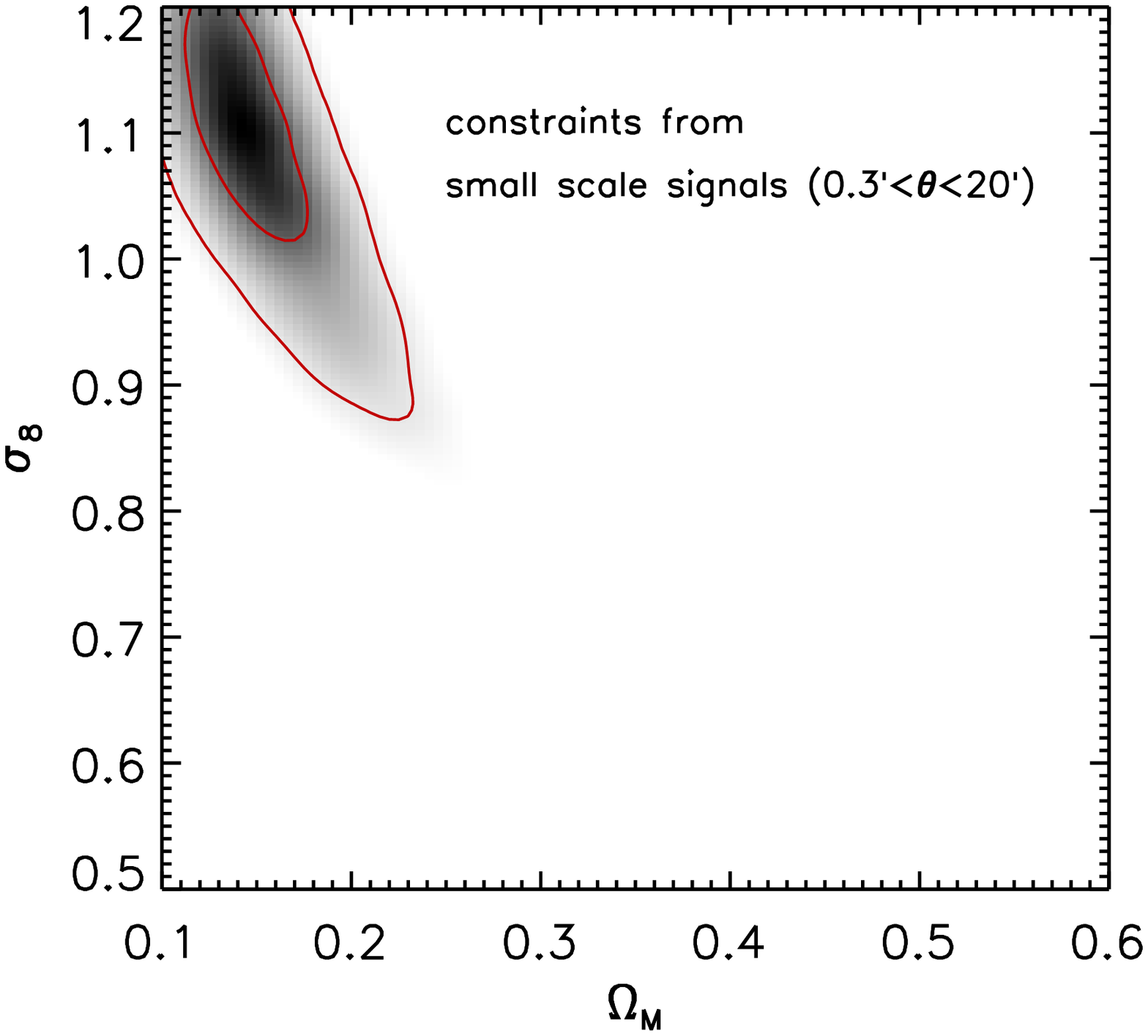} 
\includegraphics[width=9.cm]{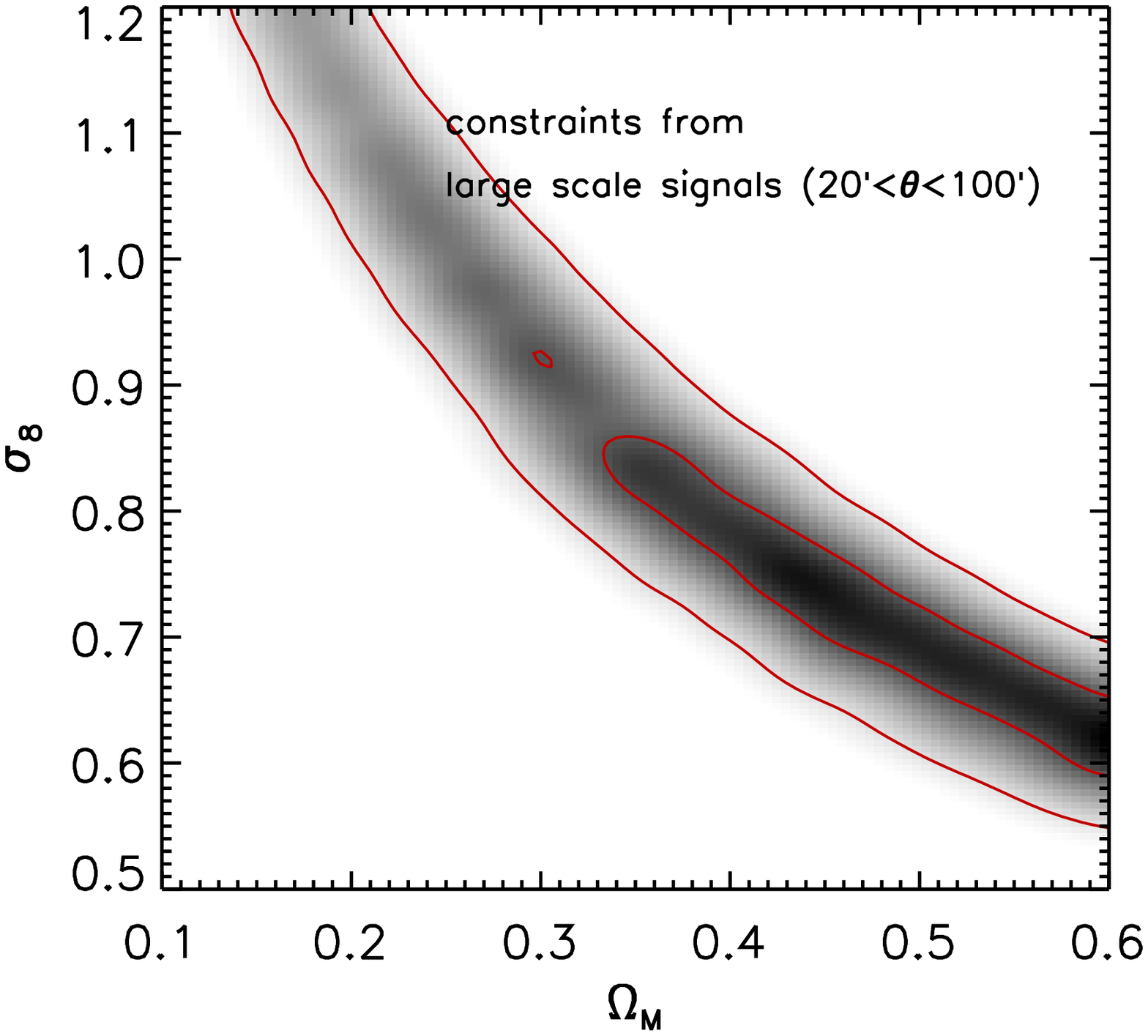}
\caption{Contribution of cosmic shear signals on different scales. 
When we use the data vectors only at $\theta<20\arcmin$,
we cannot rule out the combination of an extremely high $\sigma_8$ value and an extremely low $\Omega_M$ value. Such a combination
predicts a relatively large cosmic shear signal at $\theta \gtrsim20\arcmin$ (this is in disagreement with our data), however,
without significantly violating the data points at $\theta<20\arcmin$.
On the other hand, if we only use the data at  $20\arcmin\lesssim \theta \lesssim100\arcmin$, it is difficult to exclude the low $\sigma_8$ and
high $\Omega_M$ combination, which predicts a higher signal than the current measurements in the $\theta<20\arcmin$ range. 
Because the data points on different scales are correlated (off-diagonal elements in $\textbf{C}_{s}$),
one should not combine the results in both panels "by eye"  to obtain
the result shown in Figure~\ref{fig_omega_m_vs_sigma_8}.
  \label{fig_omega_m_vs_sigma_8_additional}}
\end{figure*}

For the joint constraints on $\Omega_{M}$ and $\sigma_8$, we only consider a flat geometry with $w_0=-1$.  
We let $\Omega_{M}$ and $\sigma_8$ vary within the ranges $[0.05,1.0]$ and $[0.2,1.5]$, respectively.
As mentioned above, the Hubble constant $h$ is marginalized over the $0.64<h<0.80$ range (flat prior).
In addition, we also marginalize over the uncertainties both in the photometric redshift estimation (\textsection\ref{section_photo_z}) and in the shear calibration (\textsection\ref{section_shear_calibration}).

We show the joint DLS  constraints on $\Omega_{M}$ and $\sigma_8$ in Figure~\ref{fig_omega_m_vs_sigma_8}.
The typical ``banana-shape" contours are seen;  this is the result of
the well-known degeneracy between the two parameters in their contribution to the amplitude of the lensing convergence power spectrum. 
The ridge of the contours can be roughly described
by $\sigma_8 \propto \Omega_M^{-0.5}$. 
In general, $\chi^2$ contours for the $\Omega_M-\sigma_8$ constraint are highly elongated and the parameter location where the mimimum $\chi^2$ occurs
has little meaning.  
However, the relatively deep and wide DLS survey yields a high S/N over a wide range of angle.
This DLS cosmic shear analysis shortens the ``banana'' to an unprecedented level and enables us to
constrain the two parameters simultaneously without relying on cosmic microwave background data, giving
a joint constraint: $\Omega_M=0.262\pm0.051$ and $\sigma_8=0.868\pm0.071$.
To more fully understand the reasons for this partial degeneracy breaking, we undertake
simulations using mock shear data.

\subsubsection{Impacts of Survey Depth on Cosmological Parameter Constraints} \label{section_mock_data}

We notice that the small size of the contours for our joint constraints on $\Omega_{M}$ and $\sigma_8$ in Figure~\ref{fig_omega_m_vs_sigma_8} is unprecedented for cosmic shear studies.
Thus, it is important to verify the result with simulations using mock data.
Of course, the most robust test is to perform end-to-end simulations, where we create simulated DLS images from an input cosmology, then carry out cosmic shear analysis, and finally compare the results. However, because here we are mainly interested in verifying the size of the contours and examining the impact of the DLS depth on constraining power, this full-scale
simulation is beyond the current scope of our paper. Instead, our simulation begins with the generation of predicted $\xi_+$ and $\xi_-$ cosmic shear signals at
the best-fit DLS parameters ($\Omega_M$,$\sigma_8$)$\equiv(0.262,0.868)$. The main goal of the current simulation is to examine the impact of the DLS survey depth
on the parameter constraining power.

These $\xi_+$ and $\xi_-$  data points are error-free and need to be perturbed by the sample variance and shape noise. We construct a full covariance matrix from
the S11 sample variance covariance, the DLS mean galaxy number density, and the mean
shear dispersion. We draw random numbers from a multivariate normal distribution with our full model covariance matrix.
These correlated random numbers are added to the ``exact" cosmic shear data vector to create the final ``mock" cosmic shear data;
the Gaussian assumption for the errors is only approximate, and this could introduce subtle errors in the comparison of the simulated data and the real data.
We run MCMC with this mock data vector in the same way that we process our real DLS cosmic shear data.

Figure~\ref{fig_mock_mcmc} shows the resulting parameter constraints with these simulated data. The solid line is obtained when we assume the current depth of DLS whereas the dashed line
results when we assume a shallower survey, for which we use a mean number density $n_{source}=12$ per sq. arcmin and the $\left < z \right > =0.8$ 
redshift distribution reported in the first CFHT cosmic shear study (Hoekstra et al. 2006). This simulation verifies that the increase in the survey depth significantly improves the constraining power.
The reason for the S/N increase is threefold. First, the higher mean redshift boosts the amplitude of the cosmic shear signal.  Second, the higher source density reduces the shot noise.
Third, the wider redshift baseline helps to break the $\Omega_M$--$\sigma_8$ degeneracy although we present here a non-tomographic study.

A close comparison of the solid contours in Figure~\ref{fig_mock_mcmc} with those in Figure~\ref{fig_omega_m_vs_sigma_8} indicates a slight difference
in size. From several runs of the above simulation, we find that the exact size and shape of the contour depends on the shape (i.e., noise from
the sample variance and the  Poissonian fluctuation) of the cosmic shear signal.
It appears that the excess of our DLS cosmic shear signal at $3\arcmin \lesssim \theta \lesssim 20\arcmin$ mentioned in \textsection\ref{section_b_mode} may
contribute to the overall S/N, making contours slightly tighter.

\subsubsection{Contribution from Different Angular Scales}

Examination of Figures~\ref{fig_dls_tophat}-\ref{fig_dls_xi_enb} shows that the statistical errors in our cosmic shear data at $\theta\gtrsim20\arcmin$ are very small.
Because our systematic errors are under good control, the signal on this large scale provides significant constraint limited only by the sample variance.
To illustrate the point, we show two additional $\chi^2$ contours in Figure~\ref{fig_omega_m_vs_sigma_8_additional}. 
When we use the data vectors only at $\theta<20\arcmin$,
we cannot rule out the combination of an extremely high $\sigma_8$ value and an extremely low $\Omega_M$ value. Such a combination
predicts a relatively large cosmic shear signal at $\theta\arcmin \gtrsim 20\arcmin$ (this is in disagreement with our data).  However, the current shape of the DLS cosmic shear signal at $\theta<20\arcmin$ favors
such high $\sigma_8$ value and an extremely low $\Omega_M$ value combinations.
On the other hand, if we only use the data at  $20\lesssim \theta \lesssim100\arcmin$, it is difficult to exclude the low $\sigma_8$ and
high $\Omega_M$ combination, which predicts a higher signal than  (thus inconsistent with) the DLS measurements  in the $\theta<20\arcmin$ range. 
Had there remained substantial systematics in our cosmic shear analysis, the contribution from  the signal in the $20\lesssim \theta \lesssim100\arcmin$ range
would have been insignificant.

\subsubsection{Impacts of Different Covariance on Parameter Constraints}

\begin{figure*}
\includegraphics[width=9cm]{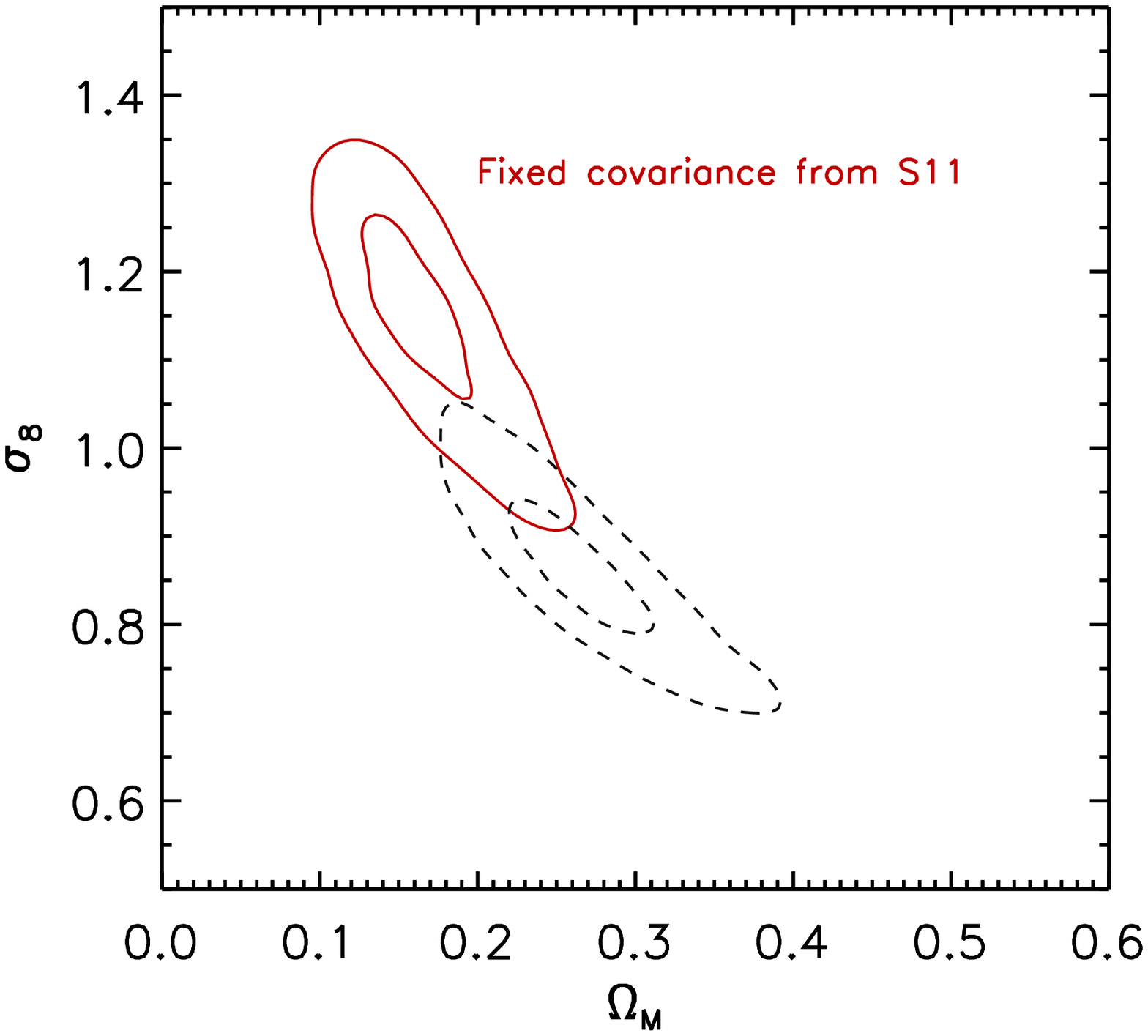}
\includegraphics[width=9cm]{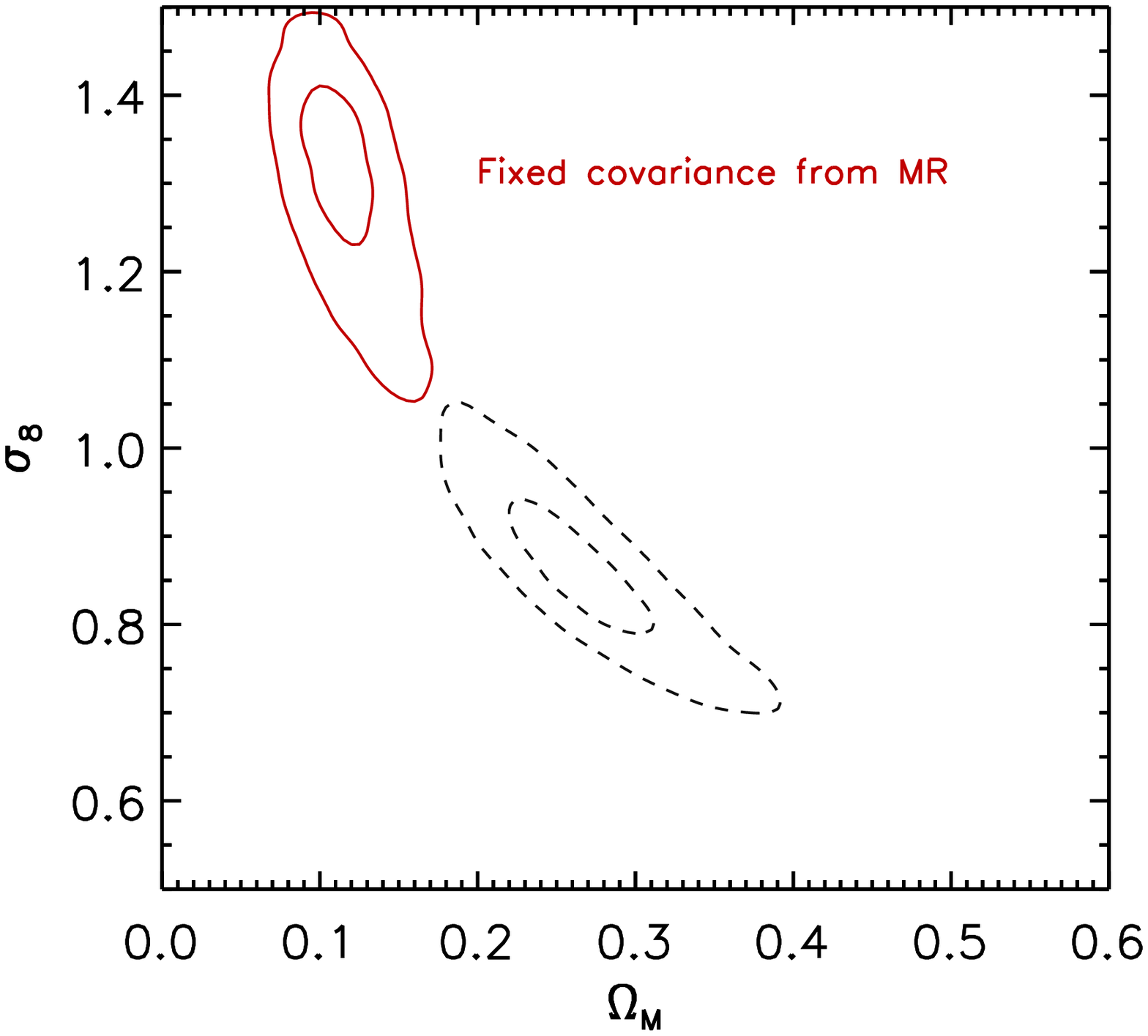} \\
\includegraphics[width=9cm]{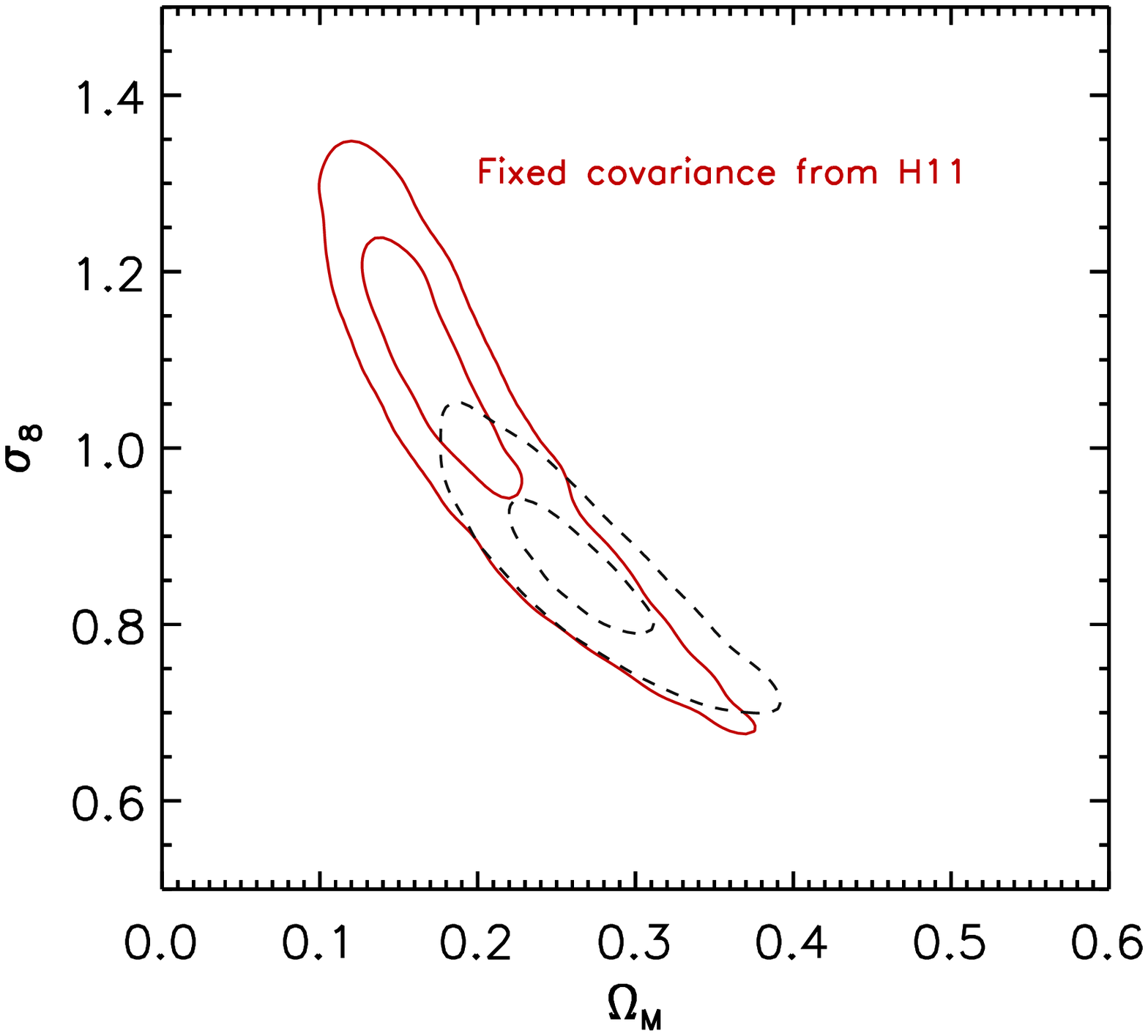}
\includegraphics[width=9cm]{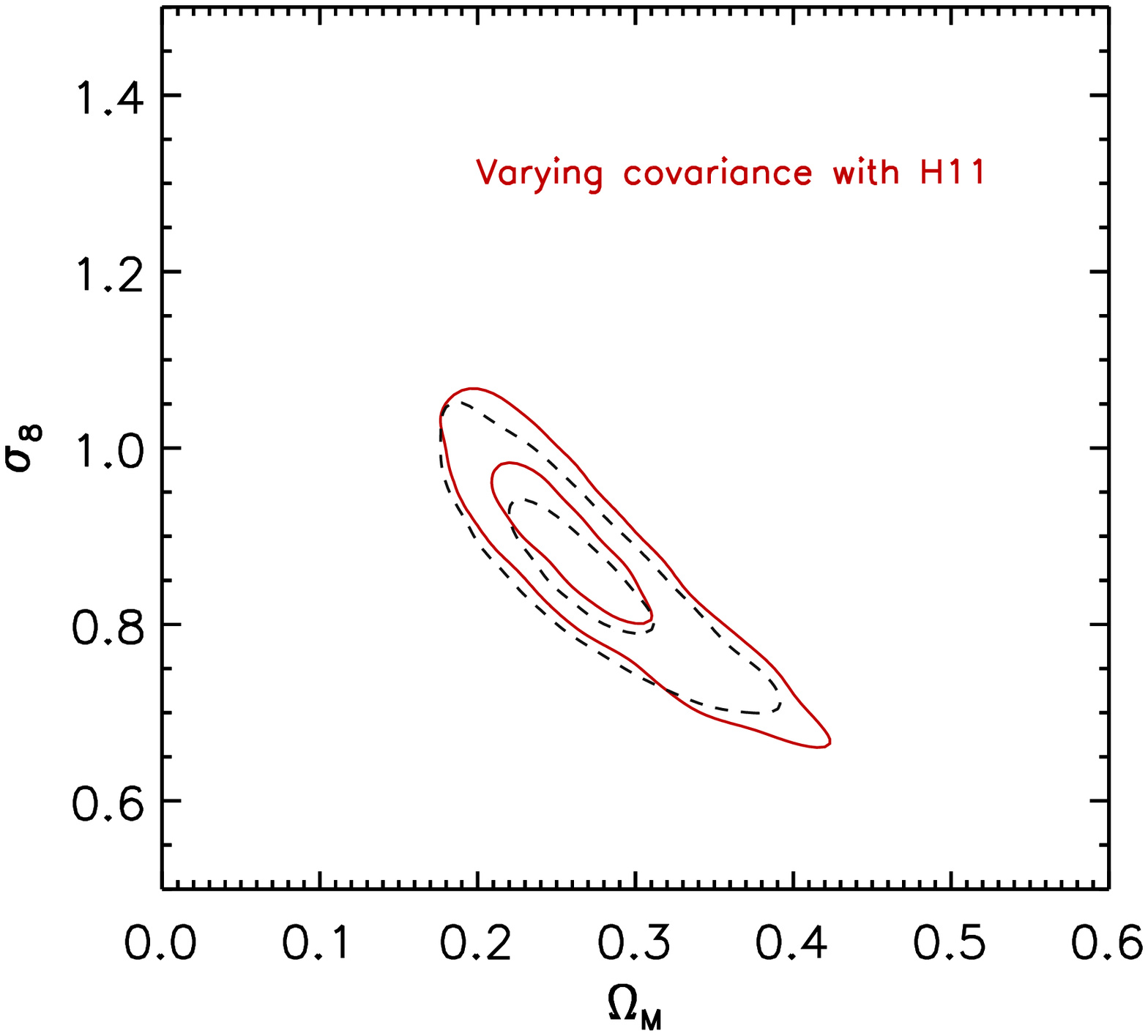}
\caption{Impact of cosmology-dependent covariance matrix on parameter constraints. Different panels show the $\Omega_M-\sigma_8$ constraint results (red) for different covariance matrices.
For easy comparison we also reproduce the result (dashed line) shown in Figure~\ref{fig_omega_m_vs_sigma_8}, which corresponds to the SCDC implemented using the S11 method.
When we fix the covariance matrix for a particular cosmology, the result favors a high-$\sigma_8$/low-$\Omega_M$ combination. By contrast, if we implement the SCDC using the H11
covariance by measuring the ratio of non-Gaussianity to Gaussianity and applying it to the cosmology-dependent Gaussian covariance, the resulting contours closely resemble the S11
SCDC case (lower right panel).
\label{fig_impact_of_cov}}
\end{figure*}

In the regime where cosmic shear data become sample-variance-limited, the role of the cosmology-dependent covariance
is critical. To facilitate this discussion, we show parameter constraints with different covariance matrices in
Figure~\ref{fig_impact_of_cov}. We consider four cases here: 1) a fixed covariance matrix obtained by ray-tracing from the S11 numerical simulation [
($\Omega_M$,$\Omega_{\Lambda}$,$\sigma_8$)=(0.238,0.762,0.76)], 2) a fixed covariance matrix obtained by ray-tracing from the Millennium
Run (MR;  Hilbert et al. 2009)  [($\Omega_M$,$\Omega_{\Lambda}$,$\sigma_8$)=(0.25,0.75,0.9)], 3) a fixed covariance matrix  analytically derived by the H11 log-normal
approximation  [($\Omega_M$,$\Omega_{\Lambda}$,$\sigma_8$)=(0.279,0.721,0.81)],  4) a cosmology-dependent covariance derived using the H11
covariance by first measuring the ratio of non-Gaussianity to Gaussianity and then applying it to the cosmology-dependent Gaussian covariance.
For the first three cases where we fix the covariance matrix for a particular cosmology, the result always favors a high-$\sigma_8$/low-$\Omega_M$ combination.
By contrast, the SCDC implementation with the H11 covariance in the last case gives the parameter constraints closely resembling the SCDC result with the S11 covariance.
As mentioned above, the low amplitude of the signal at $\theta\gtrsim20\arcmin$ plays a critical role in disallowing low $\Omega_M$/high $\sigma_8$ combinations. However, Figure~\ref{fig_impact_of_cov} shows that the parameter constraint can only benefit from this large scale cosmic shear signal when the covariance is allowed to vary during the MCMC run.
One may argue that the impact of the SCDC on the $\Omega_M-\sigma_8$ constraint shown in Figure~\ref{fig_impact_of_cov}
 is somewhat counter-intuitive because a cosmology with a large $\sigma_8$ might
reduce $\chi^2$ and thus the SCDC might favor a large $\sigma_8$ cosmology. However, we find that a large $\sigma_8$ cosmology also substantially alters
the off-diagonal elements of the covariance and the net effect is in fact toward increasing the $\chi^2$ value for that cosmology.

\begin{figure}
\includegraphics[width=8.8cm]{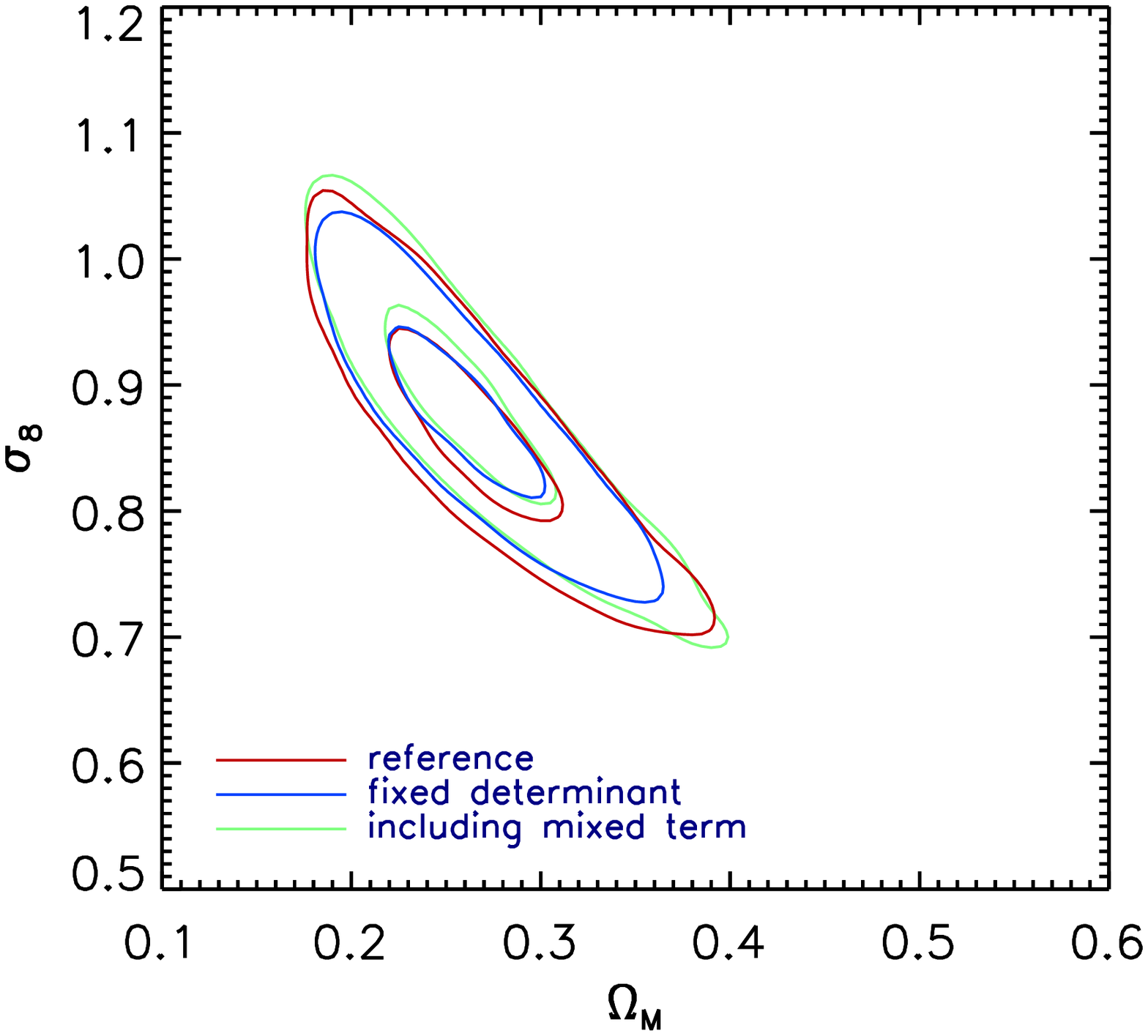}
\caption{Impact of cross-term and moving determinant on parameter constraints. The reference contours (red) are identical to the ones shown in Figure~\ref{fig_omega_m_vs_sigma_8}.
The light green contours are obtained when we add the cross terms $\textbf{C}_{\epsilon}$ (\textsection\ref{section_covariance})  between shape noise and shear correlations. The blue
contours are the results for the case when we drop the determinant $| \textbf{C} |$ in the likelihood (Equation~\ref{eqn_likelihood}).
\label{fig_mixed_term}}
\end{figure}

Note that the two SCDC implementations with H11 and S11 give a  slight difference in contour sizes (lower right panel in Figure~\ref{fig_impact_of_cov}). This is because the covariance of  H11 is higher than S11 by $\mytilde50$\% in the Gaussian regime ($\theta\gtrsim10\arcmin$) (when scaled to the same cosmology).
The difference in the Gaussian regime is mainly due to the difference in the finite field effect correction between the two methods (S11 vs. H11). Currently, no consensus has been reached 
as to which method more accurately corrects for the effect.

One should not be misled into thinking that in the absence of the cosmology-dependent covariance the result always favors a high-$\sigma_8$/low-$\Omega_M$ combination. From our ``mock" data simulations, we have seen opposite cases where the result is skewed toward a low $\sigma_8$/high-$\Omega_M$ combination.  For these cases, we also observe that the use of our SCDC tends to shift the contours in the correct direction. Of course, there are cases where both fixed and varying covariance matrices yield similar results.
However,  the number of our mock data simulation runs is not sufficiently large to generalize this observation. 

Kilbinger et al. (2012) find that the impacts of the cosmology-dependent covariance on parameter constraints are small in their CFHT cosmic shear study. Because their implementation of
the cosmology-dependent sample covariance is different from ours, our speculations on the underlying causes are limited. To begin with, we do not claim that using fixed sample  
covariance always leads to a large bias as seen in our DLS study. As mentioned above, in our mock simulation we have seen cases where both fixed and varying covariance matrices yield similar contours. 
In addition, the ``banana" of Kilbinger et al. is significantly larger than ours (the 1-$\sigma$ contour spans  from $\Omega_M\sim0.2$ to $\mytilde0.7$ while ours ranges from $\Omega_M\sim0.2$
to $\mytilde0.3$). Therefore, it is possible that the relative impact may become smaller in their study.

In Figure~\ref{fig_mixed_term}, we demonstrate that the impact of the cross-terms $\textbf{C}_{\epsilon}$ (\textsection\ref{section_covariance})  are
negligible in our parameter constraint. The green contours are obtained when we add the  cross-terms $\textbf{C}_{\epsilon}$ to the S11 covariance.
We use Equation~35 of Joachimi et al. (2008) to estimate the values for the Gaussian case and modified it for the non-Gaussian case using the S11
scaling relation. Kilbinger et al. (2012) report that in their 2D cosmic shear analysis the role of the cross-term is not negligible in contrast to our result.
To a first order, the ratio of the cross-term to the sample variance is inversely proportional to the product of the amplitude of the non-linear shear power spectrum and the source density. Hence, it is possible that both greater depth (i.e., higher amplitude of the shear power spectrum) and higher source density 
make this term insignificant in DLS cosmic shear analysis.
Also plotted in Figure~\ref{fig_mixed_term} is the result when we run our MCMC without updating the determinant
$| \textbf{C} |$ in the likelihood (Equation~\ref{eqn_likelihood}).  Although it shrinks the size of the contours slightly, the difference with respect to the reference
is insignificant. This indicates that the moving determinant is not the main driver causing the difference between the fixed and varying covariance results. 

\subsubsection{Comparison with Other Cosmic Shear Studies}

Comparison of our DLS cosmic shear results with those in previous studies requires some caution.  The following issues make it difficult
to track down origins of discrepancies.
First, previous cosmic shear studies obtained relatively long ``banana" shapes in their $\Omega_M-\sigma_8$ constraints. Therefore, comparisons should be
made at a fixed value of either $\sigma_8$ or $\Omega_M$.  Because the slope of the ``banana" can also be different, the choice of the reference will affect
the level of agreement.
Second, theoretical improvements have been made in the prediction of cosmic shear signal and the
sample variance since the first cosmic shear results were published (Bacon et al. 2000; Wittman et al. 2000; Van Waerbeke et al. 2000; Kaiser et al. 2000). Different methods of estimating
the nonlinear power spectrum affect the normalization of the cosmic shear signal non-negligibly ($\mytilde6$\% in our case).  
In addition, early cosmic shear studies use Gaussian covariance matrices, which may lead to substantial underestimation of parameter uncertainties. 
Perhaps, counteracting this underestimation is
the finite field effect discussed in \textsection\ref{section_covariance}. Absence of this correction can overestimate the parameter uncertainties.
Third, different authors marginalize over different ranges of nuisance parameters for their target parameter estimation. Both the number of nuisance parameters
and their ranges sensitively affect the parameter estimation.

The most recent cosmic shear results are reported from the analysis of the SDSS Stripe 82 field by Lin et al. (2011) and
Huff et al. (2011).  The survey area is the largest ($\mytilde275$ sq. degrees) to date for cosmic shear analysis. However, the large PSF and
relatively shallow depth provide only moderate constraints on cosmological parameters. At $\Omega_M\equiv0.272$, the result  of Lin et al. (2011)
implies $\sigma_8=0.63^{+0.08}_{-0.13}$. This is in excellent agreement with the result of Huff et al. (2011), who quote $\sigma_8=0.64^{+0.11}_{-0.15}$
from their independent study.
However, both results are in $\mytilde2\sigma$ tension with our value ($\sigma_8=0.833\pm0.034$).

Benjamin et al. (2007) performed a cosmic shear analysis after combining the shears from
the CFHTLS, RCS, and VIRMOS-DESCART, and GaBoDS surveys. At $\Omega_M\equiv0.24$, they 
quote $\sigma_8=0.84\pm0.05$, consistent with our estimate $\sigma_8=0.885\pm0.035$ for the same matter density.
From the initial 22 sq. deg CFHTLS cosmic shear analysis, 
Hoekstra et al. (2006) estimated $\sigma_8=0.85\pm0.06$ for $\Omega_M\equiv0.3$. 
Fu et al. (2007) updated the CFHTLS result using the data from a factor of two larger area and reported $\sigma_8 (\Omega_M/0.25)^{0.64} = 0.785\pm0.043$
from their aperture-mass statistic, which gives $\sigma_8=0.699\pm0.038$ at $\Omega_M\equiv0.3$.
Our DLS estimate $\sigma_8=0.804\pm0.021$ at the same matter density is well within the 1$\sigma$ range of Hoekstra et al. (2006), but in $\mytilde2\sigma$ tension
with the Fu et al. (2007) result.
Schrabback et al. (2010) measured $\sigma_8=0.75\pm0.08$ at $\Omega_M\equiv0.3$ with the COSMOS tomographic analysis. This is consistent with
our result.

\subsubsection{Joint Constraints with Cosmic Microwave Background Data}

\begin{figure}
\includegraphics[width=9cm]{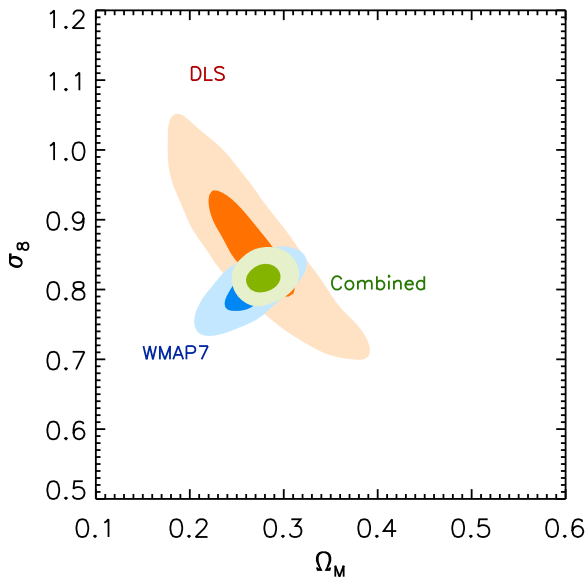}
\caption{Joint constraints on $\Omega_M$ and $\sigma_8$ with WMAP7. The prior settings are identical to the ones in our DLS-only cosmological parameter estimation.
We find $\sigma_8=0.815\pm0.020$ and $\Omega_M=0.278\pm0.018$. For $\Omega_M\equiv 0.272$ [WMAP7], we obtain $\sigma_8=0.833\pm0.034$, where the uncertainty is
estimated with a hard prior at the quoted $\Omega_M$ value from the DLS-only contours.
\label{fig_joint_with_WMAP7}}
\end{figure}

Finally, we examine the cosmological parameter constraints by 
combining  the WMAP7 results\footnote{The MCMC data are available at http://lambda.gsfc.nasa.gov.}
( Larson et al. 2011; Komatsu et al. 2011 ) and our DLS cosmic shear study in a joint analysis.
The $\chi^2$ contours from this combined analysis is displayed in Figure~\ref{fig_joint_with_WMAP7}.
The direction of the elongation of the contours from the WMAP7 results are almost perpendicular to that from our DLS cosmic shear results.
In previous studies combining cosmic shear with the CMB data, the constraint from the latter is in general so strong that cosmic shear results in fact
add only a small contribution to the joint constraint. However, note that the width of the ``banana'' in the current study
is narrower than the major axis of the WMAP7 ellipse.
Using the same settings for the priors as in our DLS only experiment, we obtain  
$\Omega_M=0.278\pm0.018$ and $\sigma_8=0.815\pm0.020$ from the joint analysis with WMAP7.

\section{SUMMARY AND CONCLUSIONS} \label{section_conclusion}
We have presented a cosmic shear study from the DLS, which is the deepest wide-field ($>10$  sq. deg) cosmic shear survey to date.
Because of the non-trivial aberration pattern of the telescope optics, the PSF modeling for the stacked image was
challenging. We overcome the difficulty by modeling the PSF exposure to exposure and CCD to CCD with the PCA method, and
stack the PSFs by closely mimicking the image stacking procedure. 
We found that our initial PSF model had to be tweaked in two steps to remove the PSF-induced anisotropy completely from galaxy images.
The first tweak was required to improve the agreement between the PSF and data PSFs, and the second
tweak removed the residual centroid bias.

Using the star-galaxy ellipticity correlation and the B-mode studies, we 
demonstrate that the systematics in our cosmic shear signal are under control, which is an important verification before we utilize the cosmic shear
measurement on large scales where the statistical errors are very small. Other systematic tests of this shape catalog using galaxy-galaxy lensing are described
in Choi et al. (2012).

Our photometric redshifts are calibrated using $>10,000$ spectroscopic redshifts in our DLS fields. We employ the photometric
redshift probability distribution $p(z)$ instead of single-point best estimates. We show that the use of $p(z)$ mitigates the
systematics caused by source galaxies with multi-peaked, broad, or skewed photometric redshift probability distributions.

The shear multiplicative factor was obtained based on realistic image simulations. We experimented with both real and artificial galaxy images to
derive calibration parameters and considered the sample variance in galaxy population. We found that the dilution of the signal
increases with magnitude, and this determines the number of available galaxies for lensing analysis.

For cosmological parameter estimation, we marginalize over shear calibration error, photometric redshift uncertainty, and the  Hubble constant.
In addition, we use semi-cosmology-dependent covariances to avoid possible bias in cosmological parameter estimation when constant
covariances at a fiducial cosmology are assumed.
We obtained an unprecedentedly tight joint constraint on $\Omega_M$ and $\sigma_8$.
Without relying on other cosmological measurements,  our DLS analysis alone yields $\Omega_M=0.265\pm0.041$ and $\sigma_8=0.865\pm0.077$, which are  consistent
with the WMAP7 results.
However, we find that when we use a cosmology-independent covariance matrix instead, the resulting parameter contours shift
substantially from the results obtained from the use of the cosmology-dependent
covariance. This confirms our belief that for a sample-variance-limited cosmic shear survey, it is critical to implement cosmology-dependence in the covariance estimation.

Combining the current results with the WMAP7 likelihood data in a joint analysis, we obtain $\Omega_M=0.276\pm0.016$ and $\sigma_8=0.816\pm0.019$.
Since our current cosmic shear study is non-tomographic, we expect that our future tomographic study will improve this constraint.

\acknowledgments
{
\noindent
{\bf ACKNOWLEDGEMENTS}

M. James Jee acknowledges support for the current research from the TABASGO foundation in the form of the Large Synoptic Survey Telescope Cosmology Fellowship.
We thank Russell Ryan and Ami Choi for useful discussions. We thank Perry Gee for relentless efforts to carefully manage the DLS database.
The development of the StackFit algorithm and its application to the DLS was funded in part by Department of Energy (DOE) grant DE-FG02-07ER41505. The DLS 
and our systematics reduction R\&D have received major funding from Lucent Technologies and from the NSF
(grants AST-0134753, AST0441072, AST-1108893, and AST-0708433).  
Stefan Hilbert acknowledges support by the NSF grant number AST-0807458-002.
Part of this work performed under the auspices of the U.S. Department of Energy by Lawrence Livermore National Laboratory under Contract DE-AC52-07NA27344.
This work is based on observations at Kitt Peak National
Observatory and Cerro Tololo Inter-American Observatory, which are operated by the Association of Universities for Research in Astronomy (AURA) under a cooperative agreement with the National
Science Foundation.}

\end{document}